\documentclass[traditabstract]{aa} 
\usepackage{graphicx}
\usepackage{amsmath,amsfonts}
\usepackage{txfonts}
\usepackage[authoryear]{natbib}
\bibpunct{(}{)}{;}{a}{}{,}
\usepackage{color}
\usepackage{flafter}
\begin{document}
   \title{An Anisotropic Propagation Model for Galactic Cosmic Rays}

   \author{I. Gebauer
          \inst{1}
          \and
          W. de Boer\inst{1}
          }

   \institute{Institut f\"{u}r Experimentelle Kernphysik, Karlsruhe Institute of Technology, P.O. Box 6980, 76128 Karlsruhe, Germany\\
  \email{[gebauer;deboer]@kit.edu}
}

 \date{\it Submitted to A\&A}


  \abstract{
 Isotropic diffusion models for Galactic cosmic ray transport put tight constraints on the maximum convection velocity in the halo. For a half halo height of 4 kpc the maximum convection speed is limited to 40 km/s in the halo, since otherwise the constraints from local secondary to primary ratios and radioactive isotopes cannot be met. The ROSAT Galactic wind observations of wind speeds up to 760 km/s therefore constitute a problem for diffusion models.\\
It is shown that such wind speeds are possible, if the diffusion coefficient in the halo is different from the diffusion coefficient in the disk. The radial dependence of the wind velocity was taken to be proportional to the source strength, as expected from winds which are sustained by cosmic ray pressure. In this case the cosmic ray density and with it the diffuse $\gamma$-ray production from nuclear interactions are suppressed near the sources. This solves in a natural way the problem of the soft gradient in the radial dependence of the $\gamma$-ray flux.
Furthermore, the large bulge over disk ratio in positron annihilation as observed by INTEGRAL, can be explained by positron escape from the disk in such a model.}


   \maketitle
%

\section{Introduction}\label{s_introduct}
High energy cosmic rays (CRs) are thought to be accelerated in the shock waves of supernovae remnants (SNRs), as can be found in standard text books \citep{Ginzburg:1990sk,Longair,Schlickeiser:2002pg}, but critical reviews exist \citep[see e.g. ][]{kirk:2001}. Although CRs are accelerated to almost the speed of light they are confined to our Galaxy for about $10^7$ yrs, as is known from the presence of radioactive nuclei with decay times of a few times $10^6$ yrs,
see e.g. \citep{Cesarsky:1980}. Such ''cosmic clocks'' require averaged drift speeds of CRs of only  a fraction of the speed of light. This is possible if CRs continuously scatter on magnetic turbulences \citep{Ginzburg:1990sk} and as a consequence perform a random walk which can be modeled by diffusion. Possible scattering centers are the turbulent magnetic fields generated by the dilute plasma of CRs themselves. In the simplest case diffusion is assumed to be isotropic with the same diffusion coefficient in the halo and the disk \citep{Strong:2007nh}. The parameters of the diffusion equation can be obtained by taking a source distribution proportional to the SNR distribution and measuring the local density and energy spectra of the CRs, which depend on the transport parameters, gas densities and magnetic fields between the source and the solar system.

An isotropic diffusion model has been implemented in the publicly available GALPROP code\footnote{The GALPROP code is publicly available under http://galprop.stanford.edu}  \citep{Strong:1998pw, Moskalenko:1998id}. Details about the underlying propagation and the parameter determination can be found in a review by \citet{Strong:2007nh}.

In addition to diffusion convective transport modes may play a role for CRs \citep{Jokipii:1976jk}. Supernovae (SNs) eject hot gas into the interstellar medium (ISM) which can expand into the halo, presumably driven by the CR pressure from supernovae remnants (SNRs)\citep{Breitschwerdt:2008na}. This  leads to a reduction of the CR density and CR interaction rate in the Galactic disk, most efficiently at radii with high source density. This mechanism has been proposed as an explanation for the relatively small  production of diffuse $\gamma$-rays near the sources in comparison to the $\gamma$-ray production at larger radii \citep{Breitschwerdt:2002vs}, known as the ''soft $\gamma$-ray gradient (SGRG) problem''. \\
Unfortunately, the maximum allowed convection speed in isotropic propagation models is restricted to a few tens of km/s \citep{Strong:2007nh}, because otherwise the constraints from local secondary to primary ratios
and radioactive isotopes cannot be met.
Wind velocities as small as this will not solve the soft-$\gamma$-ray-gradient problem. Within the isotropic diffusion models alternative solutions have been investigated; e.g. putting more sources in the outer Galaxy \citep{Strong:1998pw} or putting more gas there by assuming that the tracer for molecular hydrogen, the radiation from CO molecules, has a different proportionality factor in the outer Galaxy \citep{Strong:2004td}, as motivated by the change in metallicity.
A recent analysis of  the ROSAT data on x-rays implies wind speeds of up to 760 km/s in the halo \citep{Everett:2007dw,Breitschwerdt:2008na}. Speeds like this are sufficient to solve the SGRG problem, but they would require to loosen the constraint of isotropic diffusion.\\

An additional problem for isotropic diffusion models, is the large bulge-over-disk (B/D) ratio of the 511 MeV positron annihilation line, as observed by the INTEGRAL satellite \citep{Knodlseder:2005yq,Weidenspointner:2007rs}. These low energy positrons are thought to originate predominantly from the decay of radioactive nuclei, produced in SNIa explosions in the bulge and in the disk. Predictions of the B/D ratio from the expected number of SN explosions are well below one, because of the high rate of SNIa explosions in the disk \citep{Prantzos:2005pz}. However, INTEGRAL found a B/D ratio of a few. In a diffusion model without convection MeV positrons hardly propagate, because diffusion is proportional to the velocity and energy of the particle. In this case positrons annihilate close to their sources leading to a small B/D ratio. Convective transport in Galactic winds is independent of energy, so low energy positrons in the disk can be convected to the halo, where there are hardly electrons at rest to annihilate with. In this case the escape fraction of  positrons from the disk can be sufficient to explain the observed large B/D ratio.

      In this paper an anisotropic transport model, which allows for ROSAT compatible convection and still meets all the constraints from primary and secondary cosmic rays, cosmic clocks, $\gamma$-rays and the INTEGRAL B/D ratio, is introduced. The model features different diffusion coefficients in the halo and in the disk. By increasing the diffusion coefficient towards the halo boundary a smooth transition to free escape is obtained, so the model becomes
      insensitive to the precise position of the boundary. This is in strong contrast to isotropic propagation models, which require a precise size of the halo for a given diffusion constant.

The paper is organized as follows: in section \ref{s_iso} we summarize
the main features of the current isotropic diffusion models for CR
transport. Section \ref{s_limits} discusses the limits of the isotropic transport models in the light of observational constraints which refine our picture of CR transport.
In section \ref{s_aniso} we introduce the basic ideas of an anisotropic propagation model (aPM).
Section \ref{s_perf} discusses the parameter determination of the aPM.
The performance of the aPM with respect to charged CRs and diffuse
Galactic $\gamma$-rays is discussed in sections \ref{s_perf_CR} and \ref{s_perf_gamma}, respectively.
Section \ref{s_conc} summarizes the results.


\section{Isotropic propagation models}{\label{s_iso}}
\begin{figure*}
\includegraphics[width=0.5\textwidth,clip]{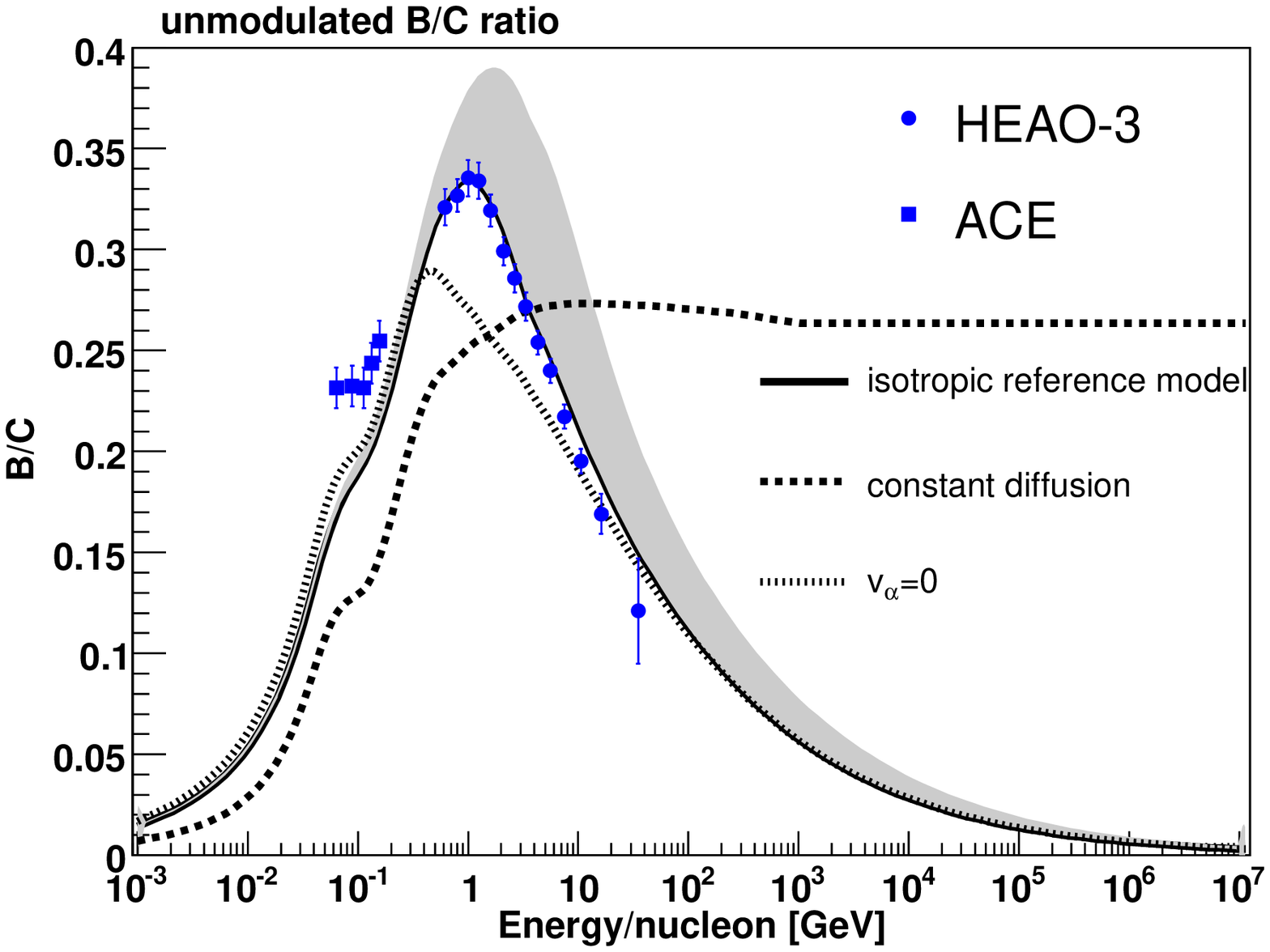}
\includegraphics[width=0.5\textwidth,clip]{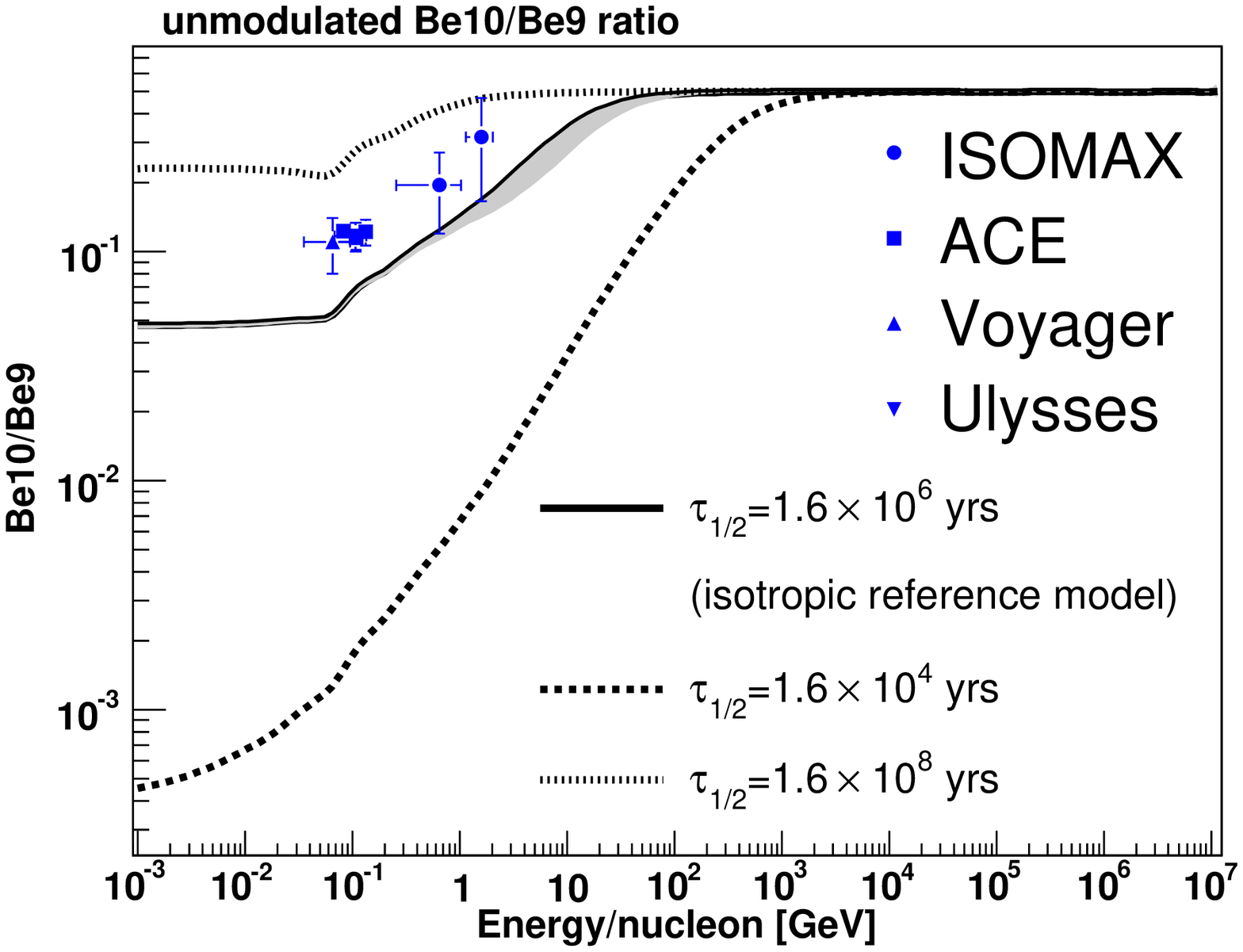}
\caption{{\bf Left} Demodulated local B/C ratio for different variations of an isotropic model: isotropic reference model (full line), isotropic model with constant (energy independent) diffusion (dashed line), isotropic model without diffusive reacceleration (dotted line). For energy independent diffusion (which means the escape time does not depend on energy), the $B/C$ ratio is independent of energy once the nuclei become relativistic. Setting the Alfv\'en velocity $v_\alpha$ to zero eliminates diffusive reacceleration, which  therefore shifts the peak in the $B/C$ ratio to lower energies. The grey band refers to a run with $z_h=5.3$ kpc. {\bf Right} Demodulated local $^{10}Be/^9Be$ ratio for different lifetimes of $^{10}Be$: $\tau_{1/2}=1.6 \cdot 10^6~\mathrm{yrs}$ (full line),
 $\tau_{1/2}=1.6 \cdot 10^4~\mathrm{yrs}$ (dashed line),
 $\tau_{1/2}=1.6 \cdot 10^8~\mathrm{yrs}$ (dotted line). The grey band refers to a run with $z_h=5.3$ kpc.}\label{f_BC_BE_iso}
\end{figure*}
CRs are thought to be accelerated in shock waves of supernovae explosions up to velocities close to the speed of light or energies up to  $10^{15}$ eV. However, they do not escape with such high speeds, but are trapped in the
magnetic field of the Galaxy for a time of the order of $10^7$ years. During this time they diffuse randomly through
the Galaxy and cross the disk many times. During each crossing secondary CRs, like B, Be, $\gamma$ rays, positrons, antiprotons and others can be produced. More details can be found in the review by \citet{Strong:2007nh} and references therein.

The main features of our Galaxy are a barred central bulge with a diameter of a few
kpc and a large spiral disk with a radius of about 15 kpc and an  density falling  exponentially in R with a scale length of about 2.5 kpc and in $z$ with a scale height  of  about 0.25 kpc. Most of the gas is distributed in the
disk with a broad maximum between $R=$4 kpc and 6 kpc for molecular hydrogen, while the distribution of ionized hydrogen is nearly constant between $R=$4 kpc and 13 kpc. The supernovae remnants (SNR) are also distributed in the disk, but
peak at a distance of a few kpc from the center with a slow fall off to larger
radii \citep{Case:1998qg}, as will be shown later. The CRs form a plasma of ionized particles, in which the
electric fields can be neglected by virtue of the high conductivity and the
magnetic fields form Alfv\'en waves, i.e. traveling oscillations of
ions and the magnetic field, where the ion mass density provides the
inertia and the magnetic field line tension provides the restoring
force.
If the wavelength of the Alfv\'en waves equals a multiple of a
particle gyroradius, resonant scattering occurs, leading to a change in
the CRs pitch angle without energy losses. Such a process leads
to a random walk, which can be described by a diffusion equation (see e.g. \citep{Ginzburg:1990sk,Longair,Schlickeiser:2002pg})
\begin{equation}
\frac{\partial f}{\partial t}=\nabla_i D_{ij}\nabla_j f -(\vec{u}\nabla)f+\frac{p}{3}\nabla\cdot \vec{u}\frac{\partial f}{\partial p}+\frac{1}{p^2}\frac{\partial}{\partial p}p^2 D_{pp} \frac{\partial f}{\partial p},\label{e_trans}
\end{equation}
where $f(\vec{r},\vec{p},t)$ is the CR phase space density, $D_{ij}$ are the components of the diffusion tensor for spatial diffusion, $D_{pp}$ the diffusion coefficient in momentum space and $\vec{u}$ is the convection velocity. Convective transport is possible either by a large scale motion of the interstellar medium with velocity $\vec{V}$ or by the effective velocity of the Alfv\'en waves $\vec{u_w}$. Assuming that the energy density in Alfv\'en waves propagating in opposite directions is the same $\vec{u_w}$ vanishes and $\vec{u}=\vec{V}$.
In addition to spatial diffusion, diffusion in momentum space can be caused by CRs scattering on moving Alfv\'en waves, which leads to diffusive reacceleration. One finds the corresponding diffusion coefficient to be $D_{pp}\propto p^2 v_\alpha^2 /D_{zz}$,  where the Alfv\'en velocity $v_\alpha$ is introduced
as a characteristic velocity of weak disturbances propagating in a magnetic field \citep{Strong:2007nh}.

Locally CR diffusion is highly anisotropic and occurs along the magnetic
field lines. However, if the B-field has no preferred direction, i.e. if
the turbulent small scale component ($\sim 100 \mathrm{pc}$) is larger than the regular large
scale component, CR diffusion is assumed to be isotropized by pitch angle
scattering on these turbulences. This is in agreement with the fact that the flux of CRs is the same from all directions
to a high degree.\\
It is usually assumed that the steady state condition is reached, i.e. ${\partial f}/{\partial t}=0$,
which implies that the injection rate of CRs by sources equals the loss rate.  CRs can be lost either by fragmentation, decay or escape from the Galaxy.
In the steady state case the diffusion equation for CRs
can be solved numerically for given boundary conditions. Usually one requires the CR density to become zero above a certain halo height.

Most primary nuclei show a power law spectrum falling with rigidity
like $E^{-2.54}$. This can be easily reproduced by selecting the injection spectrum of
the primary particles, the rigidity dependence of the diffusion
coefficient and the energy gains due to diffusive reacceleration
accordingly. During their journey CRs may
interact with e.g. the gas in the Galaxy and produce secondary particles.  This
changes the ratio of secondary/primary particles, like the $B/C$ ratio. From this
ratio one can determine that the amount of matter traversed (grammage) by a CR
during its residence time $t_{CR}$. Grammage is given by $\rho_g c t_{CR}$, where
$ct_{CR}$ is the path length for a particle traveling with the speed of light $c$ and $\rho_g$ is the gas density.
It was found to be of the order of $ 10~\mathrm{g/cm}^{-2}$ \citep{Schlickeiser:2002pg,Ginzburg:1990sk},
which corresponds to a density of about $0.2~\mathrm{atoms/cm}^{3}$. This is significantly
lower than the averaged density of the disk of about $1~\mathrm{atom/cm}^{-3}$. Under
the assumption of a homogeneous gas distribution this suggests
that CRs travel a large time in low density regions, like the halo. \\
For relativistic energies the inelastic cross sections for
secondary particle production usually do not strongly depend on the
e\-ner\-gy of the particle. This would lead to rather flat spectra for the secondary/primary ratios
in contrast to the observed $B/C$ ratio, which shows a maximum at about 1 GeV/nucleon
and decreases as $E^{-0.6}$ towards higher energy, as shown by the dashed line in the right hand side of Fig. \ref{f_BC_BE_iso}. This can be accommodated by
assuming that energetic particles diffuse faster out of the Galaxy, i.e. the
energy dependence of the diffusion coefficient and the energy gains due
to diffusive reacceleration are chosen accordingly (see the full line in the left hand side of Fig.\ref{f_BC_BE_iso}).
 The decrease  at low energies can be accommodated by both diffusive
reacceleration, which shifts the spectrum to higher
energies, as well as convective transport mechanisms \citep{Strong:2007nh}.
Alternatively, one could assume a strong increase of the diffusion coefficient at low
energies due to damping of the Alfv\'en waves \citep{Ptuskin:2005ax}.

From the ratio of unstable/stable secondary nuclei (like $^{10}Be/^9Be$) one obtains the average
residence time of CRs in the Galaxy to be of the order of $t_{CR}=10^7$
yrs \citep{Cesarsky:1980}. Fig. \ref{f_BC_BE_iso} shows the local $^{10}Be/^9Be$-fraction assuming different lifetimes of $^{10}Be$. The data require residence times between $1.6\cdot10^6$ and $1.6\cdot10^8$ years.

A widely-used program providing a numerical solution to the diffusion equation
is the publicly available GALPROP code \citep{GalpropMan}. The transport equation in GALPROP is not written in the form of a phase space density $f(\vec{r},\vec{p},t)$, but in the form of density per unit of total particle momentum $\Psi(\vec{r},p,t)$ defined by
\begin{equation}
\Psi(p)/dp=4\pi p^2 f(\vec{p})\label{e_form}.
\end{equation}
The third term in Eq.\ref{e_trans} then can be rewritten as follows
\begin{eqnarray}
\frac{1}{3}p(\vec{\nabla}\cdot \vec{u})\frac{\partial}{\partial p}\left(\frac{\Psi}{p^2}\right)=\frac{1}{3}p(\vec{\nabla}\cdot \vec{V})\frac{\partial}{\partial p}\left(\frac{1}{p^3}p\cdot\Psi\right)=&\nonumber\\
=-\frac{1}{p^2}(\vec{\nabla} \cdot\vec{V})\Psi+\frac{1}{3p^2}(\vec{\nabla}\cdot\vec{V})\frac{\partial}{\partial p}(p\Psi).&\label{e_trans2}
\end{eqnarray}
Using \ref{e_trans2} and \ref{e_form} and multiplying the transport equation \ref{e_trans} by $p^2$ one obtains
\begin{equation}
\frac{\partial \Psi}{\partial t}=\vec{\nabla}\cdot (D\vec{\nabla}\Psi-\vec{V}\Psi)+\frac{\partial}{\partial p}p^2D_{pp}\frac{\partial}{\partial p}\frac{1}{p^2}\Psi+\frac{1}{3}\frac{\partial}{\partial p}\left[\left(\vec{\nabla} \cdot \vec{V}\right)p\Psi\right],
\end{equation}
where it is assumed that the diffusion coefficient $D$ is a scalar quantity with the same value everywhere and in all directions, i.e. the tensor $D_{ij}$ has only diagonal components, which are all equal.
The full transport equation used in GALPROP can then be written as
\begin{eqnarray}
\frac{\partial \Psi}{\partial t}=q(\vec{r},t)+\vec{\nabla}\cdot (D\vec{\nabla}\Psi-\vec{V}\Psi)+\frac{\partial}{\partial p}p^2D_{pp}\frac{\partial}{\partial p}\frac{1}{p^2}\Psi-&\nonumber\\
-\frac{\partial}{\partial p}\left[\dot{p}\Psi-\frac{p}{3}\left(\vec{\nabla\cdot \vec{V}}\right)\Psi\right]-\frac{1}{\tau_f}\Psi-\frac{1}{\tau_r}\Psi,&\label{e_GP}
\end{eqnarray}
where $q(\vec{r},t)$ is the source term, $\tau_f$ is the time scale for fragmentation and $\tau_r$ is the time scale for radioactive decay.\\
 The basic
parameters for GALPROP are the injection spectrum parameters, the diffusion coefficient $D$,
the convection velocity $\vec{V}$, Alfv\'en velocity $v_{\alpha}$, which enters $D_{pp}$, and the size of the transport box, needed as a boundary to solve the differential equation.
 Usually one assumes the density of scattering centers outside the diffusion box to be so small
 that CRs propagate freely with the speed of light to outer space (free escape). Consequently the CR density becomes zero at the boundary. For an isotropic propagation model one has to assume that the density of scattering centers is constant inside the diffusion box and abruptly drops to zero outside. Of course a smooth decrease in the scattering centers would be a much more natural solution, but this requires an increase in diffusion coefficient towards the boundary. This is exactly what is needed in anisotropic propagation models, as will be discussed in Sect. \ref{s_aniso}.
  In such anisotropic models the boundary of the diffusion box can be in principle at infinity in contrast
  to isotropic models, where the residence time increases with the size of the diffusion box for a constant diffusion
  coefficient.

The propagation parameters are usually tuned to
the secondary/primary ratio and the unstable/stable ratio of locally observed charged particles.
The production of secondary and tertiary particles is calculated in GALPROP using a network with
more than 2000 cross sections. To check  the validity of such a self-consistent propagation model
at scales outside the kpc-scale of the collection volume of charged particles the fluxes of diffuse
gamma rays are calculated by GALPROP using the emissivity of the complete diffusion box. The absorption of $\gamma$-rays
in the GeV range is small, so information on the CR density and gas density even in the Galactic Center (GC) can be obtained.
The present GALPROP models with isotropic diffusion indeed describe the $\gamma$-ray fluxes remarkably well, if one ignores
 the anomalous excess above 1 GeV, as measured with the EGRET spectrometer on board of the CGO \citep{Hunter:1997apj}. This excess appears anyway to be not confirmed by the preliminary Fermi-LAT data \citep{Porter:2009sg,Portertalk:2009}.
Note, that the motivation to drop the assumption of isotropic propagation is entirely independent of the EGRET observations.


\section{Limits of the isotropic propagation model with new observational constraints}\label{s_limits}

In isotropic transport models propagation is homogeneous and dominated by isotropic pitch-angle scattering, in particular the scattering rate is assumed to be the same in the halo and in the disk. In the following we will discuss three of the main observations that cannot be explained in a straightforward way by isotropic transport: the Galactic winds as observed by ROSAT, the soft $\gamma$-ray gradient as observed by COS-B and EGRET and the large bulge/disk ratio observed by INTEGRAL.

\subsection{ROSAT: Galactic Winds}\label{ss_wind}
It has been pointed out that diffusion models are incompatible with vertical wind velocity gradients larger than 10 km/s/kpc \citep{Strong:2007nh}, because for larger wind velocities the CRs are blown into the halo and the required ratio of times spent in the halo and the disc   from the combined constraints from the local secondary to primary ratios and the radioactive cosmic clocks cannot be met.

Until recently there was no direct observational
evidence that the Milky Way's atmosphere might feature such a wind. It has been suggested that the Milky Way's SN rate is too small to build up sufficient CR pressure to overcome the Galactic gravitational potential. In so far, the fact that current transport models are incompatible with significant convective transport modes has not been considered a serious deficiency.
This changed in 2007 when it was found that the Milky Way drives a large scale wind, too: the ROSAT-satellite observed an enhancement of the diffuse soft X-ray background
emission \citep{Levenson:1997pp}.
This emission can be explained
best by a mixed CR and thermally-driven wind model \citep{Everett:2007dw}.
Their model is based on a model by \citet{Breitschwerdt:2002vs}, where the spatial shape of the wind velocity is given by the balance of the gravitational potential and CR pressure and basically follows the SNR distribution. The fitted velocities range from 173 km/s in the disk to 760 km/s in the halo. Compared to starburst Galaxies which feature wind velocities up to 3000 km/s these wind velocities are still moderate. However, the impact of even moderate convection velocities of a few 100 km/s on CR transport is significant as we will discuss in subsection \ref{ss_BC}.
Wind velocities which follow the SNR distribution will lead to an $R$-dependent
diffusion-convection boundary, thus significantly changing the R-dependence of the escape probability of CRs. In regions with a high density of CR sources the CR pressure is strong enough to overcome the gravitational forces and drives gas into the halo.
CRs driven by this wind will leave the Galaxy earlier and consequently the interaction rate will
be smaller than e.g. in the outer Galaxy. In this case grammage and escape time
depend not only particle rigidity, but become a function of the Galactocentric radius.

Following \citet{Jokipii:1976jk}  the convection-diffusion boundary $z_c$ can be defined by the parameter
$z_c(R)\sim D_{zz}(R,z_c)/V_c(R,z_c)$. Above this boundary convection dominates and the probability for a CR to return to the diffusion region below $z_c$ decreases exponentially. The $R$ and $z$-dependence of the parameters have been explicitly included. For the parameters of the aPM model including the ROSAT compatible convection, as determined in Sect. \ref{s_perf},
the R-dependence of $z_c$ is plotted in Fig. \ref{f_source_zc} for different rigidities. Clearly, for most radii the regions of confinement where diffusion dominates, i.e.  below the curves, are much smaller than the halo boundary, set at  7.5 kpc. Convection was taken to be proportional to the source distribution from \citet{Case:1998qg} (plotted in Fig. \ref{f_source_zc}. As a result the region where convection dominates peaks at $R\approx 4 $kpc for high rigidities and becomes much broader for low energies, since then  the ratio of diffusion $\propto \beta E^{0.33}$ over convection (independent of energy) becomes small. Note that near the GC the convection becomes small
by virtue of the decrease in the source distribution in our case. But such a strong decrease is also expected from first principles, because of the strong gravitational potential in the GC, which will inhibit the launching of Galactic winds.
\subsection{Soft $\gamma$-Ray Gradients }\label{ss_gradients}
\begin{figure*}
\includegraphics[width=0.45\textwidth,clip]{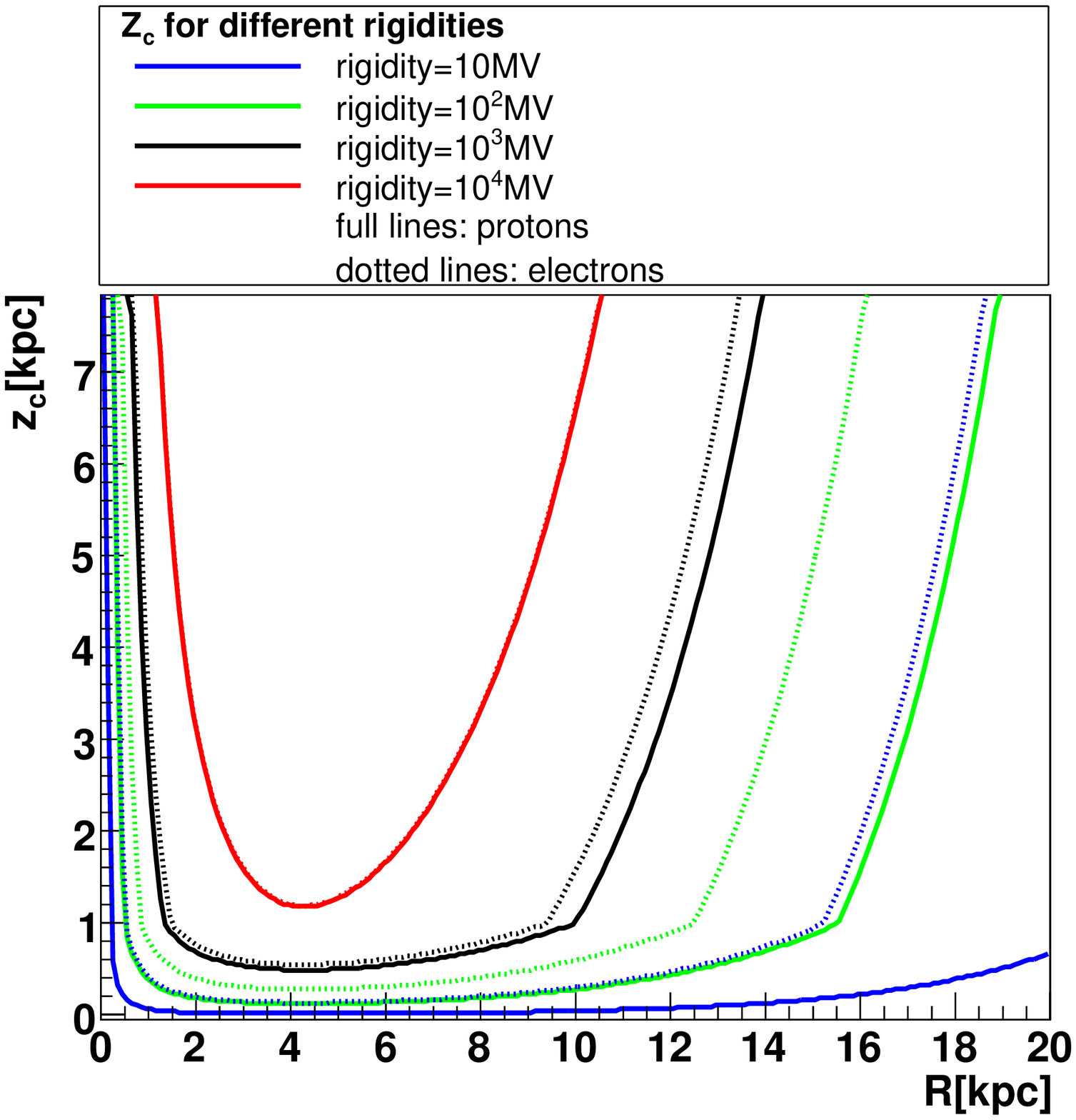}
\includegraphics[width=0.5\textwidth,clip]{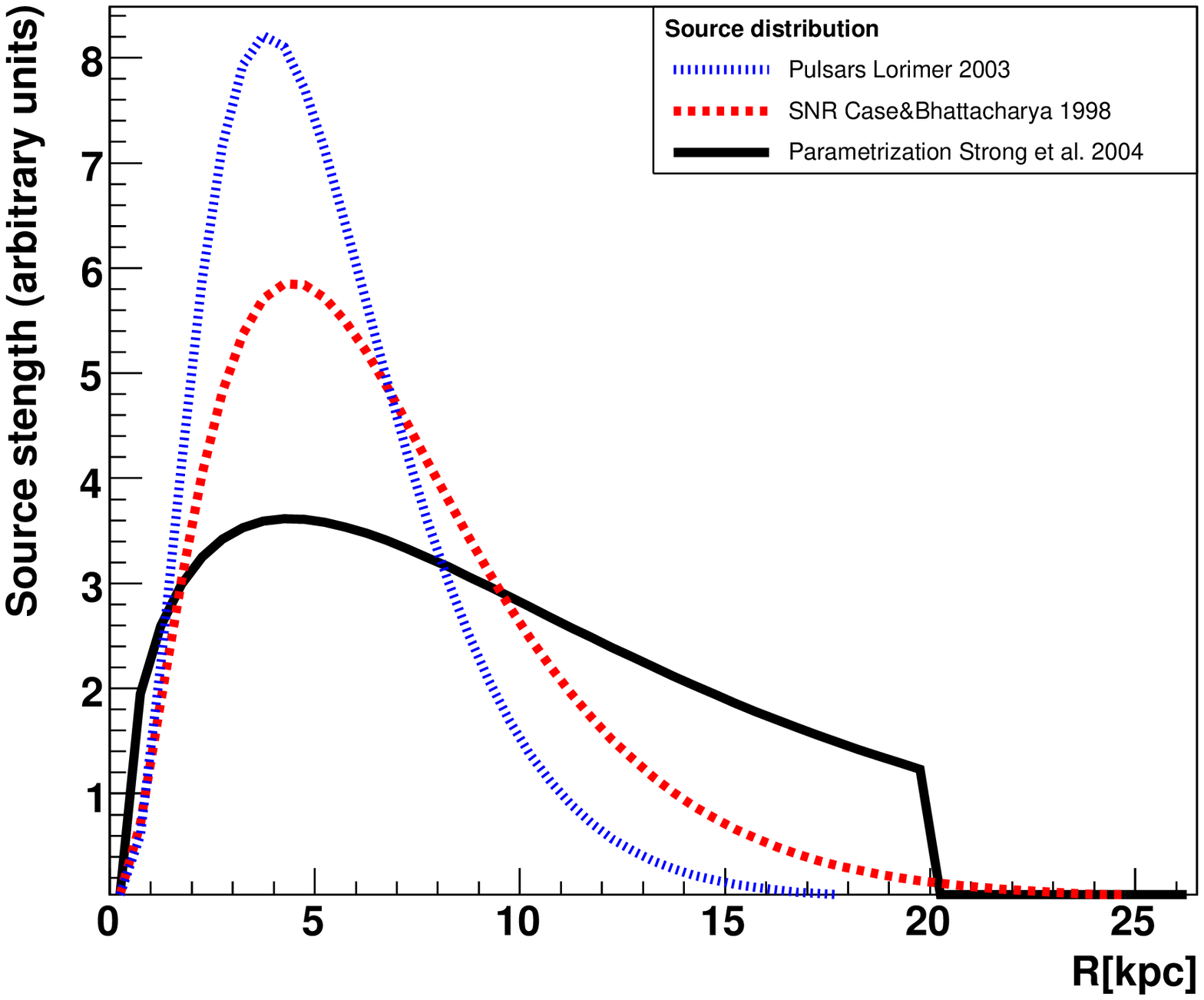}
\caption{ {\bf Left} Convection diffusion boundary as defined by $D_{zz}(R,z_c)=V_z(R,z_c)\cdot z_c$ in an aPM for different particle rigidities. The  {\it full lines} correspond to protons, the  {\it dotted lines} correspond to electrons. For rigidities larger than $10^4 MV$ $D_{zz}$ and therefore $z_c$ is the same for protons and electrons. The kink at 1kpc above the plane is caused by the sudden increase of the diffusion coefficient in this model (see section \ref{s_aniso} for details). {\bf Right} Source distribution: pulsar distribution \citep{Lorimer:2003qc} ( {\it blue, dotted}), SNR distribution \citep{Case:1998qg} ( {\it red, dashed}) and the flattened parameterization used by \citet{Strong:1998pw} ( {\it black full}).}\label{f_source_zc}
\end{figure*}
It is interesting to note that the model suggested by Breitschwerdt was not motivated by direct evidence for Galactic winds, but by another problem of isotropic transport models:
The distribution of supernova remnants (SNR), which are believed to be
the sources of CRs, peaks toward the
Galactic center, as shown in Fig. \ref{f_source_zc}. In an isotropic diffusion model the
propagated CR distribution still strongly resembles the source
distribution, leading to a peak in the radial distribution of diffuse $\gamma$-rays, i.e. one observes a strong gradient. This is incompatible with the soft $\gamma$-ray gradient as observed by COS-B and EGRET\footnote{Note, that this $\gamma$-ray gradient refers to a spatial feature of the diffuse $\gamma$-rays. A possible miscalibration of the EGRET instrument will not modify the spatial shape of the observed radiation.}.
 A solution to this SGRG-problem would be to reduce the residence time and the corresponding CR density near the sources, which is exactly what convection does, as spelled out by \citet{Breitschwerdt:2002vs}.
 This is indeed a more natural solution than an ad-hoc broadening of the source distribution, shown by black full line in Fig. \ref{f_source_zc}, as proposed by \citet{Strong:1998pw} or by increasing the gas density in the outer Galaxy by assuming a strong radial dependence of the proportionality constant $X_{CO}$ of the CO tracer for molecular hydrogen \citep{Strong:2004td}. Although an increase of $X_{CO}$ with Galactocentric radius is expected from the observed metallicity gradient, an order of magnitude increase in $X_{CO}$ seems  large and is not supported by
  all data \citep{Strong:2004td}.
In order to examine the impact of Galactic winds on CR transport, we assume a constant $X_{CO}$ factor  as a first step and allow for an increase with Galactocentric radius only if required.

\subsection{INTEGRAL: large bulge/disk ratio}\label{ss_inhom}

Low energy positrons can annihilate with electrons, preferentially bound to nuclei in order to prevent a
large momentum transfer during the collision. A detailed account of the annihilation process was given by
\citet{Guessoum:2005cb}.
Such positrons largely originate from the decay of
radioactive nuclei expelled by dying stars, especially SNIa. In the case
of SNIa the SNR
core makes up a large fraction of the mass, so it has a relatively thin layer of  ejecta, which    makes it easier for the positrons
to escape.

 Light curves, which are
sustained first by the $\gamma$-rays in the shock waves and later by the electrons and
positrons, suggest that only a few percent of the positrons escape from the ejecta
and can annihilate outside after thermalization. Positrons annihilating inside the
ejecta will also produce $\gamma$-rays, but these will not be visible as an annihilation line
 due to the successive interactions in the shock wave.

INTEGRAL found a  very strong  511 keV  positron annihilation line towards the Galactic center corresponding
 to an annihilation rate of $(1.5\pm0.1)\cdot 10^{43} s^{-1}$ \citep{Knodlseder:2005yq}. In contrast, the annihilation signal from the bulge was very weak. A  bulge/disk (B/D) ratio of a few was quoted, although
additional data show a clear disk signal as well \citep{Weidenspointner:2007rs}.
Taking the dominant source to be $\beta^{+}$ decay of $^{56}Co$ in SNIa, one would expect a B/D ratio to be well
below one, because of the higher mass in the disk and the higher rate of SNIa
explosions expected in the thick disk as compared to the bulge \citep{Prantzos:2005pz}. An
additional problem presents the observation of the 1.8 MeV line from the $^{26}Al$
radioactive isotope, which was clearly observed in the bulge  {\it and} the disk by
the Comptel detector on NASA's CGRO observatory \citep{Diehl:2006cf}. These nuclei are
thought to be produced by nucleosynthesis in massive stars and yield in their decay
on average 0.85 positrons. The observed flux of positron annihilation in the disk seems to be
saturated already by the positrons from $^{26}Al$ and $^{44}Ti$, leaving little room for additional positrons from SNIa explosions in the disk. Remind that $^{26}Al$ has a half life of the order of $7.2\cdot10^5$ years, so the positrons from their decay are not
convected away by the wind from the dying star.
In isotropic transport models this result cannot be understood, because the $\sim$ MeV positrons do not propagate ($D_{zz}=D_{RR}\propto \beta \cdot \rho^{\delta}$), so that positrons annihilate close to their sources in the bulge and in the disk.
In order to explain the observed B/D ratio one has to assume an addition positron population. These positrons have to be confined to the bulge in oder to reproduce the observed B/D ratio.
Recently a low-mass X-ray binary population showing the morphological features of the observed emission from the Galactic Center has been found \citep{Weidenspointner:2008zz}. However, the expected emission from these binaries is not sufficient to explain the complete signal from the Galactic Center and in particular it is not able to explain, why the emission observed from the bulge can be already attributed to $^{26}Al$.
\citet{Prantzos:2005pz} pointed out that B/D ratios as small as 0.5 are compatible with the INTEGRAL data if the disk positrons diffuse sufficiently away from their sources. He estimated that about 50\% of the $\sim \mathrm{MeV}$ positrons have to leave the confinement region below $z_C$ before slowing down.\\
In section \ref{ss_INT2} it will be shown that transport models including convection can explain the large B/D ratio in a natural way.\\


\section{Anisotropic propagation models}{\label{s_aniso}}
The publicly available GALPROP code was modified to allow for
anisotropic transport. The following changes were made:
\begin{itemize}
\item In GALPROP an equidistant grid is used to numerically solve the diffusion equation. However, for parameters  varying on small scales  a fine grid is required, which would dramatically increase the memory requirements and computing time. Therefore a non-equidistant user-defined spatial grid with arbitrary grid points was implemented to allow
for a course grid spacing in the halo with simultaneously a fine grid in the disk.
\item The isotropic diffusion coefficient is replaced by $D_{RR}$ for transport in the $R$ direction and $D_{zz}$ for transport in the $z$ direction. Both diffusion coefficients may depend on spatial coordinates and may have an independent energy dependence.
\item The convection velocity  depends on Galactocentric radius.
\item To model local gradients in transport parameters and gas density the user can specify two regions (possibly corresponding to the Local Bubble and the Local Fluff \citep{Frisch:2008vh}) for which the transport parameters or gas density may differ from the global parameters.
\item The spatial derivatives of the propagation parameters, which are needed  to solve the diffusion equation,  have been calculated for the additional $R$ and $z$ dependence of the diffusion and convection parameters.
    The corresponding Crank-Nicholson coefficients have been given in the Appendices \ref{a_conv} and
    \ref{a_diff}.
\end{itemize}

The parameterization of the spatial dependence of the convection and diffusion will be discussed in the following sections.

\subsection{ Parameterization for Convection} {\label{s_conv}}
\begin{figure*}
\includegraphics[width=0.45\textwidth,clip]{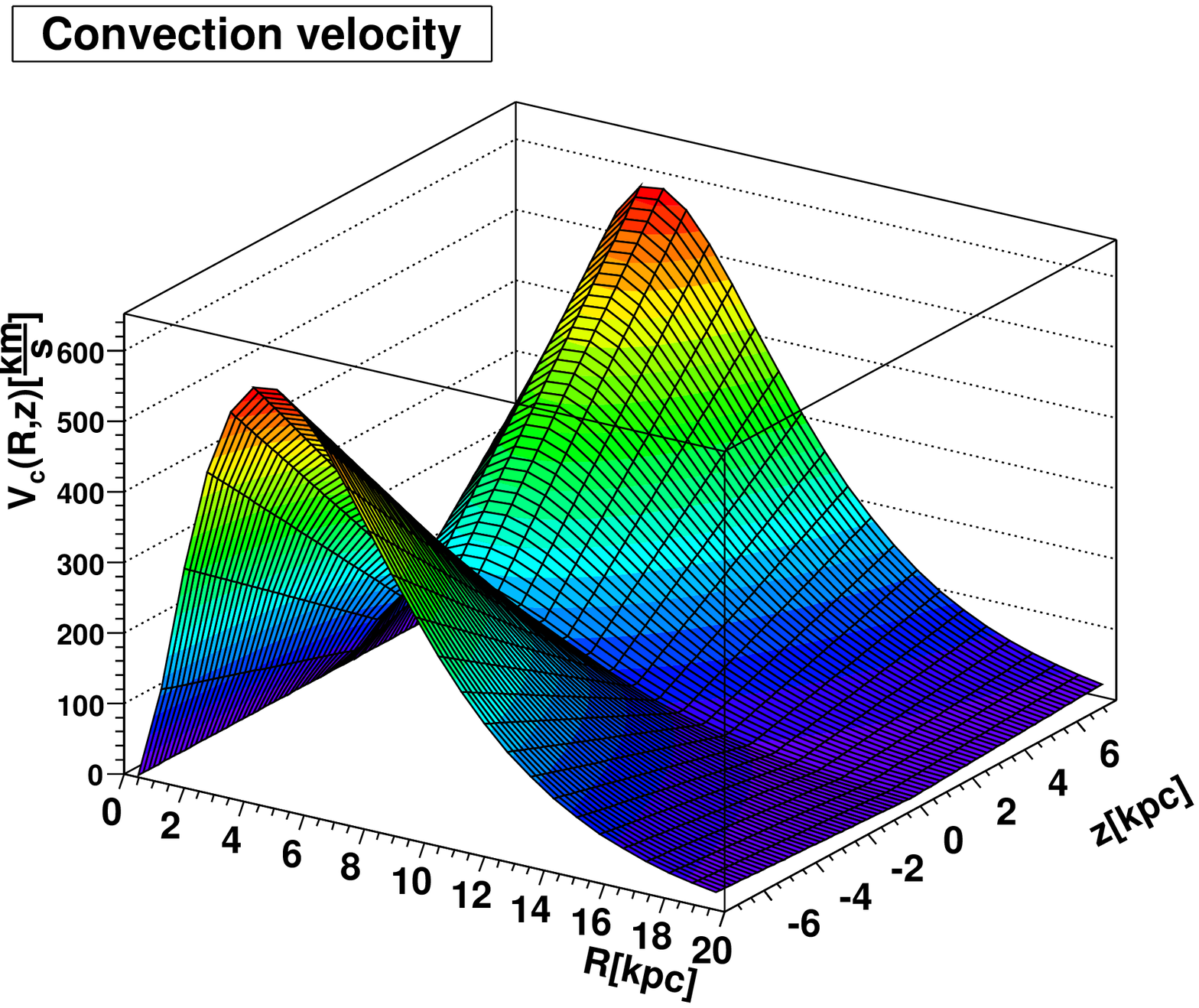}
\includegraphics[width=0.45\textwidth,clip]{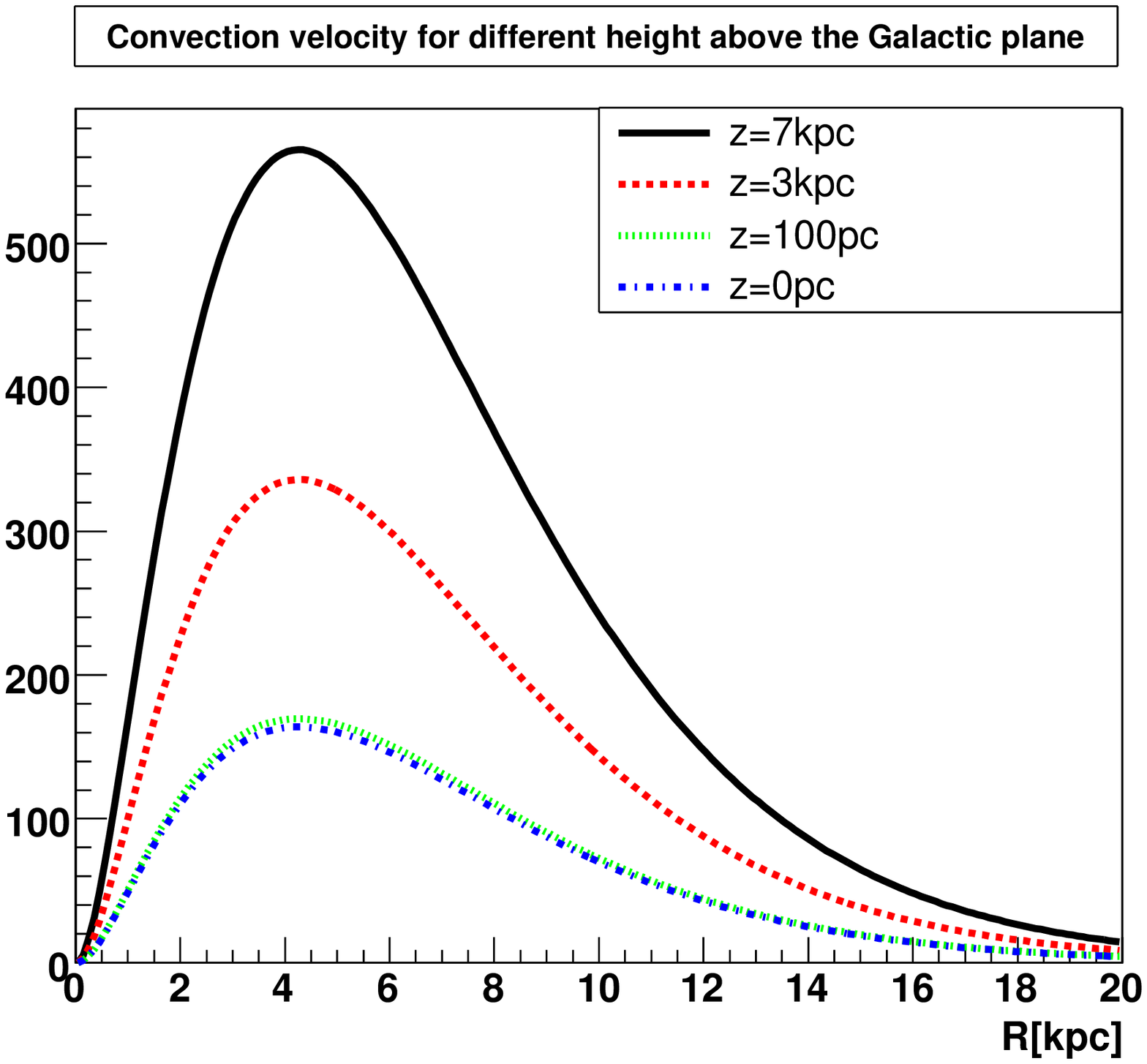}
\caption{{\bf Left} Convection velocity in an aPM.$V_{c}(R,z)= Q(R,0)(V_{0}+z(dV/dz))$, with $V_{0}=100~\mathrm{km/s}$ and ${dV}/{dz}=35~\mathrm{km/s}$ and $Q(R,0)$ given by the SNR distribution by \citet{Case:1998qg} (normalized to 1 at $R_o$). {\bf Right} Convection velocity for different distances from the Galactic plane. Below z=0.1 kpc only the contribution from $z{dV}/{dz}$ plays a role. The maximum wind velocity was chosen to be $591~\mathrm{km/s}$ (in good agreement with ROSAT).}\label{f_conv2D_conv}
\end{figure*}
 Convection was chosen to be proportional to
     the $R$-dependence of the source density to take care
of the increased CR pressure close to the maximum of the source
distribution. Note, that the exact proportionality is not a prediction of the model developed by \citet{Breitschwerdt:2002vs}. However, their predicted wind velocity is very similar to the source distribution. 
The convection velocity is parameterized as follows:
\begin{equation}
V_{c}(R,z)= Q(R,0)(\Theta(|z|-z_{0})V_{0}+\frac{dV}{dz}z)\label{e_conv}.
\end{equation}
Here $Q(R,0)$ is the $R$-dependence of the CR source distribution,
     $z_{0}$ is the initial
     height from where the wind is launched, $\Theta$ is the Heaviside step function, $V_0$ is the convection velocity at $z=z_0$ and  ${dV}/{dz}$ the gradient of the convection velocity. A linear increase in z was chosen for reasons of simplicity.
The source distribution was adopted from \citep{Case:1998qg}:
\begin{equation}
Q(R,z)=q_0 \left(\frac{R}{R_0}\right)^{\alpha} e^{-\beta \frac{(R-R_0)}{R_0}}e^{\frac{-|z|}{z_s}},\label{e_source}
\end{equation}
where $R_0$ is the Galactocentric radius of the sun, $\alpha=1.2$,
     $\beta=3.22$ and $z_s=2~\mathrm{kpc}$.\\
The ROSAT data indicate an initial velocity of 173km/s at the base of the wind \citep{Everett:2007dw}. Therefore  $z_0=0~\mathrm{kpc}$ with a local start velocity of 100km/s was chosen, which  results in a start velocity of about 170 km/s at the peak of the SNR distribution. $dV/dz$ was chosen to be 35 km/s/kpc leading to a maximum convection velocity of 591 km/s at $z_h=7.5 ~\mathrm{kpc}$ and 733 km/s at 16 kpc, which is in good agreement with ROSAT.
 The
spatial distribution of the convection velocity for these parameters is shown in Fig.
\ref{f_conv2D_conv}.

\subsection{ Parameterization for Diffusion} {\label{s_diff}}
With increasing distance from the plane the density of scattering centers will decrease until resonant scattering becomes negligible at the free escape boundary. Except for this increase in diffusion coefficients the spatial dependence of diffusion is not well constrained.
Here, a linear vertical increase in diffusion coefficients is chosen, since the CR halo density decreases in first order linearly towards the halo boundary; any possible R-dependence is neglected.
In the disk diffusion is assumed to be constant. In this model we use a rather large zone of constant diffusion with a half height of $z_d=1~\mathrm{kpc}$. The diffusion coefficients are parameterized as
\begin{eqnarray}
D_{zz}=D_{RR}=
\begin{cases}
\beta D_0 \left(\frac{\rho}{\rho_0}\right)^{\delta},& |z|<z_d\\
\beta D_0 \frac{|z|}{z_{d}}\left(\frac{\rho}{\rho_0}\right)^{\delta}, &|z|\geq z_d,
\end{cases}
\end{eqnarray}
where $\rho_{0}$ is the reference rigidity, $D_0$ is a proportionality constant, which is treated as a free parameter, $\delta$ is the slope of the power law describing the rigidity dependence of the diffusion coefficient and $\beta=v/c$ is the particle velocity. The parameter $\delta$ is taken to be the same for all rigidities, i.e. there is no break in the power law of the rigidity dependence. The scattering is locally assumed to be isotropic, i.e.  $ D_{zz}=D_{RR}$, but the scattering rate, or equivalently the mean free path, depends on the position in the halo.

In this paper we assume that the regular magnetic field plays no role, although this is clearly a questionable assumption.
In principle one would expect an anisotropy in diffusion coefficients,
     because even in the presence of strong scattering, the large scale
     component of the Galactic magnetic field is non-negligible \citep{DeMarco:2007eh, Codino:2007nm}.
     In the model presented here we choose $D_{zz}=D_{RR}$, which implies locally an isotropic scattering, but the overall scattering rate decreases towards the boundary by the positive gradient in $D_{zz}$ and $D_{xx}$. 
For a large enough gradient in convection the convection will dominate over diffusion above a certain $z$ value.
The distance from the disk above which convection dominates over diffusion can be estimated from
the convection-diffusion boundary, which has been shown before in Fig.  \ref{f_source_zc}
 for different rigidities using the parameters discussed in the following section.

\section{Parameter determination for the Anisotropic Propagation Model (aPM)}\label{s_perf}
\begin{table}
  \begin{center}
      \begin{tabular}{|c|c|c|}
\hline
Parameter& aPM&Conventional Model\\
        \hline
        \hline
  \multicolumn{3}{|c|}{Injection Spectra}\\
  \hline\hline
Protons/&&\\
nuclei&&\\
$\alpha_1/\alpha_2/\alpha_3$&1.6/1.8/2.41&1.98/2.42\\
$\rho^p_{1}/\rho^p_{2}$&2 GeV/9 GeV &9 GeV\\
\hline
Electrons&&\\
$\beta_1/\beta_2/\beta_3$& 1.6/2.54&1.6/2.54\\
$\rho^e_{1}/\rho^e_{2}$& 4 GeV&4 GeV\\
 &&\\

\multicolumn{3}{|c|}{Injection spectra are of the form}\\
\multicolumn{3}{|c|}{$\left(\frac{\rho}{\rho_i}\right)^{-(\alpha_i,\beta_i)}$}\\

\hline
\hline
 \multicolumn{3}{|c|}{Transport Parameters}\\
  \hline\hline
$D_{0}$&$5.3\cdot 10^{28}~\frac{\mathrm{cm}^2}{\mathrm{s}}$&$5.8\cdot 10^{28}~\frac{\mathrm{cm}^2}{s}$\\
$\rho_0$&4 GeV&4 GeV\\
$\delta$&0.33&0.33\\
$z_d$&1 kpc&-\\
\hline
$V_{0}$&$100~\frac{\mathrm{km}}{\mathrm{s}}$&-\\
$z_0$&0 kpc&-\\
$\frac{dV}{dz}$&$35~\frac{km}{s\cdot kpc}$&-\\
$V_c(R)$&$\propto Q(R)$&-\\

$v_{\alpha}$&$56\frac{km}{s}$&$30\frac{km}{s}$\\
$B_{0}$&$6.5~\mu\mathrm{G}$&$6.1~\mu\mathrm{G}$ \\
\hline
  \end{tabular}

    \caption{Parameters of the aPM and a conventional GALPROP model.}
   \label{table1}
\end{center}
  \end{table}

In this section we discuss the parameter tuning for the aPM proposed in section \ref{s_aniso}. It is not the aim of this study to present a fine-tuned best fit model, which anyway would be rather short-lived in the light of the upcoming Fermi-LAT and PAMELA data releases, but rather to show that the ROSAT Galactic winds in principle are compatible with local CR measurements.


 The optimization of the parameters follows the same path as for the isotropic model discussed in Sect. \ref{s_aniso},
 i.e. the diffusion
coefficients are chosen to best reproduce the local $B/C$ and $^{10}Be/^9Be$ ratio and the injection spectra for protons and electrons are chosen to fit the local proton and electron spectra.
The convection velocity parameters are taken from the ROSAT data, as discussed above in Sect. \ref{s_diff}.
The most important transport parameters for this model can be found in Table
\ref{table1} and will be discussed in more detail below.

\subsection{Injection spectra in an aPM}\label{ss_conv_aPM}
\begin{figure*}
\includegraphics[width=0.5\textwidth,clip]{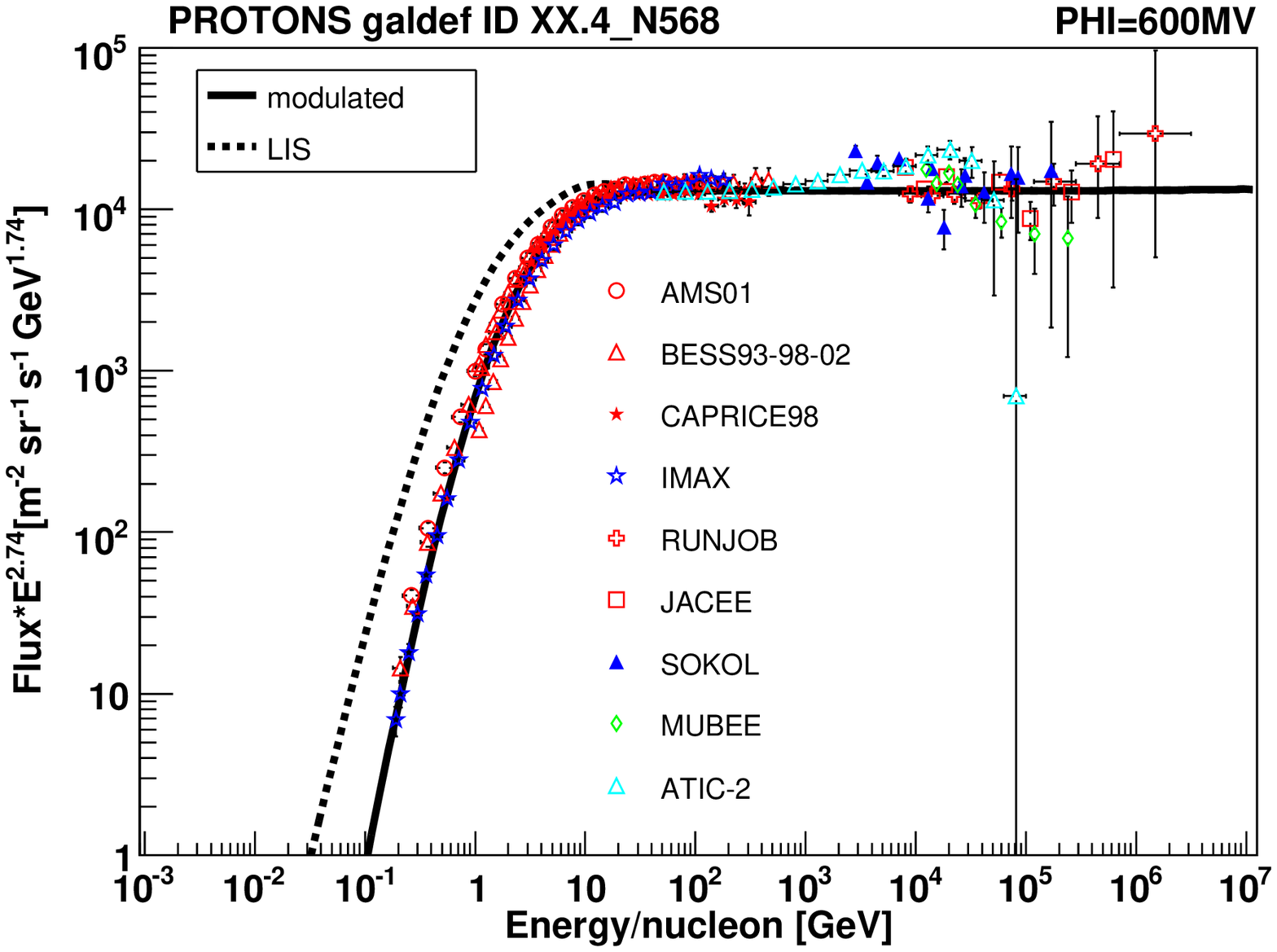}
\includegraphics[width=0.5\textwidth,clip]{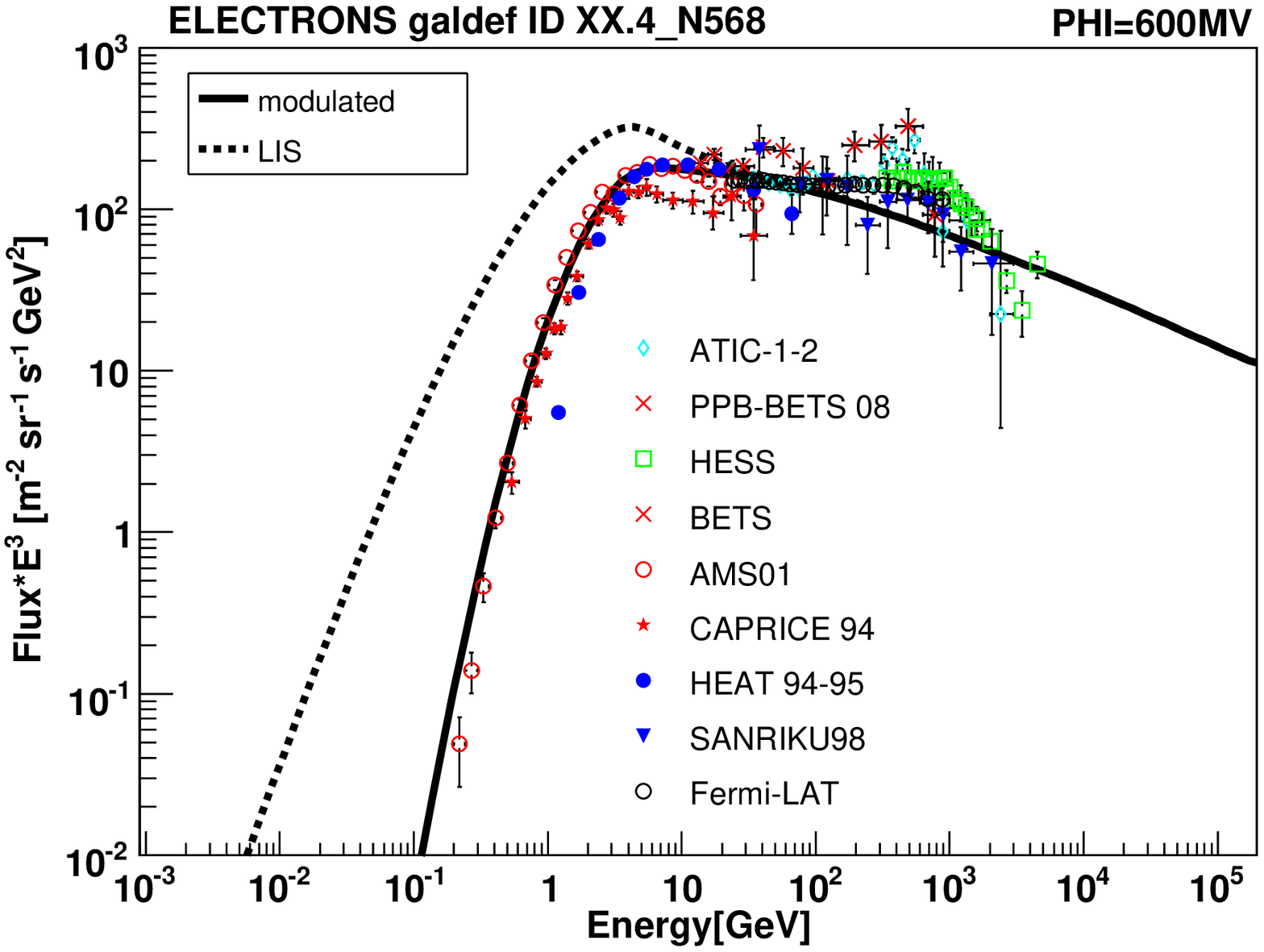}
\caption{{\bf Left} Proton flux in am aPM: local proton flux ( {\it full line}) and LIS proton flux ( {\it dashed line}). {\bf Right} Electron flux in a conventional aPM. Line coding as on the left. Data are taken from the CR database by \citet{Strong:2009xp}.}\label{f_conv_aPM_p_e}
\end{figure*}
Fig.  \ref{f_conv_aPM_p_e} shows the local and local interstellar proton (LIS) spectrum, where the local one is the observed spectrum and the LIS spectrum is the observed one  corrected for solar modulation.
 The correction for solar modulation was done in the force-field
approximation \citep{Gleeson:1968}, as implemented in the GALPLOT program \footnote{The GALPLOT
routine is available from http://www.mpe.mpg.de/$\sim$aws/propagate.html.}.

Fig.  \ref{f_conv_aPM_p_e} shows the local electron spectrum.
The electron injection index was chosen to give a propagated electron spectrum with an index of roughly 3.3. This yield a somewhat softer spectrum than   observed by Fermi \citep{abdo:181101}, but agrees with an extrapolation of the low energy electron data. The reasoning behind this is that for high energies local sources may contribute significantly to the local electron flux, because the large electron synchrotron losses cool electrons from distant sources efficiently. These sources are not included in our calculations, which means that one  {\it expects} our propagated electron spectrum to be somewhat softer than the data indicate.

\subsection{Diffusion in an aPM}\label{ss_BC}

\begin{figure*}
\includegraphics[width=0.5\textwidth,clip]{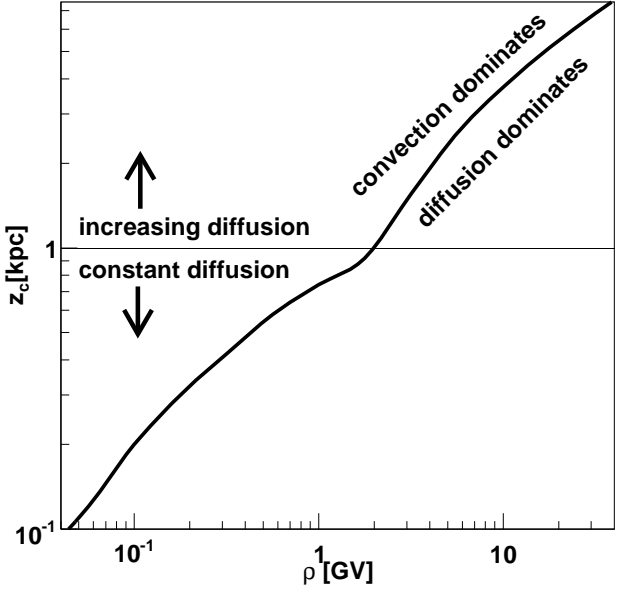}
\caption{ The rigidity dependence of the convection-diffusion boundary $z_{c}$ for $R=8.3$ kpc as defined by $D_{zz}(8.3~\mathrm{kpc},z_c,\rho)=V_c(8.3~\mathrm{kpc},z_c)\cdot z_c$. The change in slope at 1 kpc corresponds to the transition from constant diffusion to z-dependent diffusion.}\label{f_rhocrit}
\end{figure*}

\begin{figure*}
\includegraphics[width=0.5\textwidth,clip]{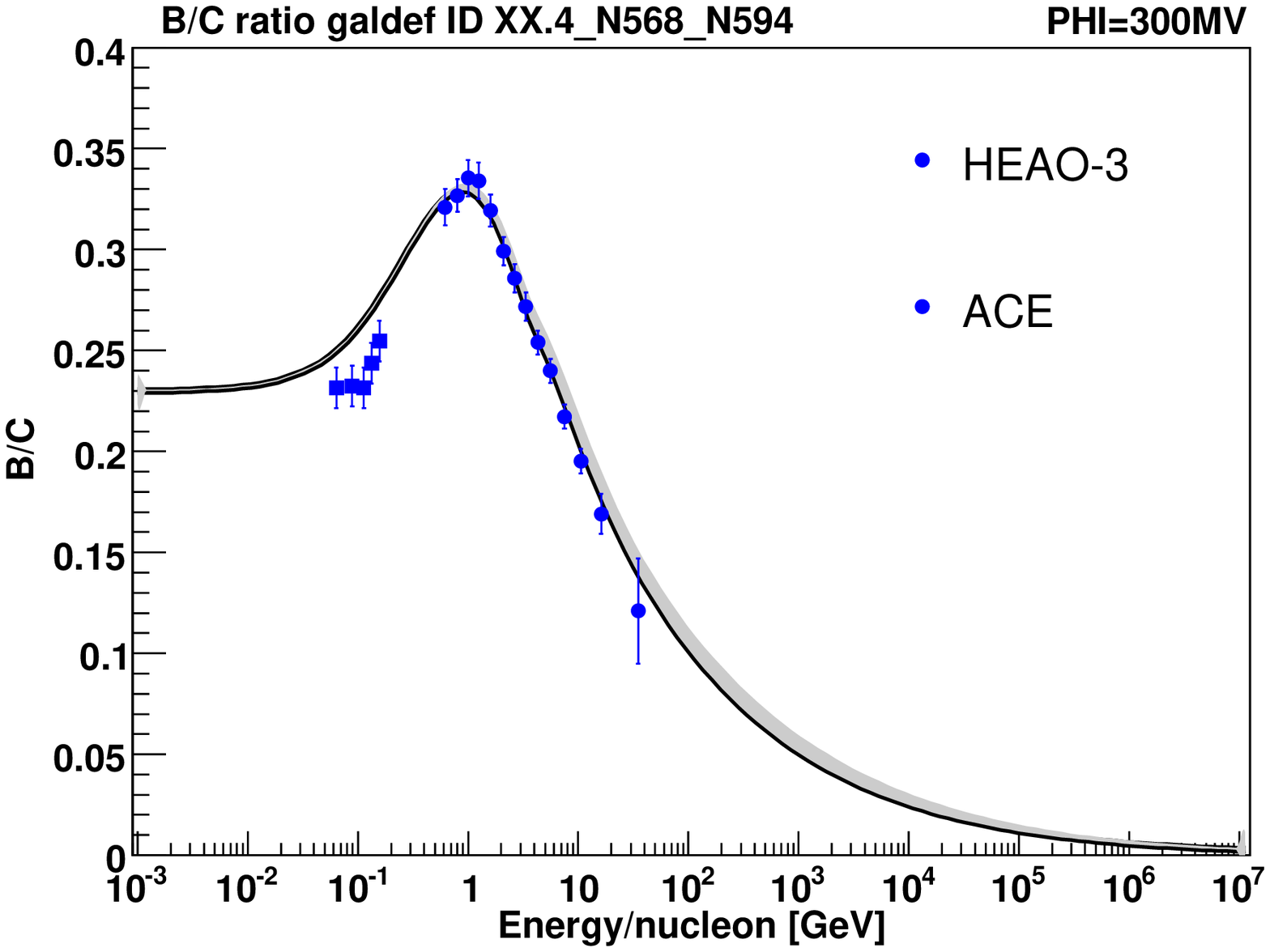}
\includegraphics[width=0.5\textwidth,clip]{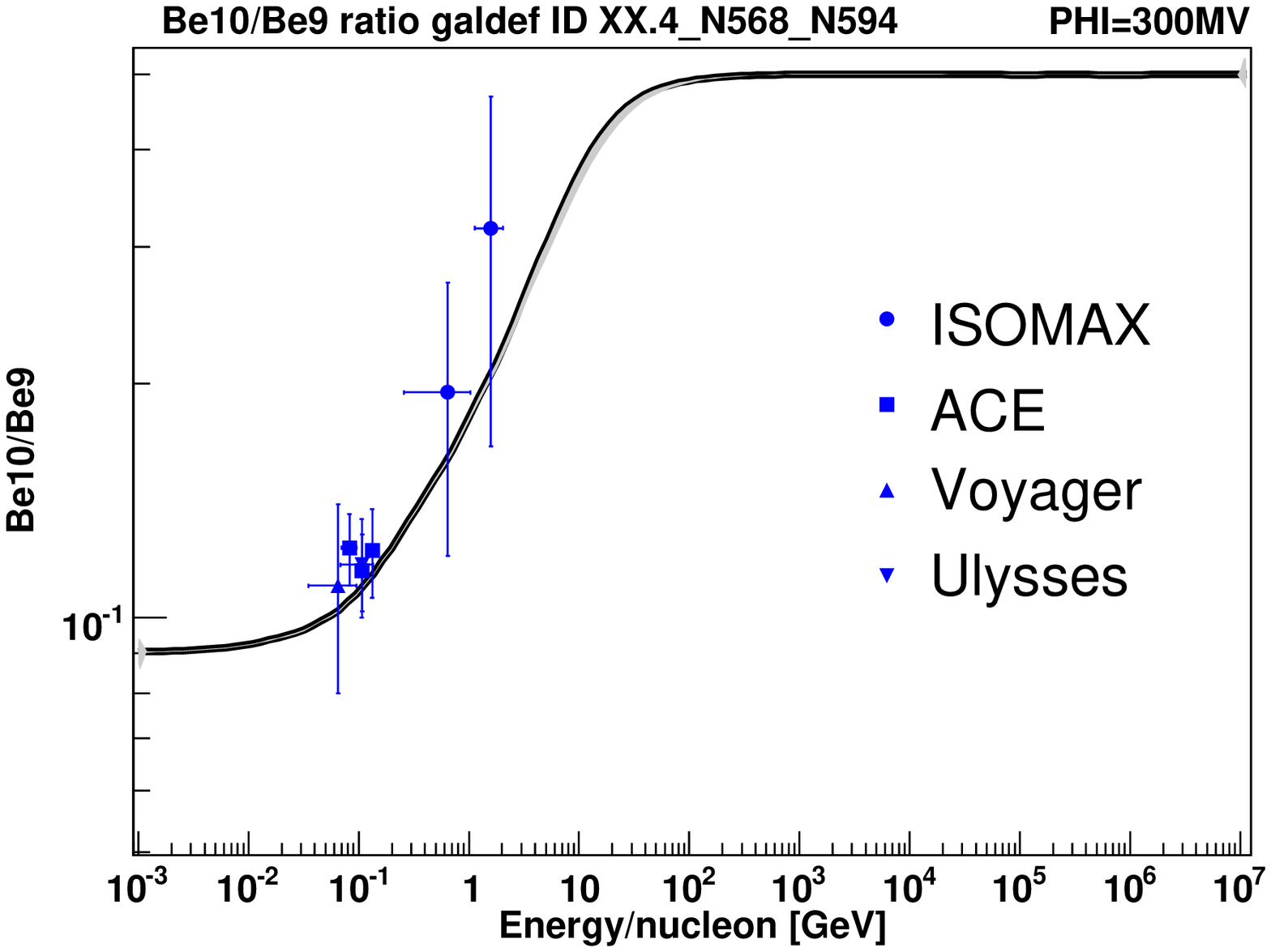}
\caption{{\bf Left} Local $B/C$ ratio in a conventional aPM ( {\it full line}) and LIS $B/C$ ratio ( {\it dashed line}).  {\it Data:} HEAO-3 \citep{Engelmann:1990} ACE \citep{Davis:2000}. {\bf Left} Local beryllium fraction in a conventional aPM. Line coding as on the left.  {\it Data:} ISOMAX \citep{Hams:2004}, ACE \citep{Yanasak:2001}, Voyager \citep{Lukasiak:1999}, Ulysses \citep{Connel:1998}. The narrow gray  band on top of the black line refers to a model with an increased halo height $z_h$=30 kpc (see section \ref{ss_escape} for a discussion ). }\label{f_conv_aPM_bc_be}
\end{figure*}

In order to cope with the large CR residence time and the comparably small amount of secondaries, CRs have to spend a certain time in the halo.
The times spent in the halo and the disk, denoted by  $t_h$ and $t_d$, respectively, are constrained by the ratio of secondary/primary CRs, which is most precisely measured for the B/C ratio and the ratio of unstable/stable CRs, which is most precisely measured for the $^{10}Be/^9Be$ ratio.
Isotropic transport models can reproduce the observed $B/C$ and $^{10}Be/^9Be$ ratios, because the diffusion coefficient and halo size can always be chosen in such a way that $t_{d}/(t_{h}+t_{d})$ agrees with the measurements.\\
In the presence of Galactic winds the probability for a CR to return to the disk drops fast with increasing height above the disk, leading to a relative reduction of $t_d$. Thus, if diffusion is isotropic and homogeneous, there is a limit for the convection speed which still allows to reproduce the correct ratio of $t_d$ and $t_h$. If convection speeds larger than this value are considered a mechanism which efficiently returns CRs from the halo is required. This can be a larger diffusion coefficient in the halo or a generally larger $D_{zz}$. The gradient in $D_{zz}$ can be tuned to reproduce the correct ratio of $t_{h}/(t_{h}+t_{d})$ for a given wind velocity, which can be seen as follows.
With $V_c=V_0+dV/dz\cdot z$ and Eq. \ref{e_GP}, CR transport along $z$ can be written as
\begin{equation}
\left[\frac{\partial D_{zz}}{\partial z} -V_o-\frac{dV}{dz}z\right]\frac{\partial \Psi}{\partial z}+D_{zz}\frac{\partial^2 \Psi}{\partial z^2}-\frac{d V}{d z}\Psi=\left(\frac{d \Psi}{d t}\right).\label{e_conv_diff}
\end{equation}
The first and the last term on the l.h.s. in Eq. \ref{e_conv_diff} describe the drift of CRs due to convection and diffusion. Drift due to diffusion occurs when CRs are diffusing from regions with large diffusion coefficient to regions with small diffusion coefficient where they are confined more efficiently. If the sign of $\partial D_{zz}/\partial dz$ is chosen to be positive, i.e. $D_{zz}$ increases in the halo, then the drift due to convection can be compensated by the drift due to diffusion (at least for a certain energy range).
Unlike the vertical gradient $dV/dz$ the initial wind velocity $V_0$ only occurs in the first term. An {\it increase} in $V_0$ can be interpreted as an {\it decrease} in $\partial D_{zz}/\partial z$, which just corresponds to the reduction of the forward mean scattering length in a moving reference frame.

Of course, in reality the situation is more complicated due to energy losses, the energy dependence of the diffusion coefficient and because $\partial D_{zz}/\partial z=0$ for $|z|\leq 1$, but the basic idea is that CRs can easily return to the disk via diffusion. 

 Since diffusion is energy dependent and convection energy independent the
return probability depends on energy.
Figure \ref{f_rhocrit} gives an impression of the rigidity dependence of the diffusion convection boundary $z_c$ above which convective transport dominates. Above $z=1$ kpc, where $\partial D_{zz}/\partial z \ge 0$,
 the critical rigidity increases from 2 GeV to more than 30 GeV.
This energy dependence describes quite well the $B/C$ ratio, while keeping the residence time in agreement with the  $^{10}Be/^9 Be$ ratio, as shown in Fig.  \ref{f_conv_aPM_bc_be}.
 Here the parameters given in Table \ref{table1} were used.
 The predicted $^{10}Be/^9Be$ ratio is slightly below the data indicating that the residence time is slightly too large, but well within the errors of the data.
 The slightly too high value of the $B/C$ ratio  below 1 GeV CRs also indicates that low energy CRs stay slightly too long in the Galaxy, which again indicates that $dV/dz$ is slightly too low (note that an increase in the base velocity $V_0$, will not improve this situation, but on the contrary, an increase in $V_0$ will help to keep CRs longer in the Galaxy). A refined tuning  can be performed as soon as the PAMELA $B/C$ data will be released, since  the ROSAT data allow for $dV/dz$ up to $\sim 43~\mathrm{km/s/kpc}$, which is somewhat higher than the value of $35~\mathrm{km/s/kpc}$ from Table \ref{table1}.

\subsection{Halo size}\label{ss_escape}

\begin{figure*}
\includegraphics[width=0.5\textwidth,clip]{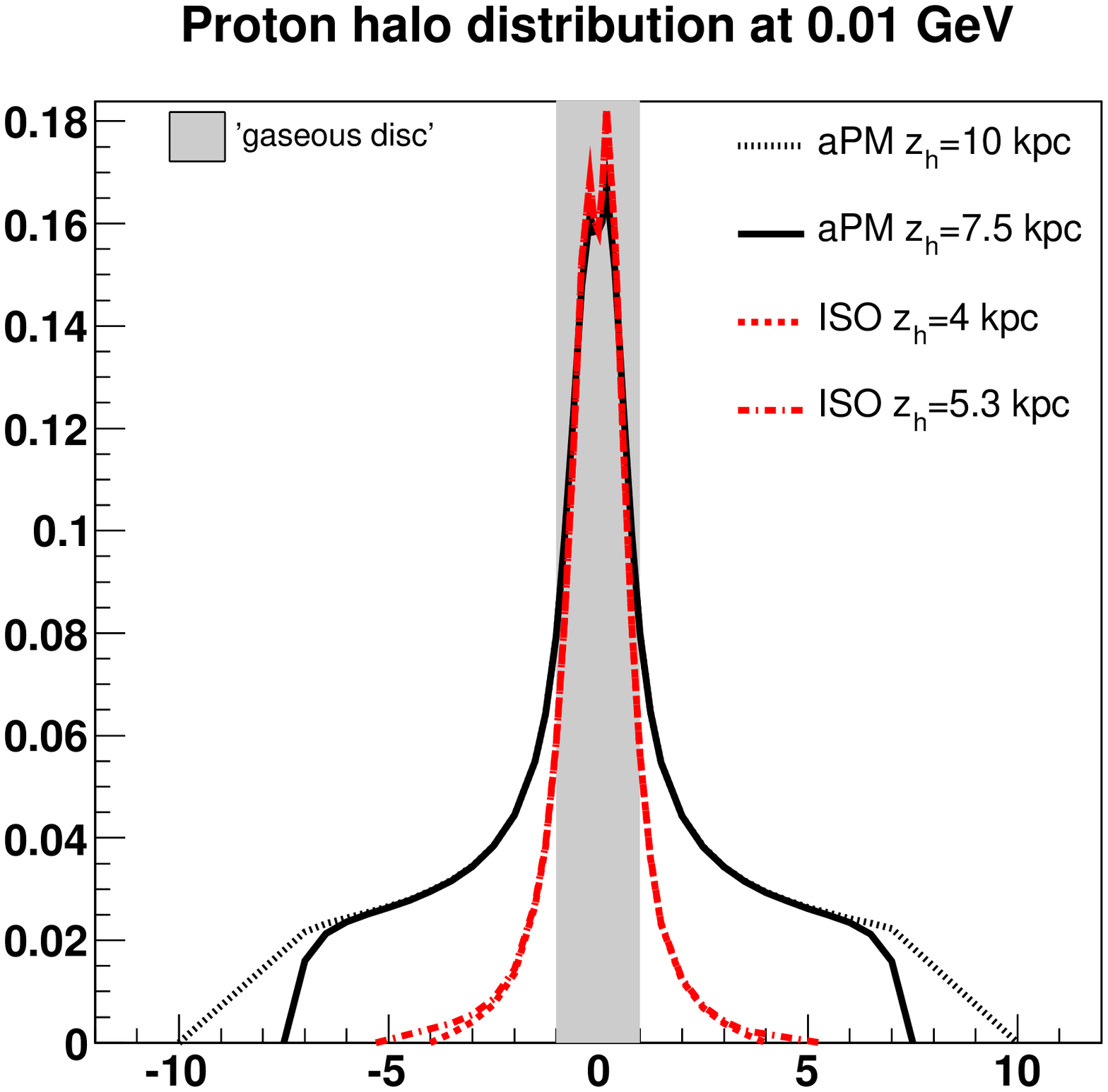}
\includegraphics[width=0.5\textwidth,clip]{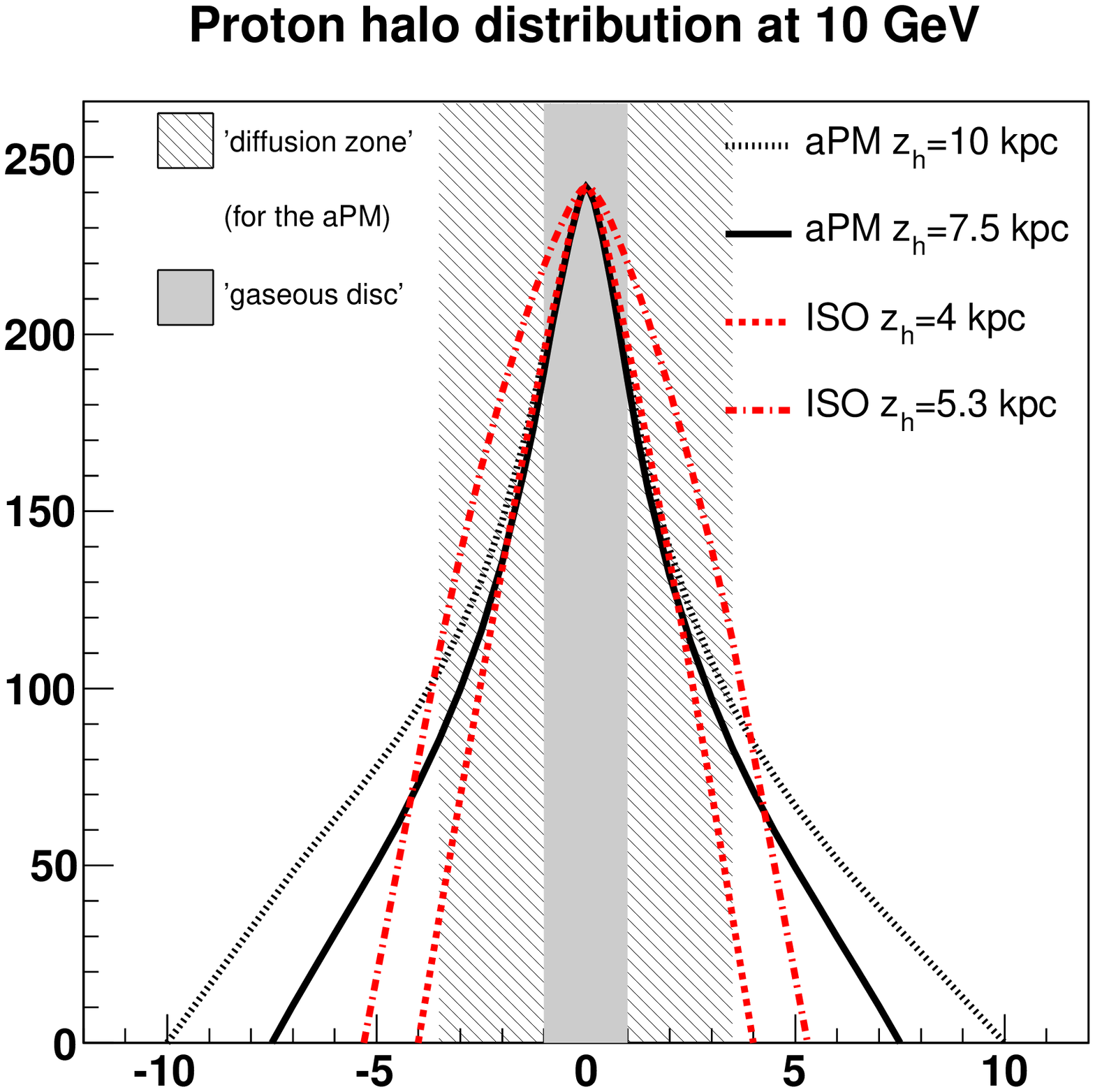}
\caption{The halo density profile of protons for rigidities of 0.01 GeV and 10 Ge at $R_o$ for an aPM (black line) and an isotropic model (red lines).
In order to allow for easy
comparison, the proton density for each energy bin has been normalized to
the aPM with $z_h=7.5~\mathrm{kpc}$ in this figure.}\label{f_p_distro1}
\end{figure*}

As discussed before, the diffusion coefficient increases with increasing $z$, thus providing a natural transition to free escape independent of the boundary condition. E.g. increasing the boundary box from $z$=7.5 to 30 kpc does not change the
residence time or the ratio  $t_{h}/(t_{h}+t_{d})$, as can be seen from the gray  band on top of the black line for the B/C ratio  in Fig. \ref{f_conv_aPM_bc_be}, which is hardly distinguishable  from the black line for a boundary of  $z$=7.5 kpc.
This is in strong contrast to isotropic transport models, which are highly sensitive to the size of the diffusion region, as discussed in Sect. \ref{s_iso} (see Fig. \ref{f_BC_BE_iso}). Given this large difference in sensitivity to the boundary condition
between the isotropic and anisotropic models it is interesting to compare the CR density profiles for protons in the halo.
This is done in Fig.  \ref{f_p_distro1} for two halo sizes in the aPM ($z_h=7.5$ and 10 kpc) and two halo sizes in an isotropic model ($z_h=4$ and 5.3 kpc) at rigidities of  0.01 GeV and 10 GeV at the Sun's Galactocentric radius. Most sources are located in the disk, so the source density at $z=0$ kpc is highest.
In the isotropic model low energy CRs are almost at rest and the CRs stay for a large fraction inside the gaseous disk (indicated by the vertical gray band in Fig. \ref{f_p_distro1}) with tails towards the halo boundary, where the density drops to zero.  For the aPM this distribution is broadened at larger $z$-values by convection.
For high energies the situation changes, because the diffusion starts to become more important and the diffusion-convection boundary $z_c$ moves to $z$-values of a few kpc, as indicated by the hatched area on the right-hand side of Fig. \ref{f_p_distro1}. The ratio of times spent in the disc and in the halo is approximately given by the area below the curves in the disk and the halo (a disk height of 1 kpc is chosen for illustrative purposes, note that this is not identical to the significantly smaller scaleheight of 250 pc). One observes that this changes for the isotropic model significantly, if one moves the boundary from 4 to 5.3 kpc, while for the aPM the changes are mainly in the convection zone from which only few CRs return to the disk. 
Convection therefore allows us to reproduce the vertical distribution of the CR distribution in such a way that the requirements from local $B/C$ and $^{10}Be/^9 Be$ are met and at the same time reduces the model's sensitivity to the exact position of the boundary condition, provided that the diffusion-convection boundary $z_c$ is small enough. Clearly, with increasing energy the convection-diffusion boundary also moves further into the halo until it becomes comparable to the halo height $z_h$. With increasing energy any aPM will therefore become more and more sensitive to the position of the boundary condition. At the same time however, CRs become less and less confined and for very high energies the diffusion approximation will break down. Figure \ref{f_conv_aPM_bc_be} shows that for convection speeds as expected from ROSAT a halo height of 30 kpc still does not yield any significant change on the local $B/C$ and $^{10}Be/^9 Be$ ratio for all energies of interest, i.e. the energy range in which the diffusion approximation holds.\\
The radial difference between the two models is less significant, as shown by
 the two-dimensional proton distribution at 5 GeV in the $Rz$ plane in Fig.  \ref{f_p_distro2}. Both distributions are  similar up to Galactocentric radii of 4 kpc. For larger radii the aPM falls off less steep than the isotropic model. The reason for this is that the diffusion is assumed to be independent of radius, while the convection decreases at larger radii, so diffusion becomes more important there, thus widening the distribution in $R$.
This helps in solving the soft $\gamma$-ray gradient problem, as will be discussed in more detail in section \ref{ss_gradient}.

\begin{figure*}
\includegraphics[width=0.5\textwidth,clip]{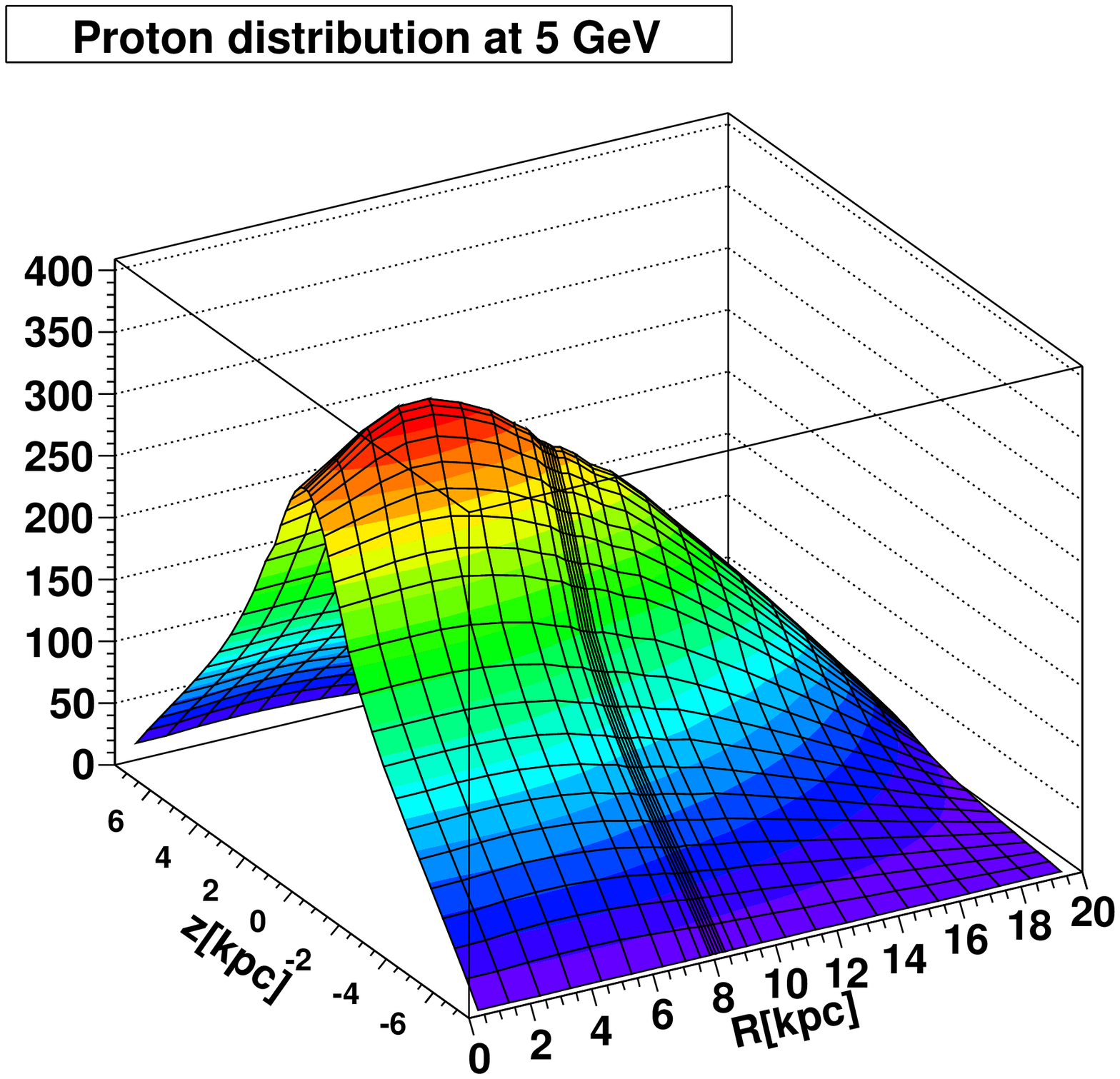}
\includegraphics[width=0.5\textwidth,clip]{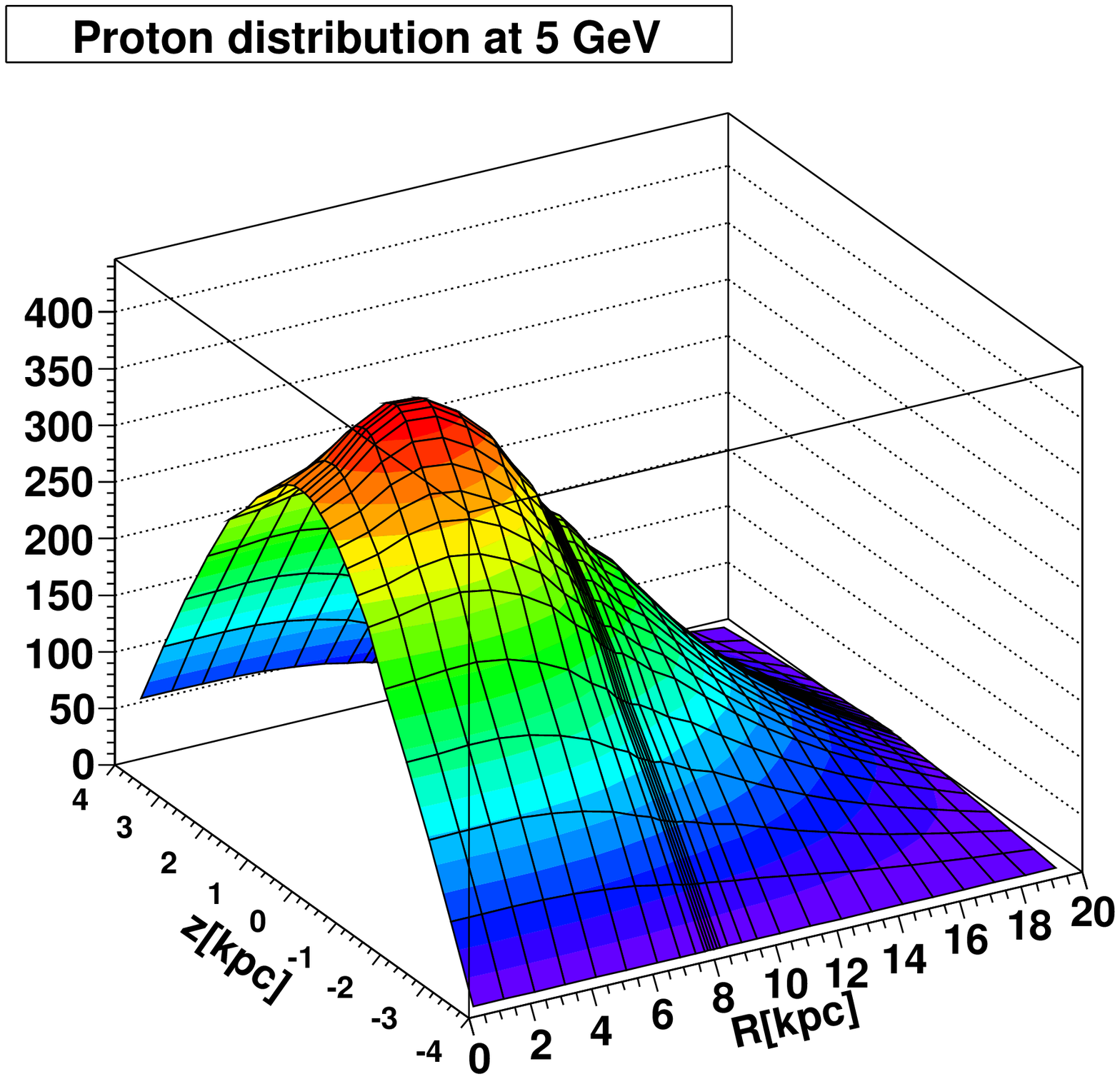}\\

\caption{The density distribution of 5 GeV protons  in the $Rz$ plane for an aPM (left) and an isotropic model (right).} Note the different  boundaries in $z$ of 7.5 and 4 kpc, respectively. The boundary in $R$ is 20 kpc in both models. \label{f_p_distro2}
\end{figure*}

\subsection{Antiprotons and positrons in an aPM}\label{ss_pb_aPM}

\begin{figure*}
\includegraphics[width=0.5\textwidth,clip]{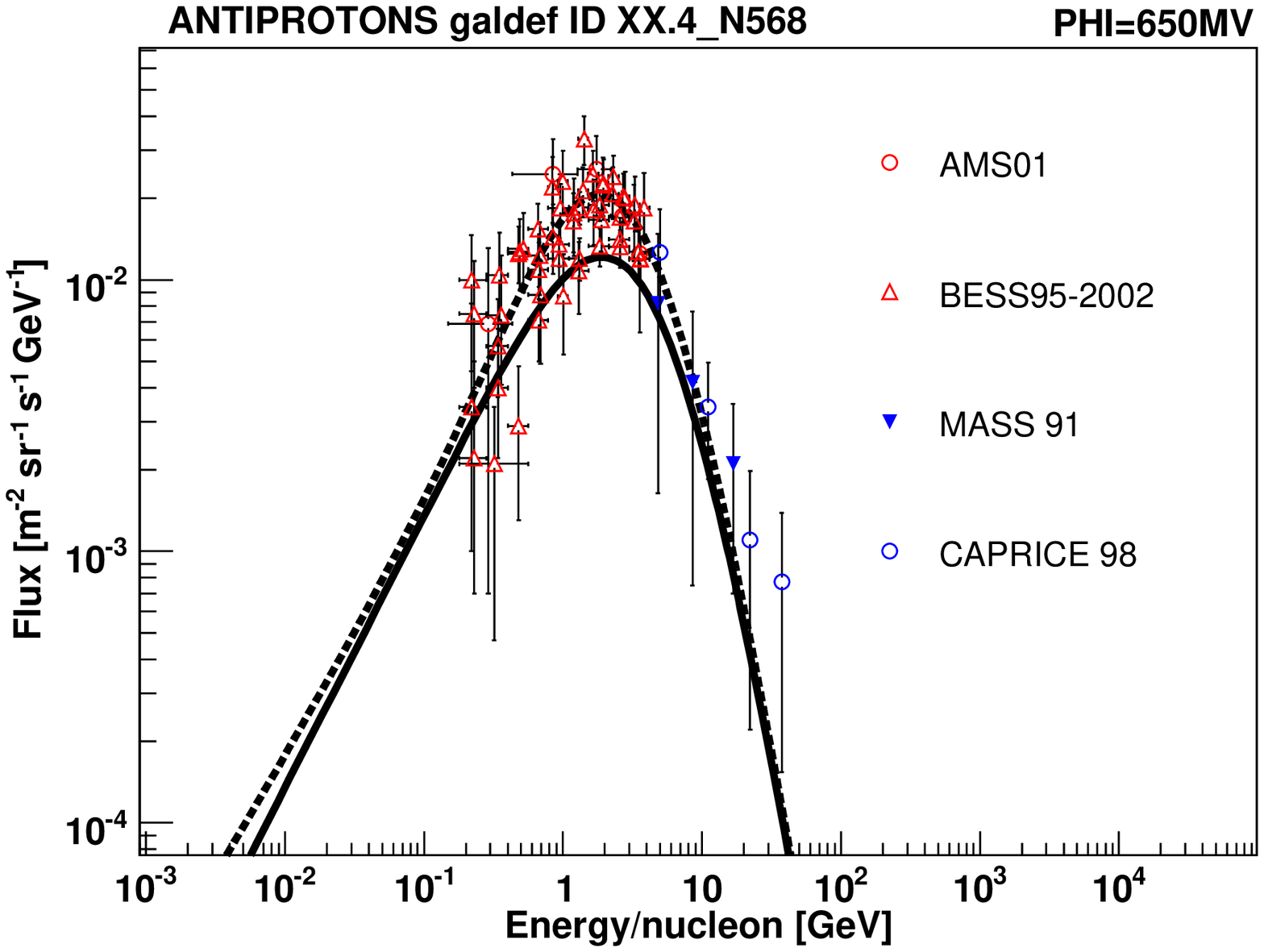}
\includegraphics[width=0.5\textwidth,clip]{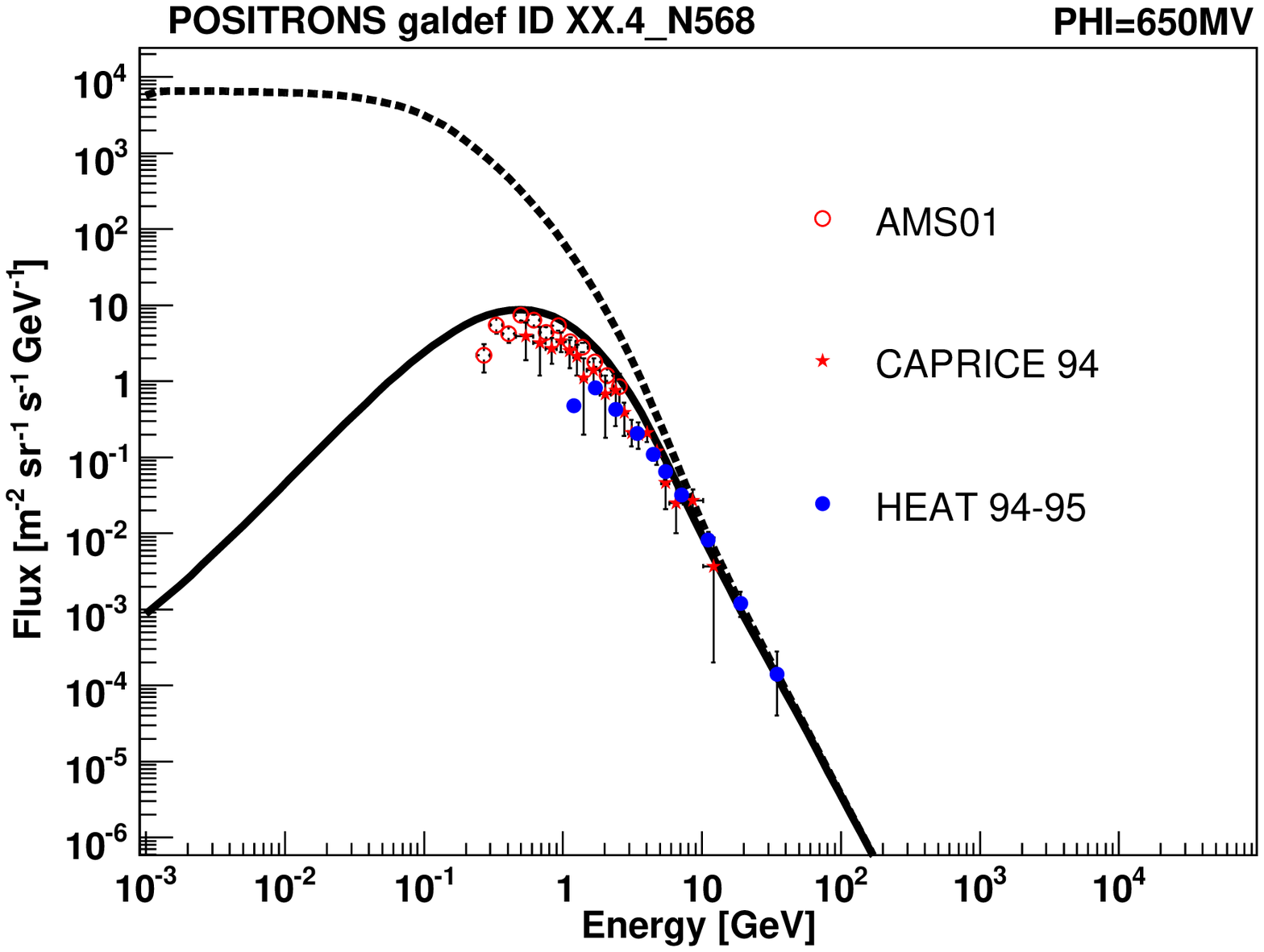}
\caption{{\bf Left} The antiproton flux in the aPM, where the solid line denotes the local flux and the dashed line the flux corrected for solar modulation.  {\it Data:} BESS 95-97 \citep{Orito:1999re}, CAPRICE 98 \citep{Boezio:2001ac}, MASS91 \citep{Basini:1999hh}. {\bf Right} The positron flux in the aPM. {\it Data:} AMS I \citep{Alcaraz:2000vp}, CAPRICE 94 \citep{Boezio:2000te}, HEAT 94 \citep{DuVernois:2001bb}}\label{f_conv_aPM_pb_ep}
\end{figure*}

With the basic transport parameters fixed by the local measurements of $B/C$ and $^{10}Be/^9Be$ and the CR injection spectra given by the local proton and electron spectra the model parameters are basically settled. Fine-tuning of the transport parameters can be done by considering additional secondary particles, like antiprotons and positrons, which originate from nucleon-nucleon collisions in the ISM.\\
The left side of Fig.  \ref{f_conv_aPM_pb_ep} shows the local antiproton and positron flux. Antiprotons show an excess of about 40\% to 50\% below 4 GeV. This excess is very similar to the excess seen in the isotropic transport models \citep{Strong:2007nh}. It is possible to increase the local antiproton flux by increasing the local CR interaction rate via diffusion or convection. However, this would worsen the $B/C$ ratio and the local positron flux.

The right side of Fig.  \ref{f_conv_aPM_pb_ep} shows the local positron flux. For energies below 5 GeV the model shows a slight excess in local positrons. This is probably the result of too efficient diffusive reacceleration driven by either a too large $v_{\alpha}$ or a too small diffusion coefficient. On the other hand less efficient diffusive reacceleration would worsen the local $B/C$ ratio. An improvement for both positrons and the local $B/C$ ratio is expected if convective transport was slightly more efficient. We will see in subsection \ref{ss_gamma} that a less steep radial gradient in Galactic winds can help to improve local charged CRs and diffuse $\gamma$-rays.

\section{Performance of the Anisotropic Propagation Model - charged CRs}\label{s_perf_CR}

\subsection{The INTEGRAL positron annihilation signal in an aPM}\label{ss_INT2}
\begin{table*}
\begin{center}
\begin{tabular}{|c|c|c|c|c|c|c|c|c|c|c|}\hline
Model&$D_{0}$ &$dD/dz$&$V_{0}$&$dV/dz$&$z_h$&$f_{esc}$@0.1MeV& $f{esc}$@1MeV&$f_{esc}$@4MeV&INT&ROS\\
&$[cm^2/s]$&$[cm^2/s/kpc]$&$[km/s]$&$[km/s/kpc]$&$[kpc]$&[\%]&[\%]&[\%]&&\\
\hline\hline

aPM3&$5.4\cdot 10^{28}$&0                    & 100&30&7.5  &88.5&85.3&82.1&+&+\\

aPM1&$4.86\cdot 10^{28}$&$4.86\cdot 10^{28}$ & 100&30&7.5  &89.3&87.3&83.3&+&+\\
aPM&$5.4\cdot 10^{28}$&$5.4\cdot 10^{28}$   & 100&30&7.5  &89.2&86.3&83.2&+&+\\
aPM2&$5.94\cdot 10^{28}$&$5.94\cdot 10^{28}$ & 100&30&7.5  &88.0&83.7&81.7&+&+\\
aPM4&$5.4\cdot 10^{28}$&$5.4\cdot 10^{28}$   & 100&15&7.5  &89.0&86.5&82.5&+&+\\
\hline
aPM5&$5.4\cdot 10^{28}$&$5.4\cdot 10^{28}$   & 100&0 &7.5  &89.0&85.2&82.0&+&-\\
aPM6&$5.4\cdot 10^{28}$&$5.4\cdot 10^{28}$   & 50&0 &7.5  &78.0&66.9&65.3&+&-\\
aPM7&$5.4\cdot 10^{28}$&$5.4\cdot 10^{28}$   & 30&0 &7.5  &65.9&54.4&47.7&+&-\\
aPM8&$5.4\cdot 10^{28}$&$5.4\cdot 10^{28}$ & 0  &100&7.5  &58.6&58.3&56.4&+&-\\
\hline
ISO1&$5.4\cdot 10^{28}$&$5.4\cdot 10^{28}$ & 10  &0&7.5  &29.1&8.4&0&-&-\\
ISO2&$5.4\cdot 10^{28}$&$5.4\cdot 10^{28}$ & 0  &30&7.5  &38.9&34.4&32.0&-&-\\
ISO3&$5.4\cdot 10^{28}$&$5.4\cdot 10^{28}$   & 0  &15&7.5  &24.9&18.0&13.4&-&-\\
ISO4&$5.4\cdot 10^{28}$&$5.4\cdot 10^{28}$   & 0  &7&7.5  &11.7&4.1&1.7&-&-\\
ISO&$5.4\cdot 10^{28}$&0                  & 0  &7 &4    &10.7&4.9 &3.1&-&-\\
\hline

\end{tabular}

\caption{Positron escape fraction for different convection velocities and diffusion coefficients. For details see text.}\label{t_escape}
\end{center}
\end{table*}
As discussed in Sect. \ref{ss_inhom} the INTEGRAL satellite observed a large B/D ratio for the positron 511 keV annihilation line. In an aPM this is expected, because low energy particles propagate predominantly by convection and convection is large in the disk and small in the bulge. Positrons produced in the disk are transported into the halo, but stay near the sources in the bulge.  Furthermore the bulge has a large extension in all directions, so even if the positrons are slowly transported by diffusion, they still have time to thermalize and find an electron to annihilate. In the disk positrons are produced predominantly in the region of a high source density, i.e. a region of high convection. To quantify the escape probability from the disk the SNR distribution given by equation \ref{e_source} is taken as the source distribution for MeV positrons, but a somewhat larger scale height of 300pc is used, which is the scale height of SN1a \citep{Prantzos:2005pz}.
The positron  spectrum  from $^{56}Co$ $\beta^{+}$-decays is modeled as a rectangular function between 0.1 MeV and 5 MeV.
Table \ref{t_escape} shows the fraction of positrons above $z_c$ for different parameters of the aPM.  For the aPM more than $89\%$ ($83.2\%$) of the 0.1  (4) MeV positrons escape from the Galaxy. Since CR transport in this energy range is mainly governed by convection, variations of $10\%$ in the diffusion coefficient or a constant diffusion coefficient in the halo (aPM1-aPM3) do not change these numbers significantly. In the disk $V_0$ is larger than $dV/dz$, so that a decrease in $dV/dz$ by $50\%$ (aPM4) still yields about the same positron escape fraction. When decreasing $V_0$ from $100~\mathrm{km/s}$ to $30~\mathrm{km/s}$ for a constant $dV/dz=0~\mathrm{km/s/kpc}$ (aPM5-aPM7) the positron escape fraction decreases significantly and becomes incompatible with the INTEGRAL requirements for $V_0=10~\mathrm{km/s}$ (ISO1). A strong increase in convection of $dV/dz=100~\mathrm{km/s/kpc}$ is compatible with the requirements by INTEGRAL even if the velocity at the base of the wind is zero (aPM8), however in this case the requirements by ROSAT and local charged CRs are no longer met. For a quasi isotropic model with convection velocity of $dV/dz=7~\mathrm{km/s/kpc}$ and $z_h=4~\mathrm{kpc}$ (ISO) the positron escape fraction rages from $10.7\%$ at 0.1 MeV to only $3.1\%$ at 4 MeV. Models with $V_0$ smaller than $30~\mathrm{km/s}$ are generally incompatible with the INTEGRAL requirements.

In summary,
models that meet the requirements by ROSAT automatically provide a positron escape fraction high enough to explain the large bulge/disk ratio observed by INTEGRAL. Intermediate convection velocities too small for ROSAT are already able to meet the requirements by INTEGRAL. In the keV range the scale height of the propagated positron distribution is $\sim 1-1.5~\mathrm{kpc}$ in excellent agreement with the scale height of 1 kpc adopted by \citet{Prantzos:2005pz} in his models B and D. Note, that the positron escape fraction shown here is just a  simple estimate. A more detailed modeling of the positron source spectrum would have a significant impact upon the energy dependence of the escape fraction.

It should be noted that due to the smaller source strength the number of positrons produced in the bulge is not sufficient to entirely explain the signal from the bulge \citep{Prantzos:2005pz}. Even in an aPM an additional process or an additional source population in the bulge is required to explain the observed emission from the bulge. \citet{Prantzos:2005pz} suggested that a fraction of the positrons escaping the disk may even
be channeled by the poloidal field to the bulge where the positrons then would be confined by the large magnetic field. A detailed study of the efficiency of such a channeling process would require to adapt the anisotropy in the diffusion coefficient according to the direction of the magnetic field in the halo and take care of the positron confinement in the bulge by a decreased diffusion coefficient. In this rather qualitative approach we refrain from any fine-tuning of the diffusion coefficients. \\
Recently it has been found that an additional low mass x-ray binary (LMXB) population seems to reside in the bulge region showing even the morphological features of the observed annihilation signal \citep{Weidenspointner:2008zz}. The additional positrons expected from the LMXB are not sufficient to account for the complete emission observed from the bulge and this also cannot explain why there is almost no annihilation signal from positrons from $^{56}Co$ from the disk, but added to an aPM this additional LMXB population might nicely explain the B/D ratio.\\
With wind velocities taken to be simply proportional to the source distribution we already find excellent agreement with the model presented by \citet{Prantzos:2005pz}. The problem of the large B/D ratio for positron annihilation
is thus intimately related to the propagation of positrons.

\section{Performance of the Anisotropic Propagation Model - $\gamma$-rays and radio emission}\label{s_perf_gamma}
\begin{figure*}
\includegraphics[width=0.5\textwidth,clip]{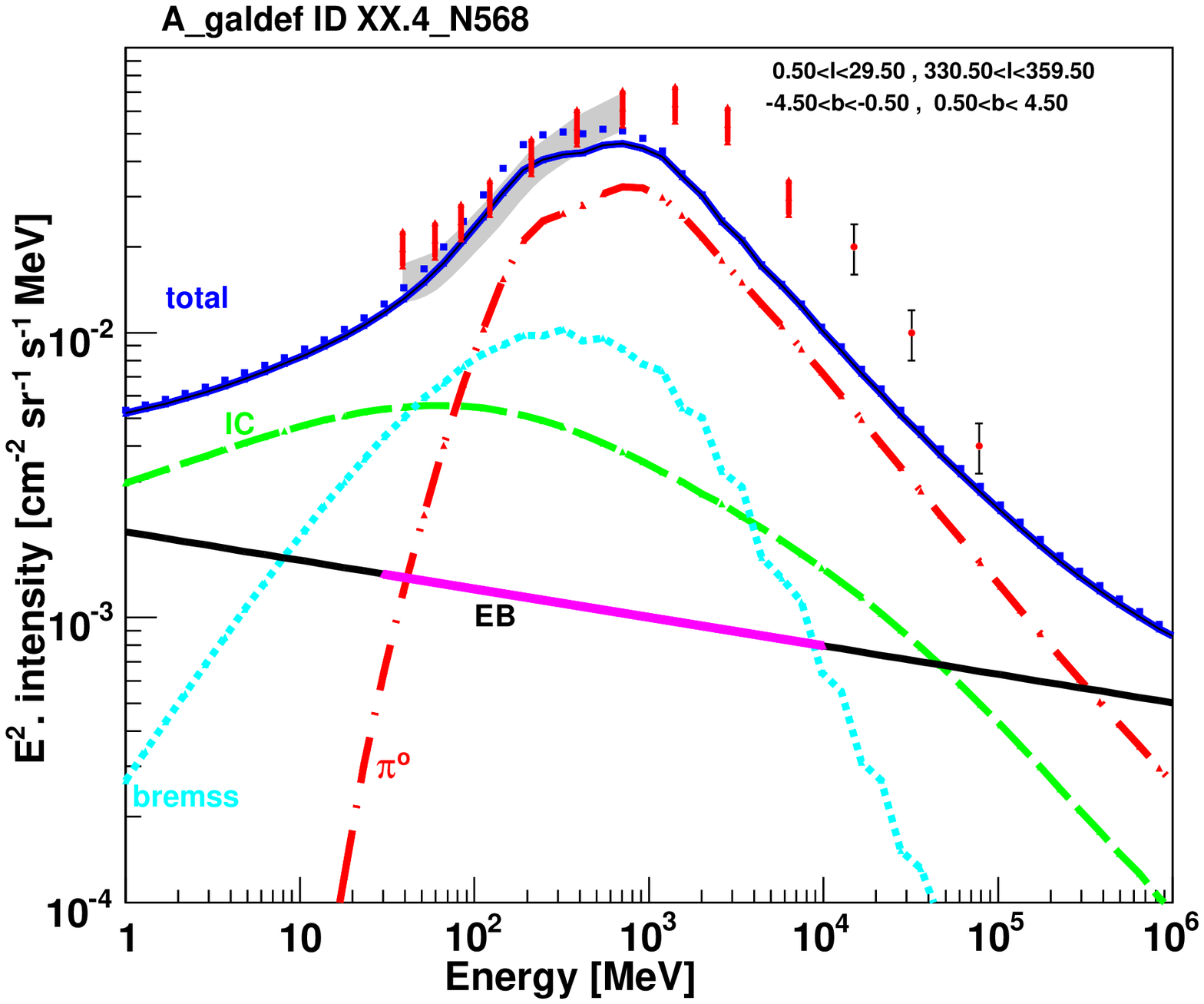}
\includegraphics[width=0.5\textwidth,clip]{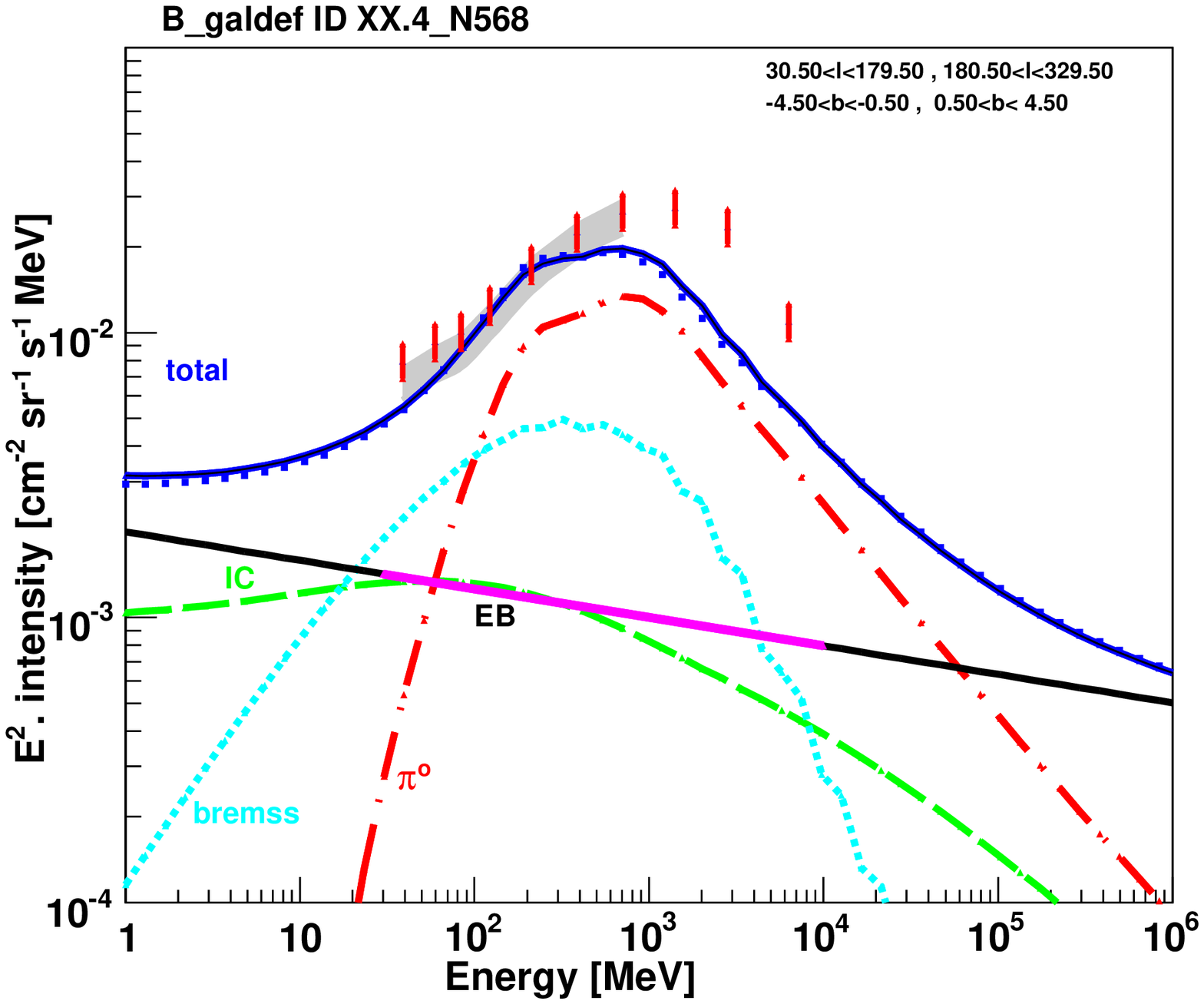}\\
\includegraphics[width=0.5\textwidth,clip]{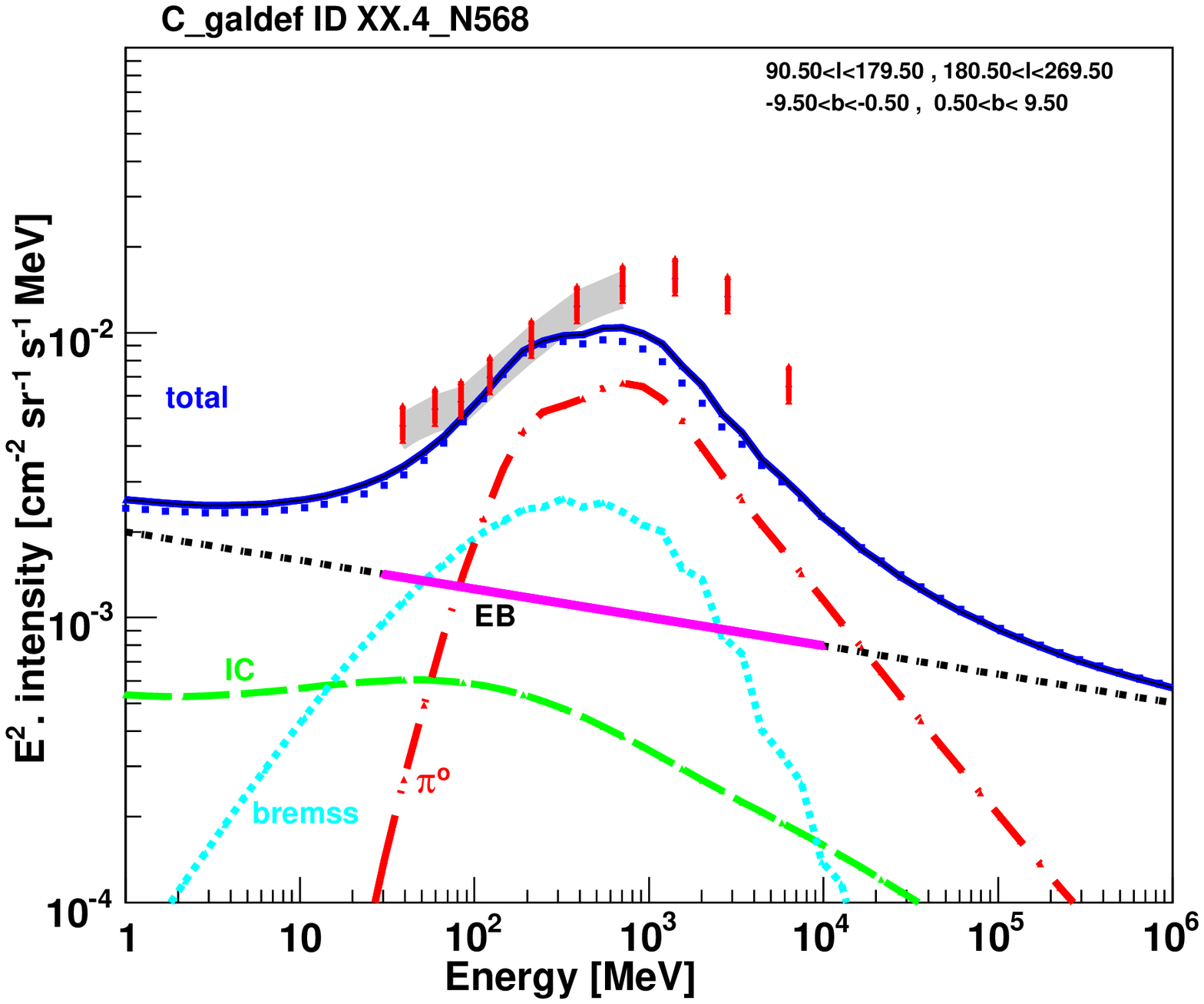}
\includegraphics[width=0.5\textwidth,clip]{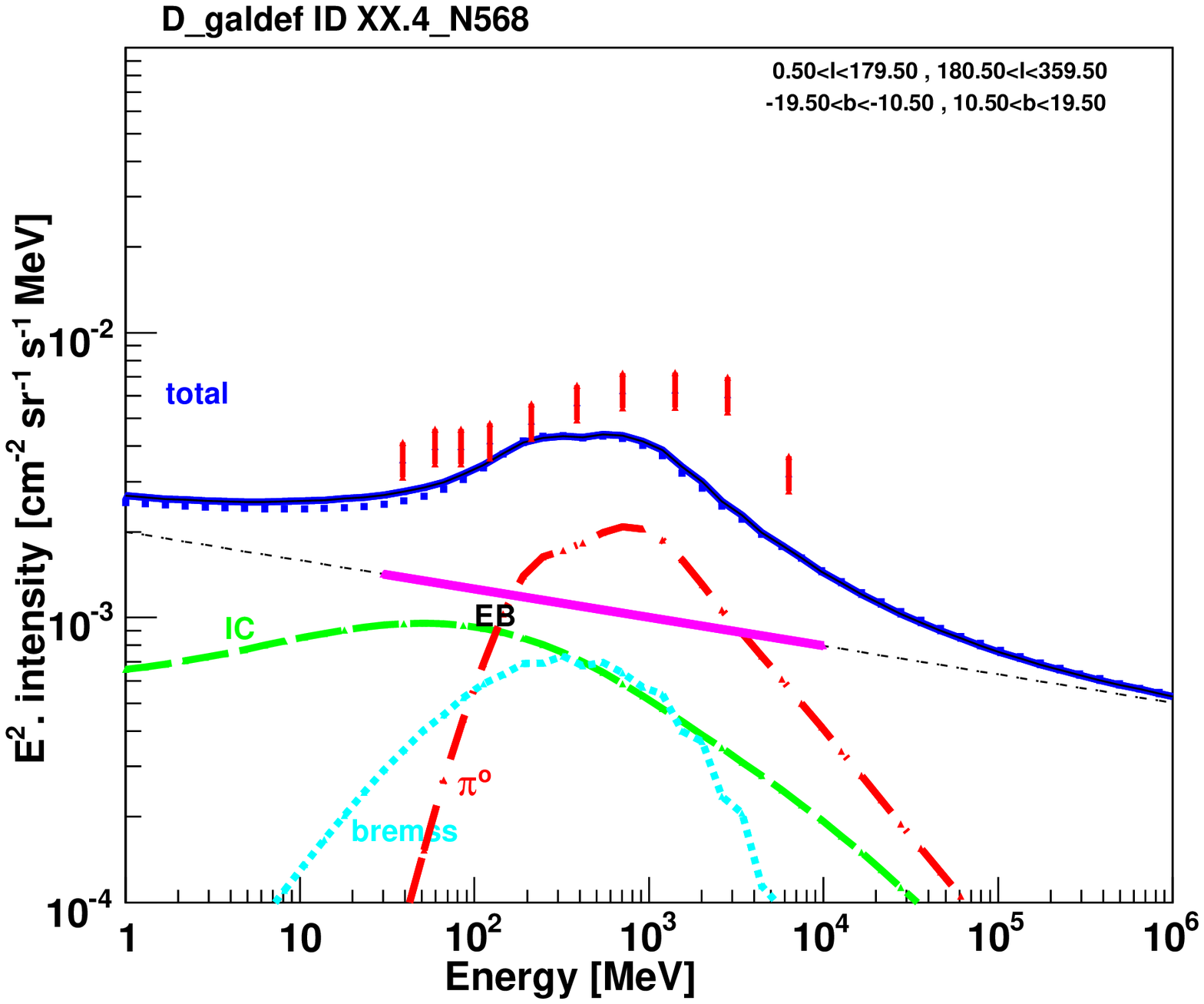}\\
\includegraphics[width=0.5\textwidth,clip]{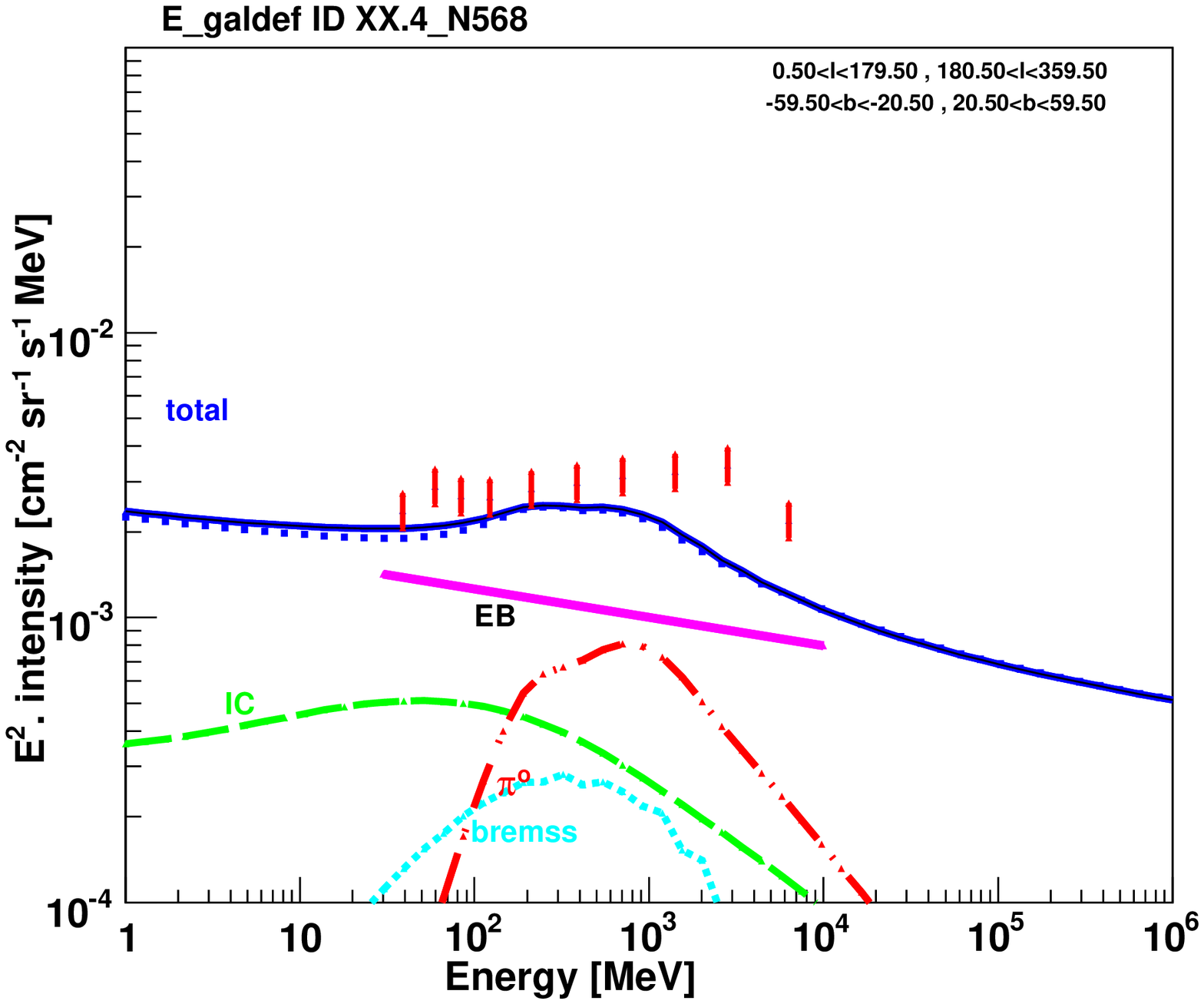}
\includegraphics[width=0.5\textwidth,clip]{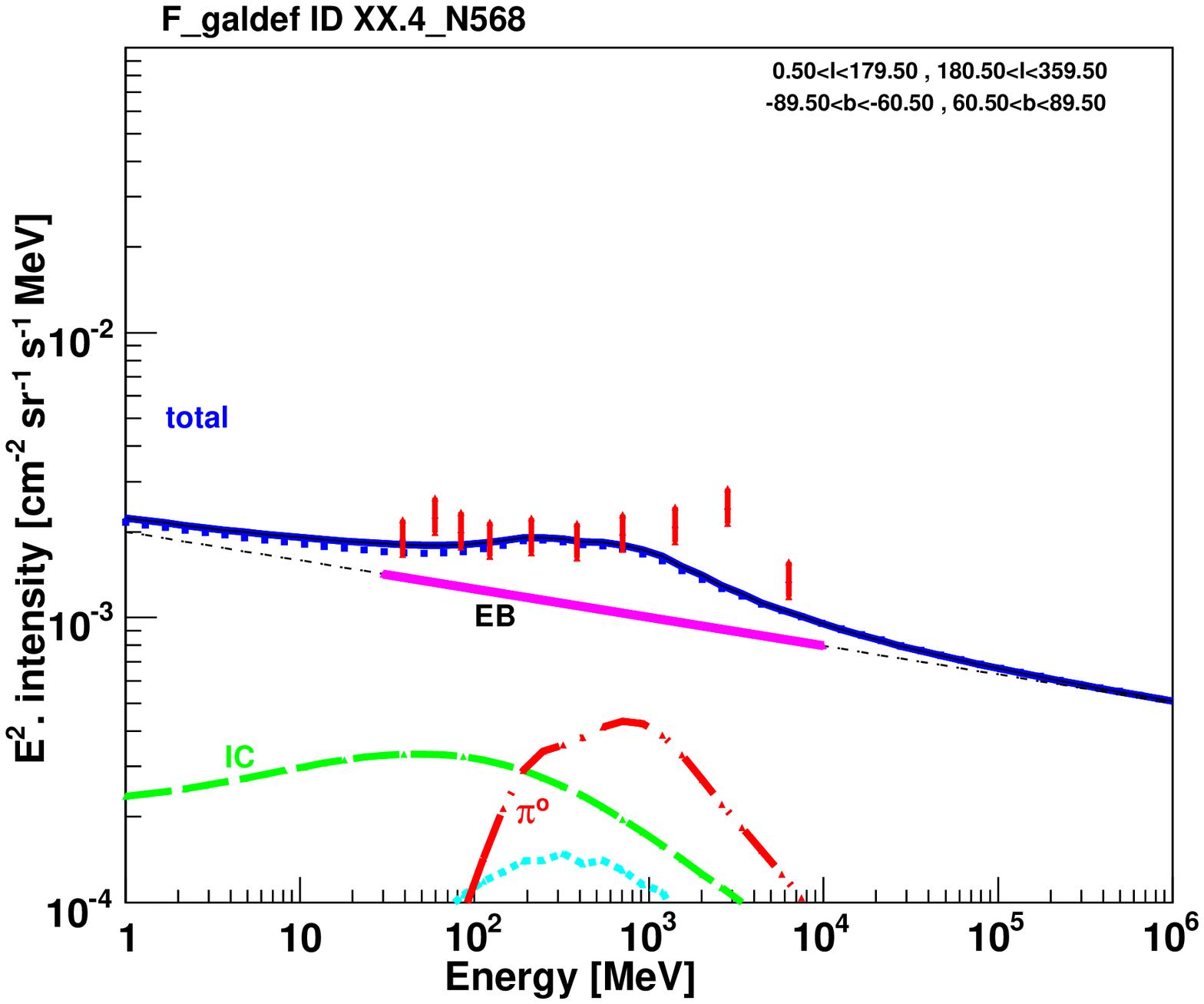}
\caption{Diffuse $\gamma$-rays for the six different sky regions as defined in \citet{Strong:2004de}. Line coding: bremsstrahlung ({\it light blue dashed}), inverse Compton ({\it green long dashed}), $\pi^0$-decay ({\it red long dashed-dotted}), total ({\it blue full}). The pink full line is the extragalactic background model according to \citet{Sreekumar:1997un}. The EGRET data are corrected for the point-spread function (PSF), for regions A, B and C the uncorrected EGRET data are shown as the grey band. Also shown is the total $\gamma$-ray flux as predicted by an isotropic model with $D_0=5.8\cdot 10^{28}\mathrm{cm}^2/\mathrm{s}$ at 4 GeV and $v_{\alpha}=30$km/s (dotted blue line).}\label{f_conv_aPM_Regions}
\end{figure*}

\subsection{Diffuse $\gamma$-rays}\label{ss_gamma}
\begin{table}
\begin{center}
 \begin{tabular}{cccc} \hline
 Region & Longitude $l$ & Latitude $|b|$ & Description\\\hline
 A & 330-30 & 0-5 & Inner Galaxy\\
 B & 30-330 & 0-5 & Disk without inner Galaxy\\
 C & 90-270 & 0-10 & Outer Galaxy\\
 D & 0-360 & 10-20 & Low longitude \\
 E & 0-360 & 20-60 & High longitude\\
 F & 0-360 & 60-90 & Galactic Poles\\
 \hline
 \end{tabular}
 \end{center}
 \caption[skyregions]{The longitude and latitude of the six sky regions shown in Fig. \ref{f_conv_aPM_Regions}.}
 \label{table3}
 \end{table}

In the light of the upcoming release of the Fermi-LAT data on diffuse $\gamma$-rays a cross-check with diffuse $\gamma$-rays is difficult. The EGRET data do not agree with the preliminary Fermi-LAT data even in the lowest energy range \citep{Porter:2009sg}. In the absence of other options we continue to use the EGRET data as a cross-check, keeping in mind that the softer Fermi data will probably require a larger contribution from inverse Compton (IC) and bremsstrahlung.

\begin{figure*}[]
\includegraphics[width=0.33\textwidth,clip]{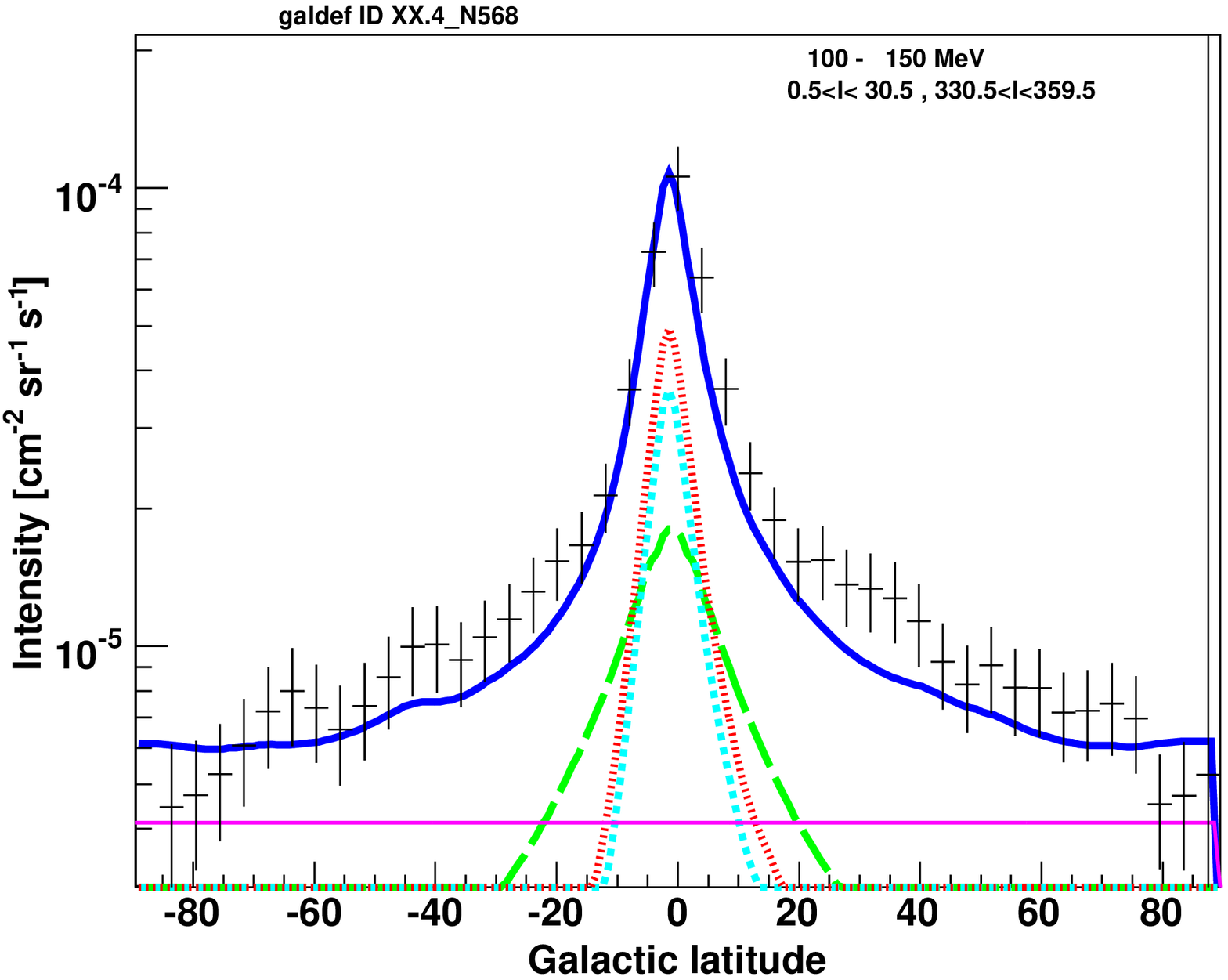}
\includegraphics[width=0.33\textwidth,clip]{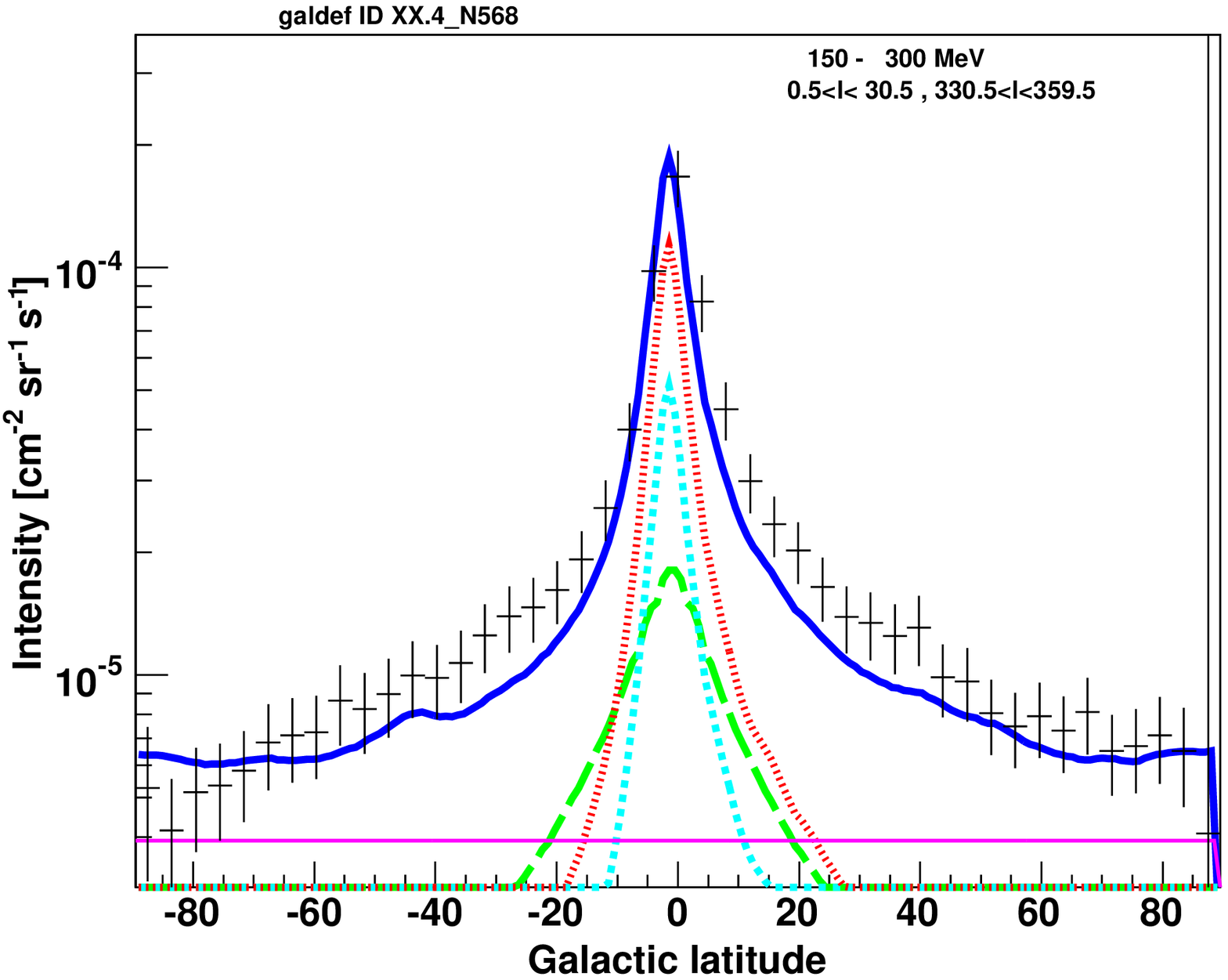}
\includegraphics[width=0.33\textwidth,clip]{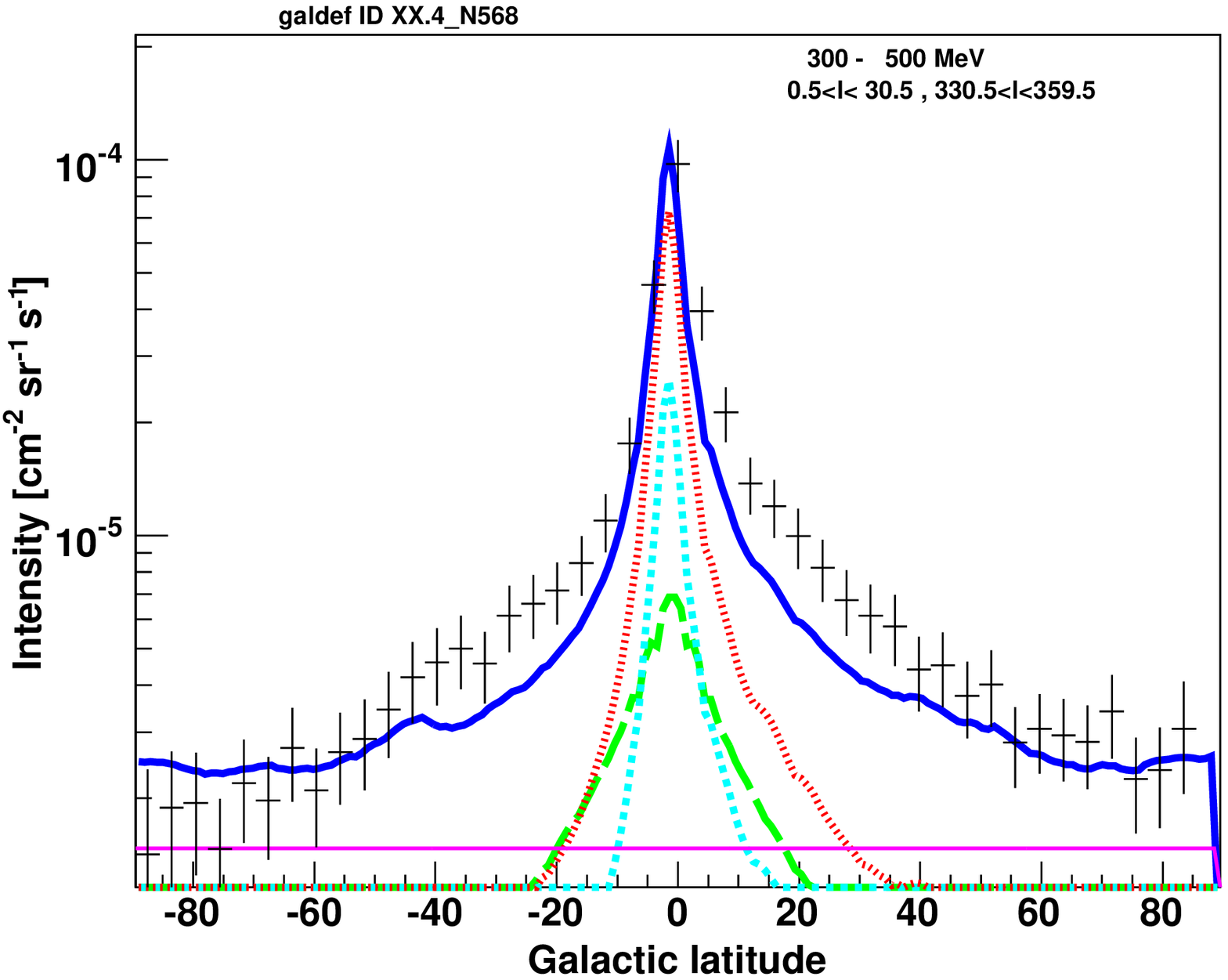}\\
\caption{Latitude profiles for the inner Galaxy ($|l|<30.5$) for the aPM compared to the EGRET data between 100 and 500 MeV.The EGRET-excess above 500 MeV  is not confirmed by preliminary FERMI data, so only data below 500 MeV are considered here. Line coding as in Fig.\ref{f_conv_aPM_Regions}}\label{f_conv_aPM_lat}
\end{figure*}

\begin{figure*}[]
\includegraphics[width=0.33\textwidth,clip]{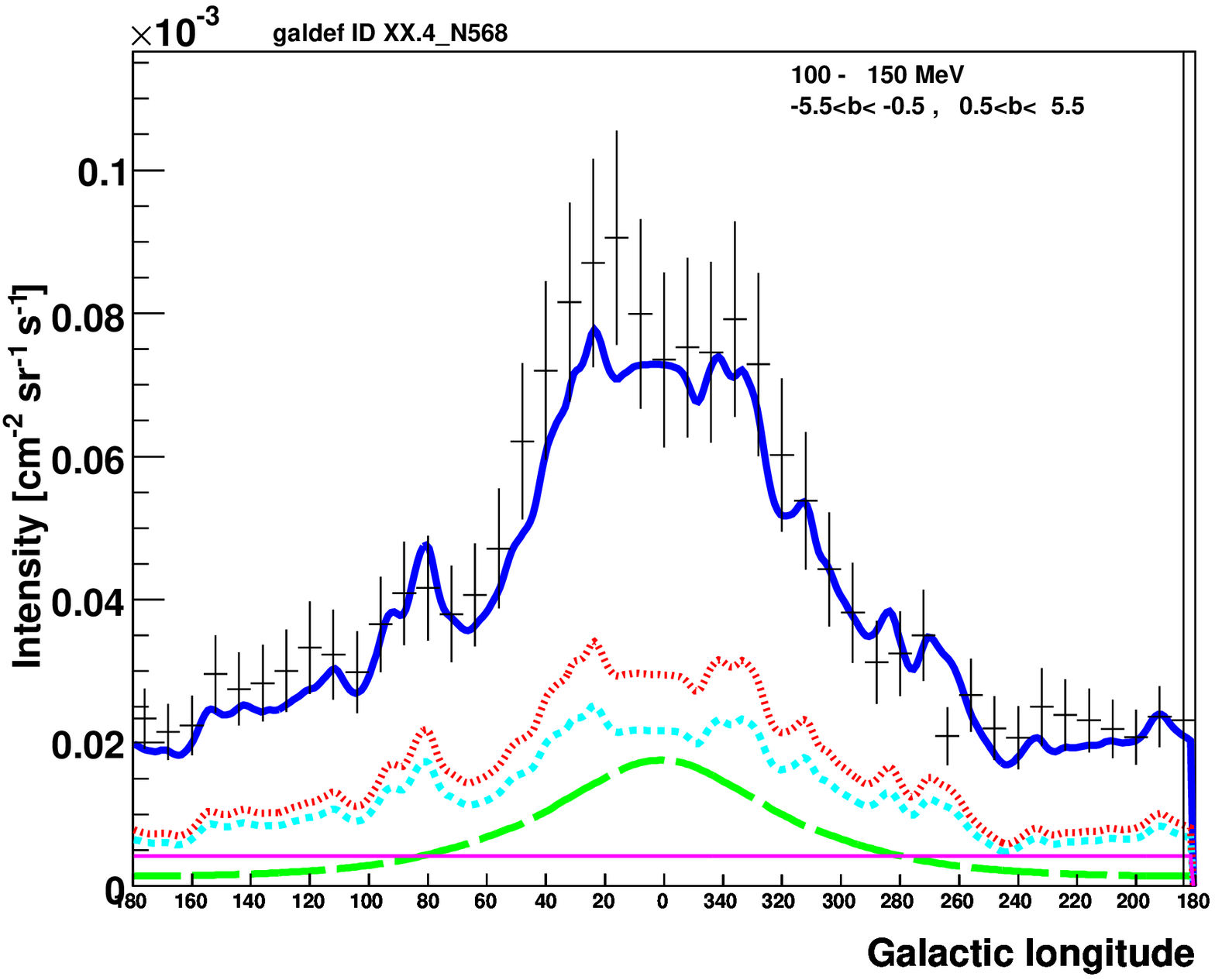}
\includegraphics[width=0.33\textwidth,clip]{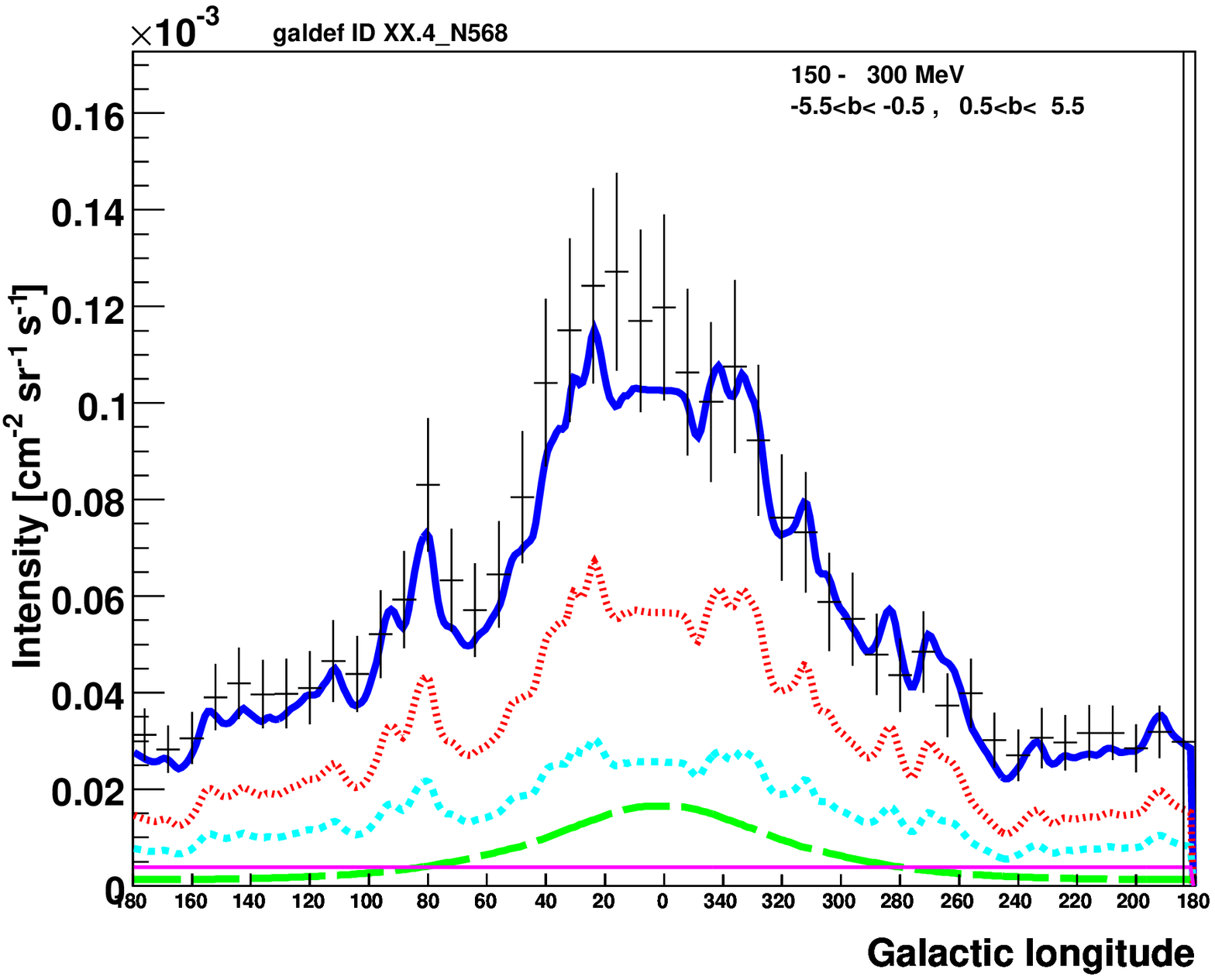}
\includegraphics[width=0.33\textwidth,clip]{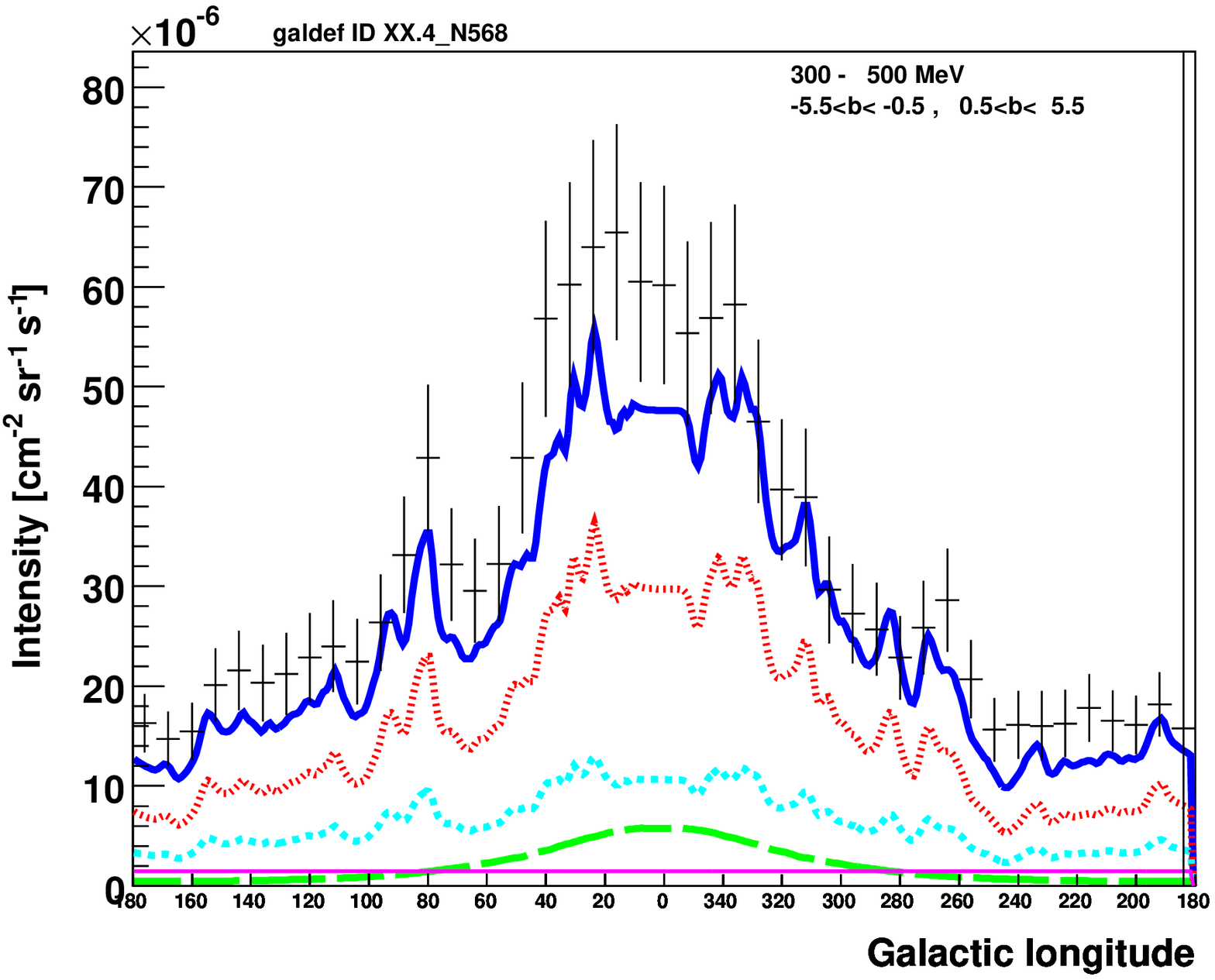}\\
\caption{Longitude profiles for the Galactic disk ($|b|<5.5$) for the aPM compared to the EGRET data between 100 and 500 MeV. Line coding as in Fig.\ref{f_conv_aPM_Regions}.}\label{f_conv_aPM_long}
\end{figure*}

\begin{figure*}
\includegraphics[width=0.33\textwidth,clip]{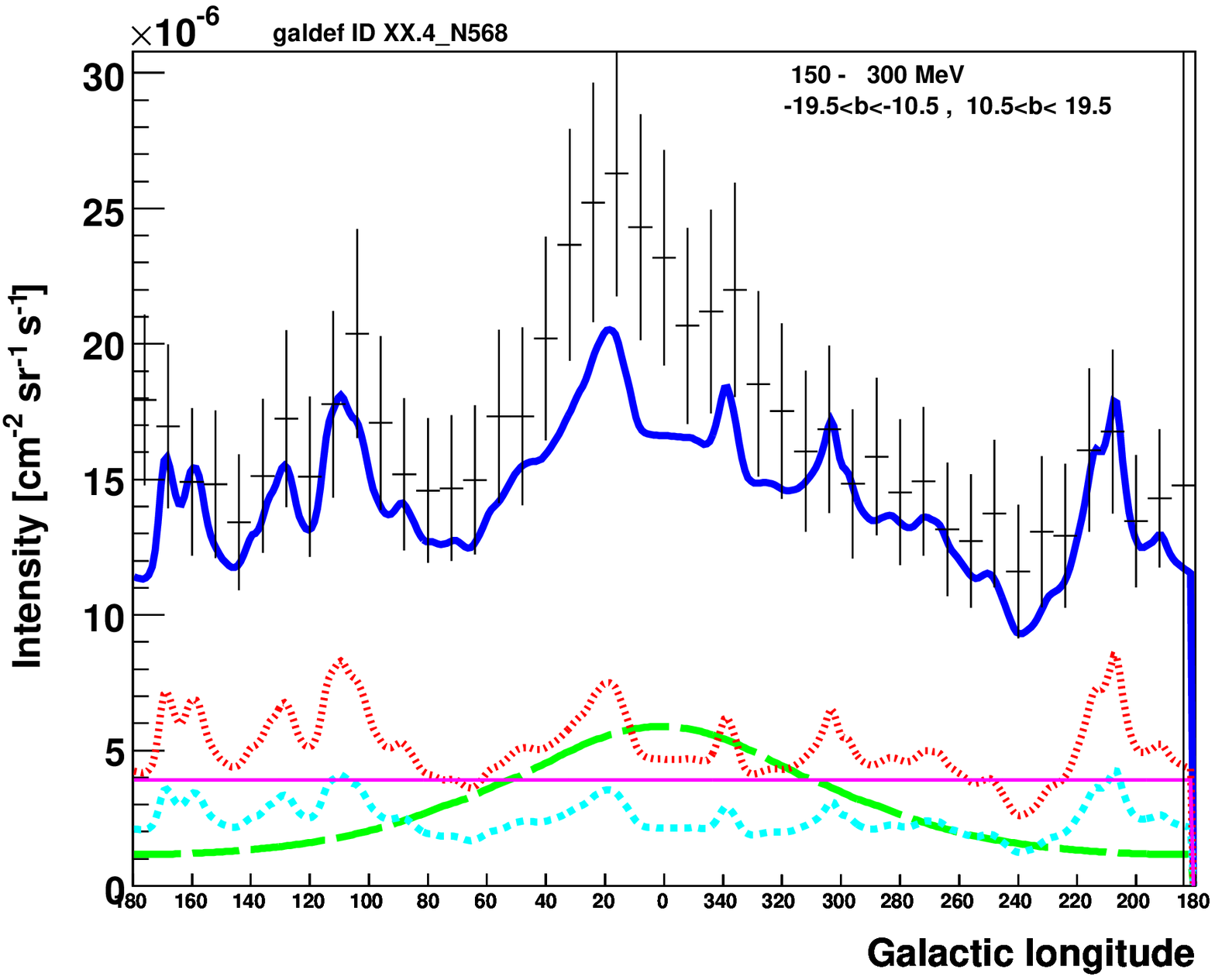}
\includegraphics[width=0.33\textwidth,clip]{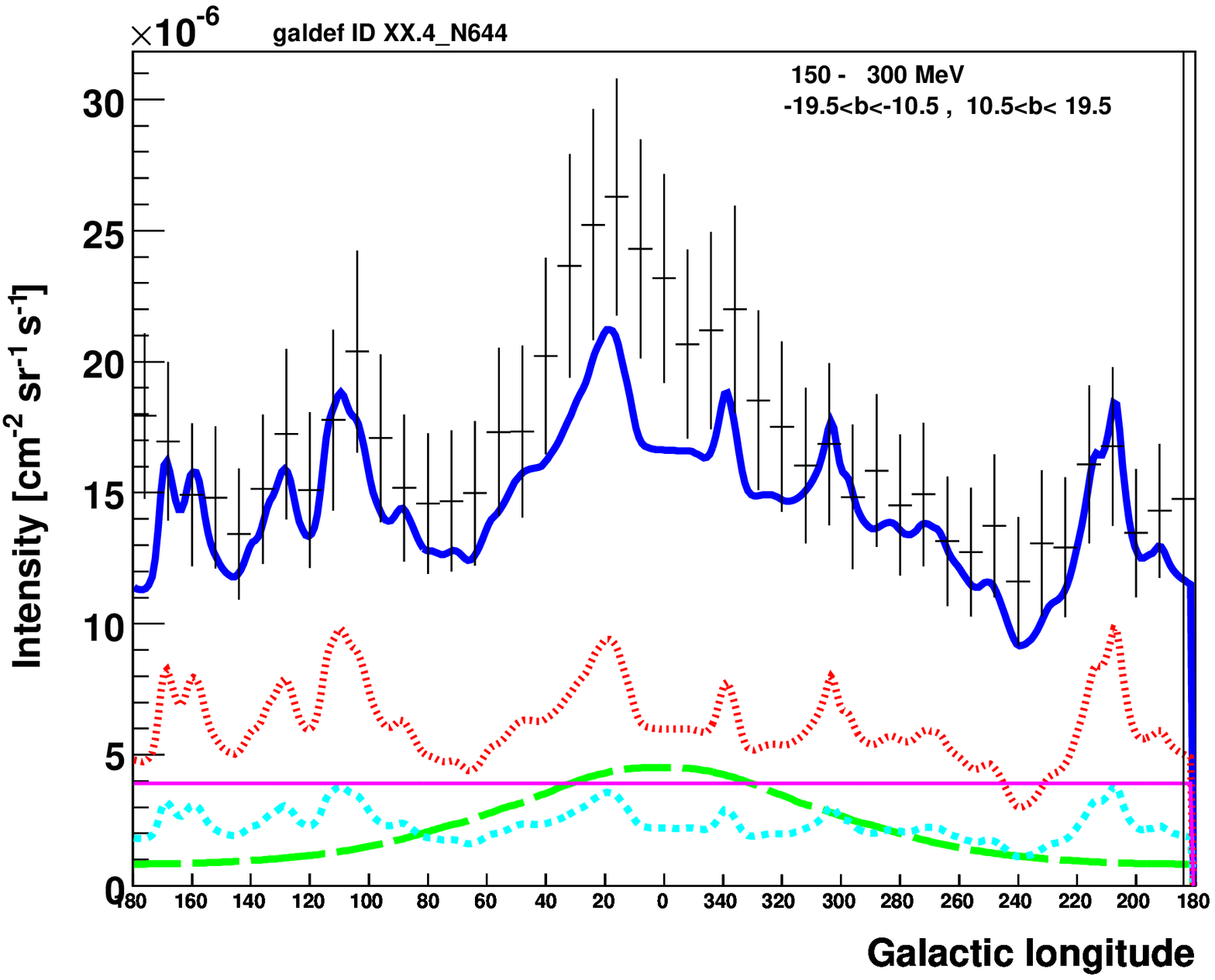}
\includegraphics[width=0.33\textwidth,clip]{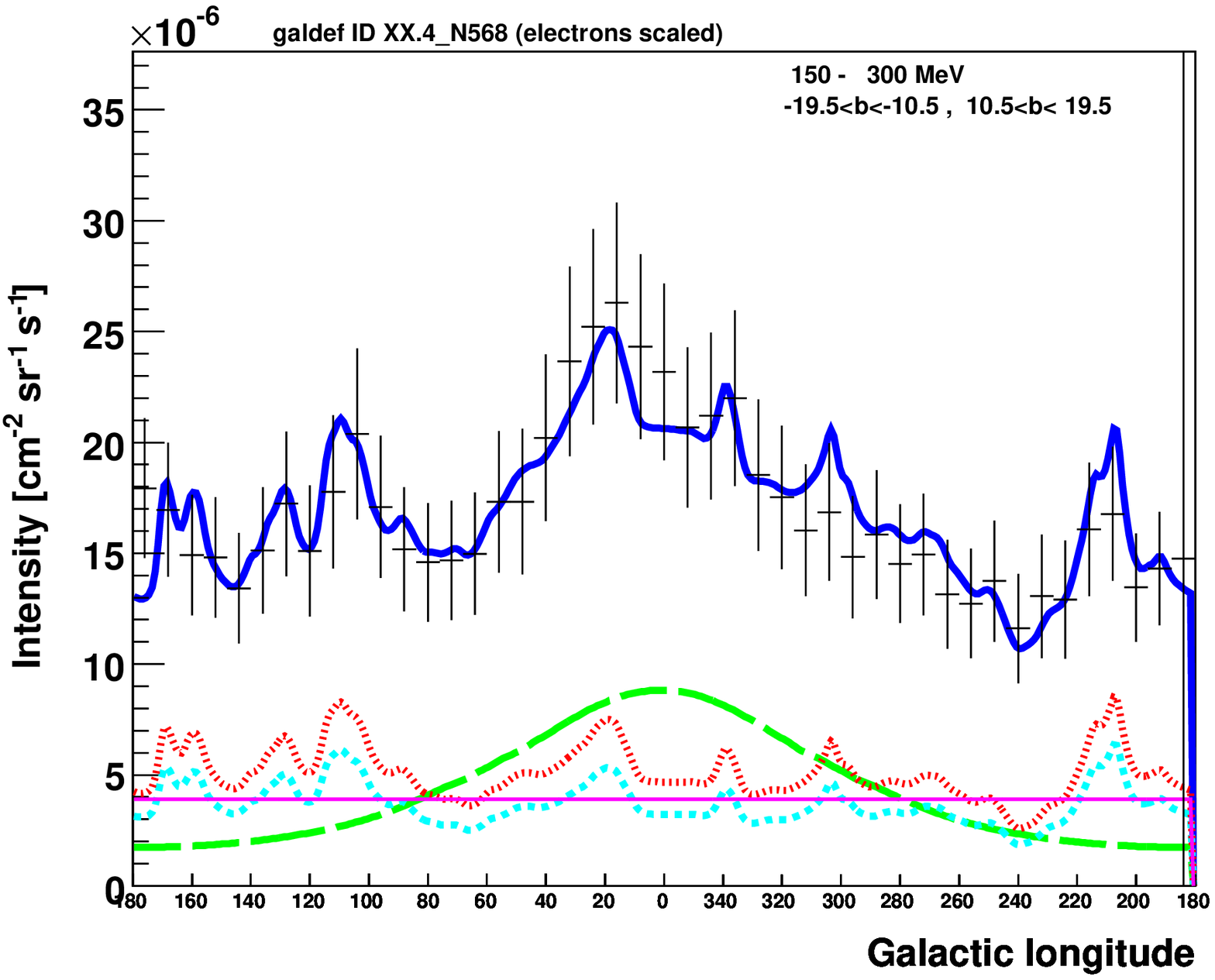}
\caption{Longitude profiles for region D in the aPM ({\bf left}), an isotropic model ({\bf middle}) and the aPM with the Galactic electron density increased by a factor 1.5 ({\bf right}) for $\gamma$-rays between 150 MeV and 300 MeV. The full blue line is the sum of the contributions from inverse Compton (green dashed), bremsstrahlung (light blue dotted), $\pi^0$-decay (red fine-dotted) and extragalactic background (purple full).}\label{long_d}
\end{figure*}

\begin{figure*}
\includegraphics[width=0.5\textwidth,clip]{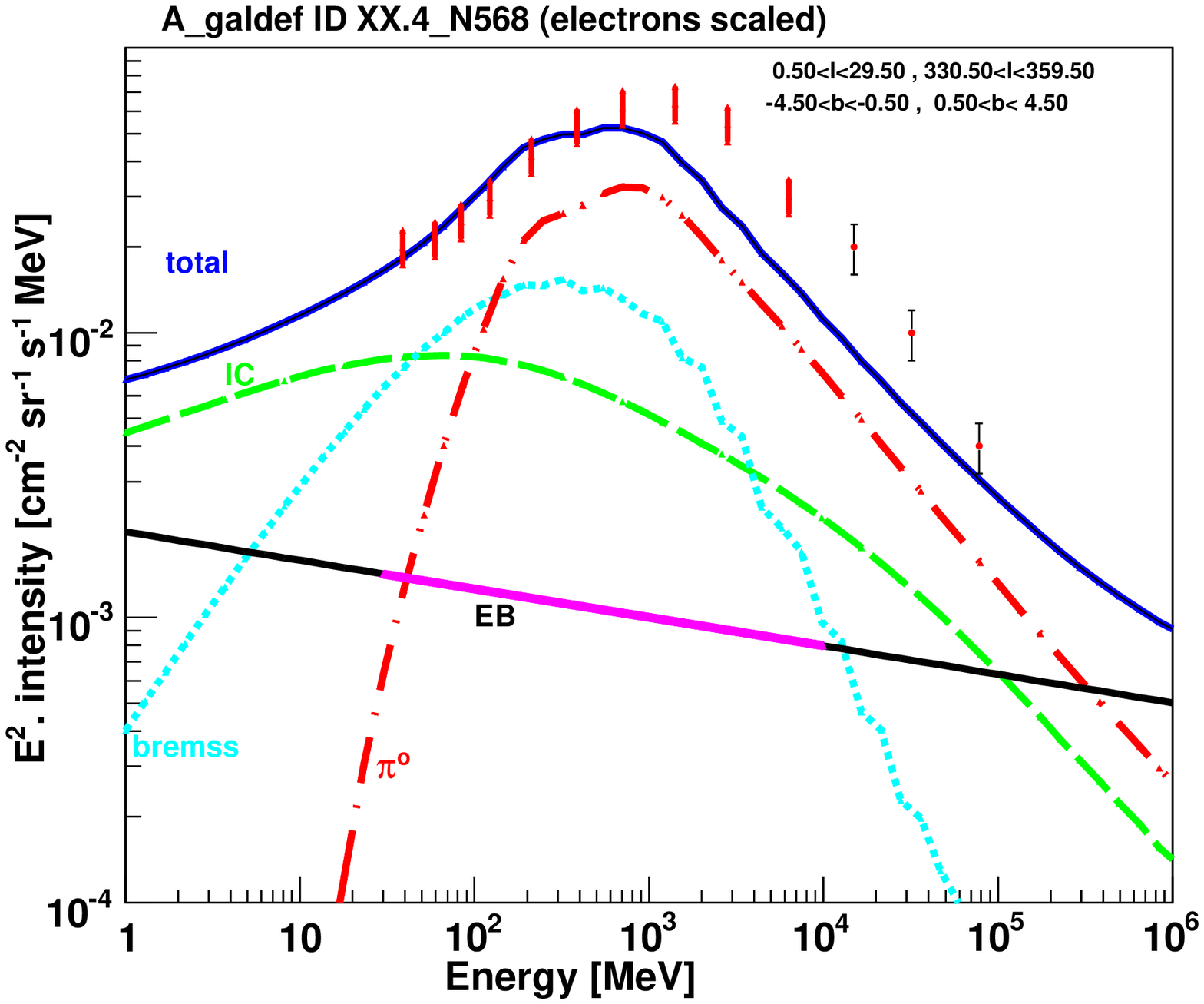}
\includegraphics[width=0.5\textwidth,clip]{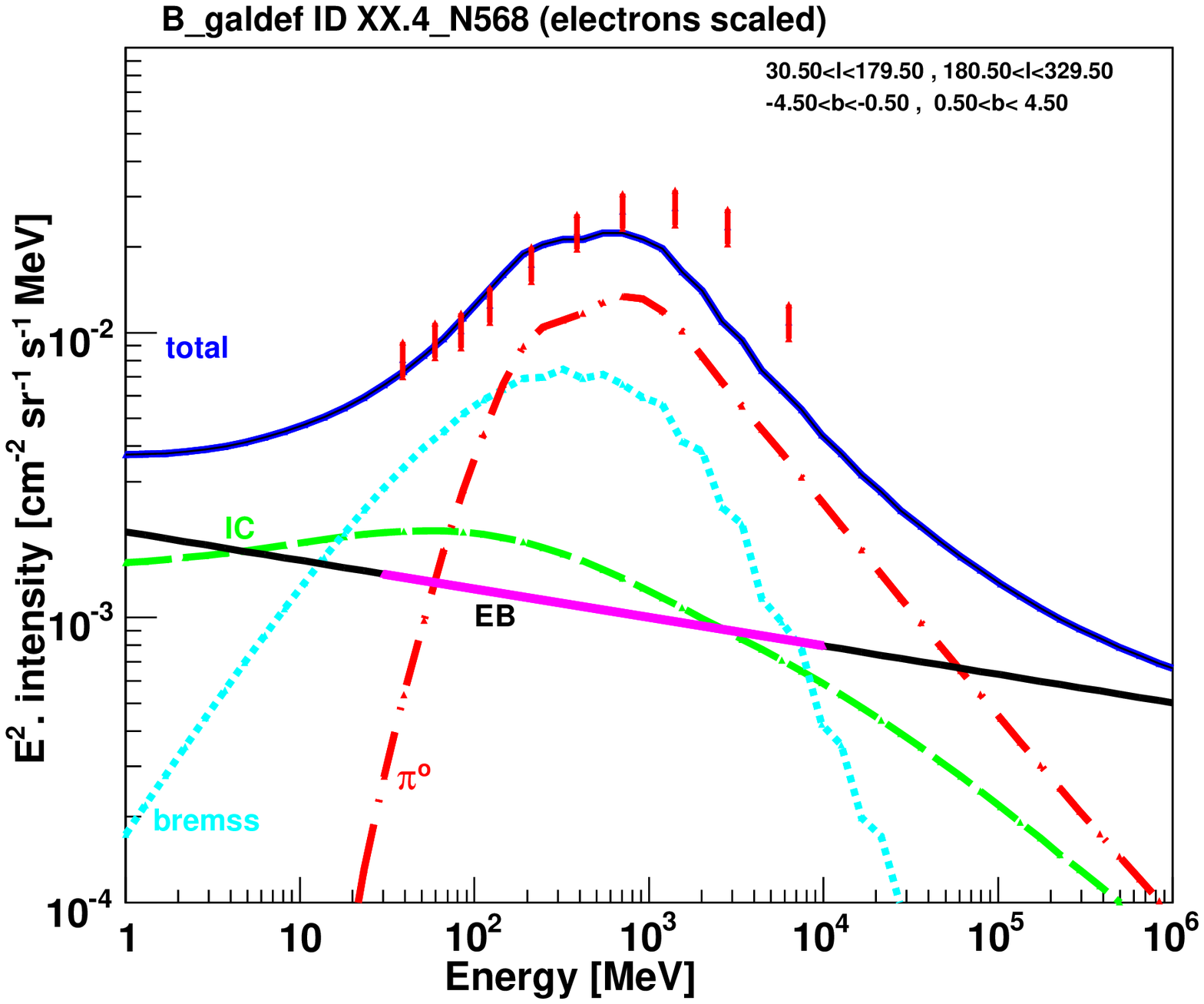}\\
\includegraphics[width=0.5\textwidth,clip]{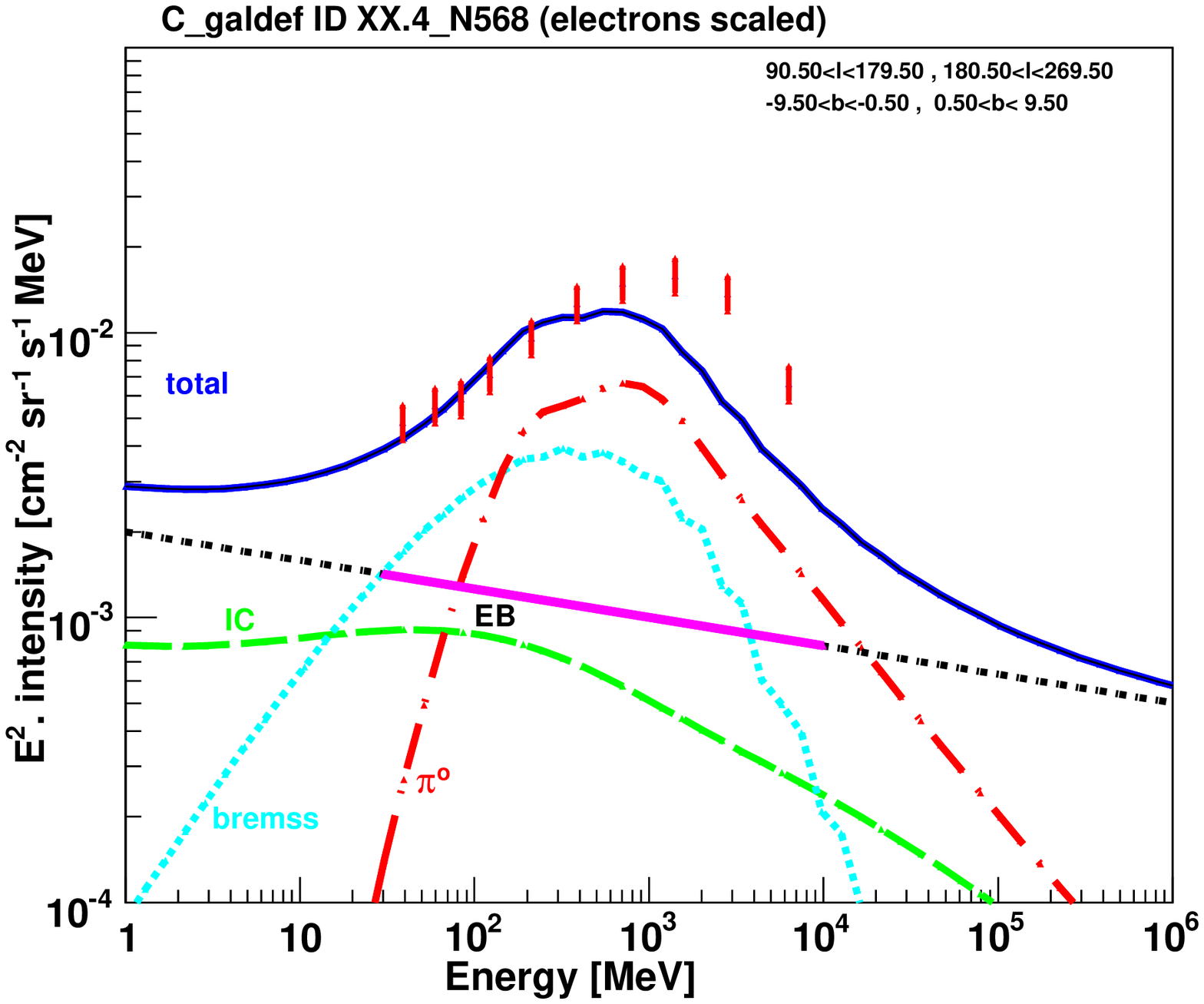}
\includegraphics[width=0.5\textwidth,clip]{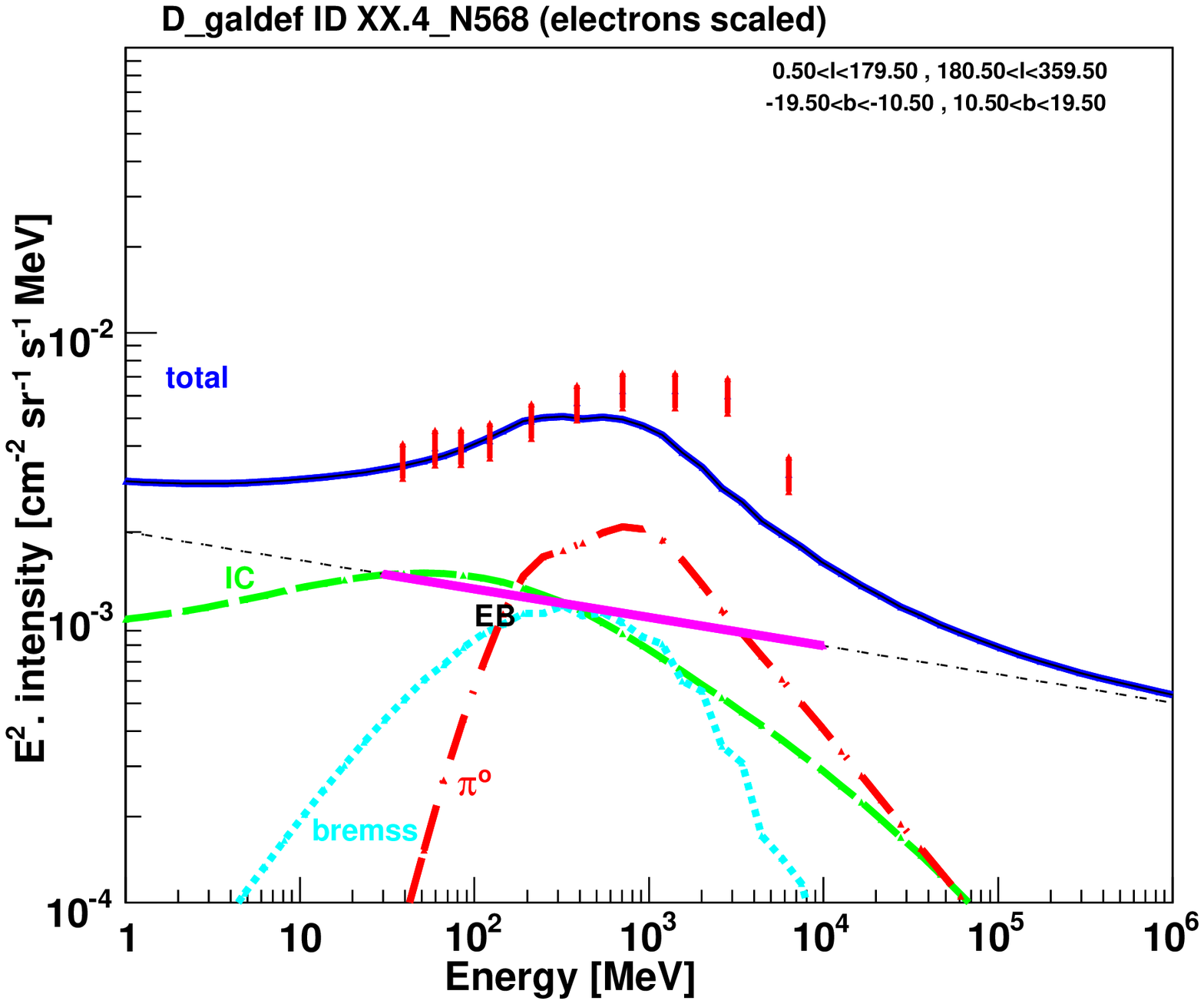}\\
\includegraphics[width=0.5\textwidth,clip]{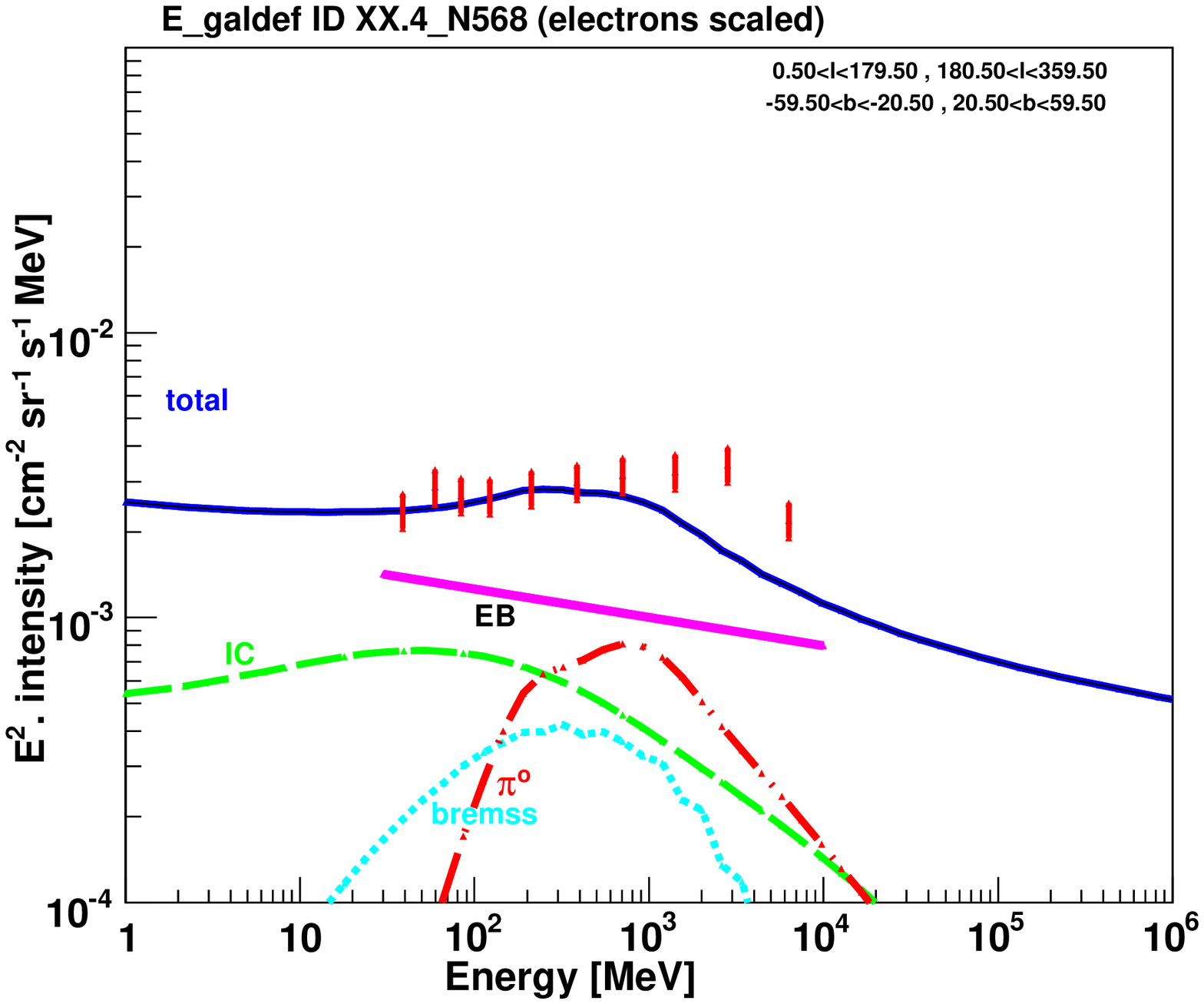}
\includegraphics[width=0.5\textwidth,clip]{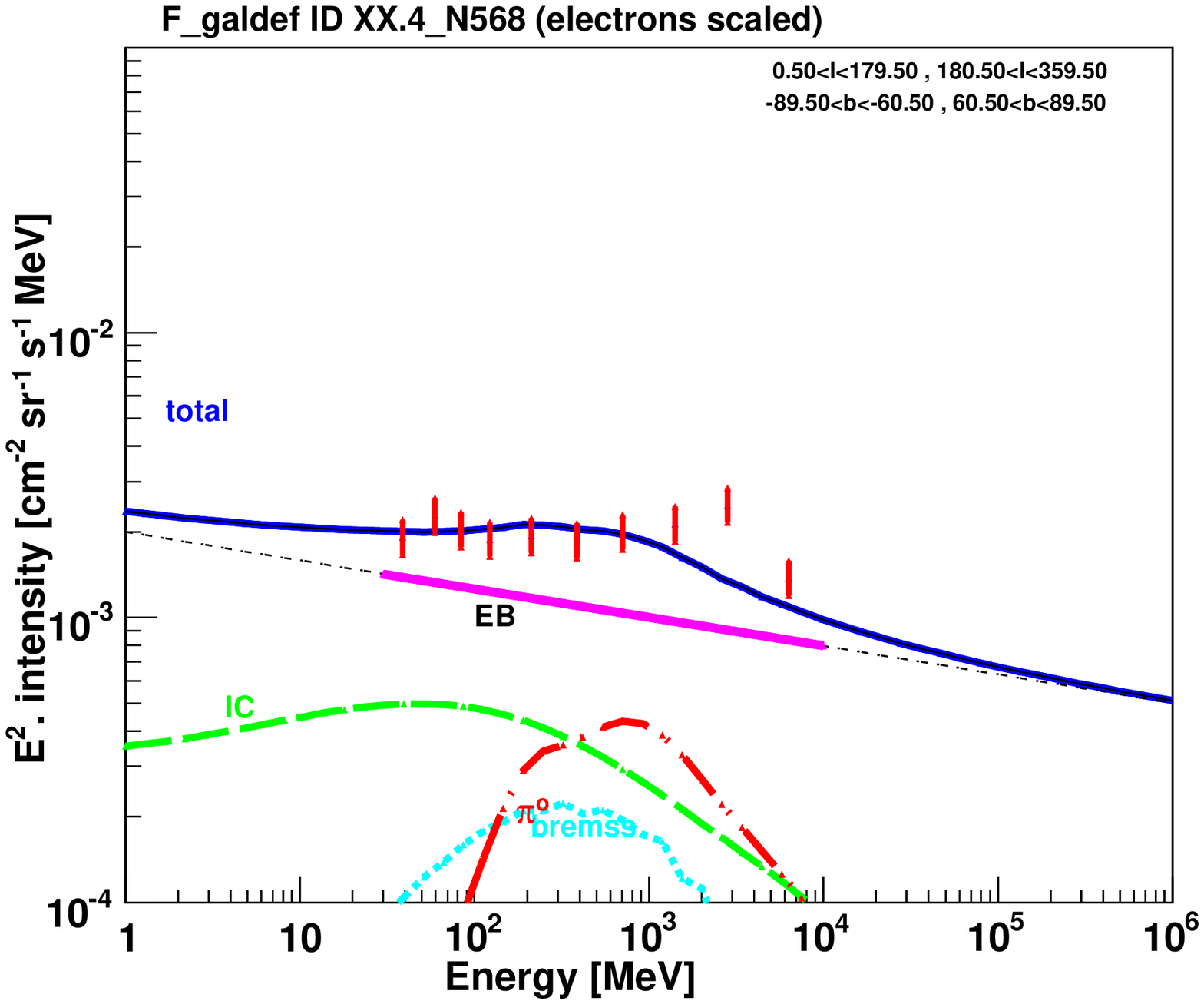}
\caption{Diffuse $\gamma$-rays for the six different sky regions as defined in \citet{Strong:2004de} for an aPM with the Galactic electron density increased by a factor 1.5. Line coding: bremsstrahlung ( {\it light blue dashed}), inverse Compton ( {\it green long dashed}), $\pi^0$-decay ( {\it red long dashed-dotted}), total ( {\it blue full}).The pink full line is the extragalactic background model according to \citet{Sreekumar:1997un}. The EGRET data are corrected for the PSF.}\label{f_conv_aPM_Regions_1.5}
\end{figure*}

\begin{figure*}[]
\includegraphics[width=0.33\textwidth,clip]{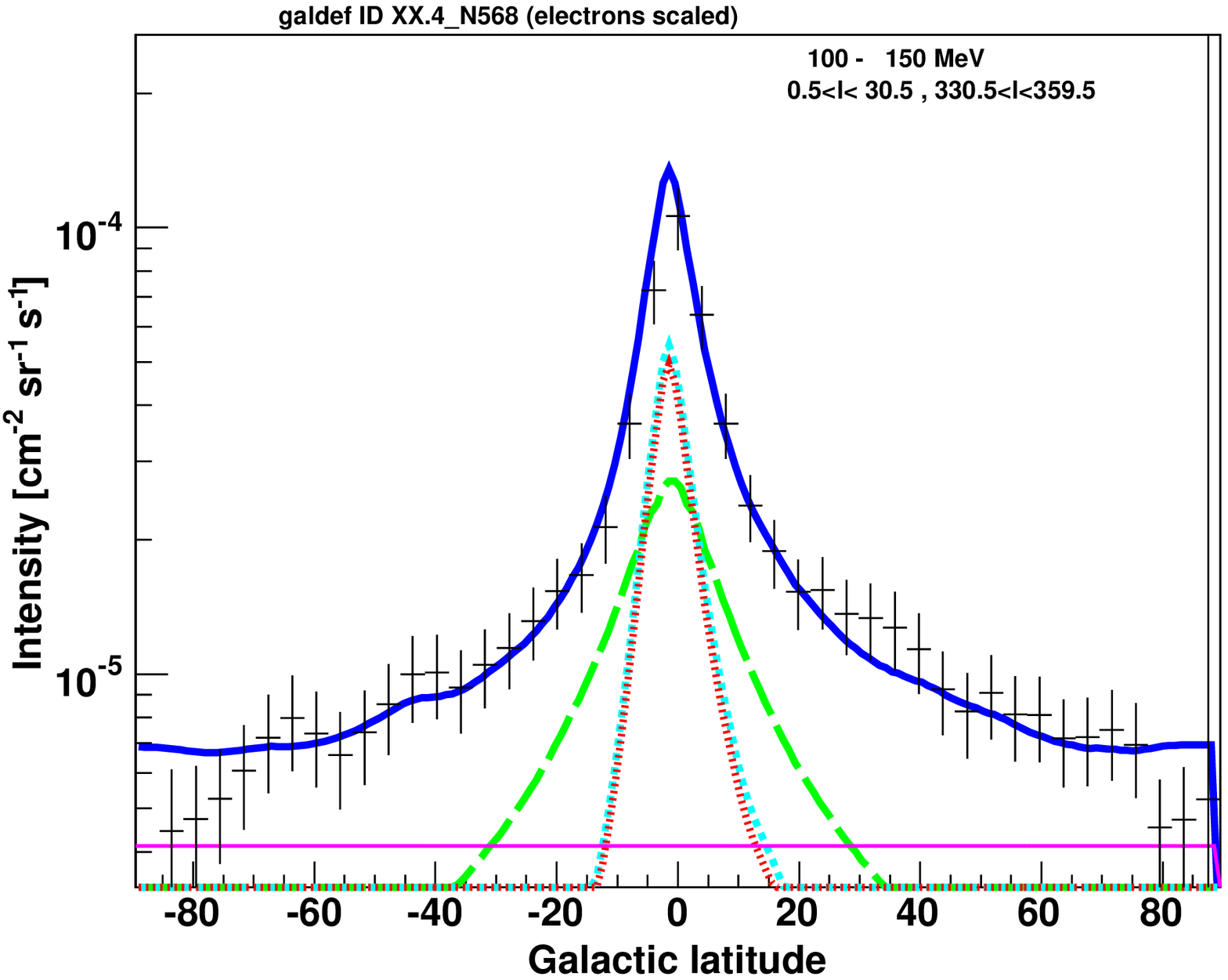}
\includegraphics[width=0.33\textwidth,clip]{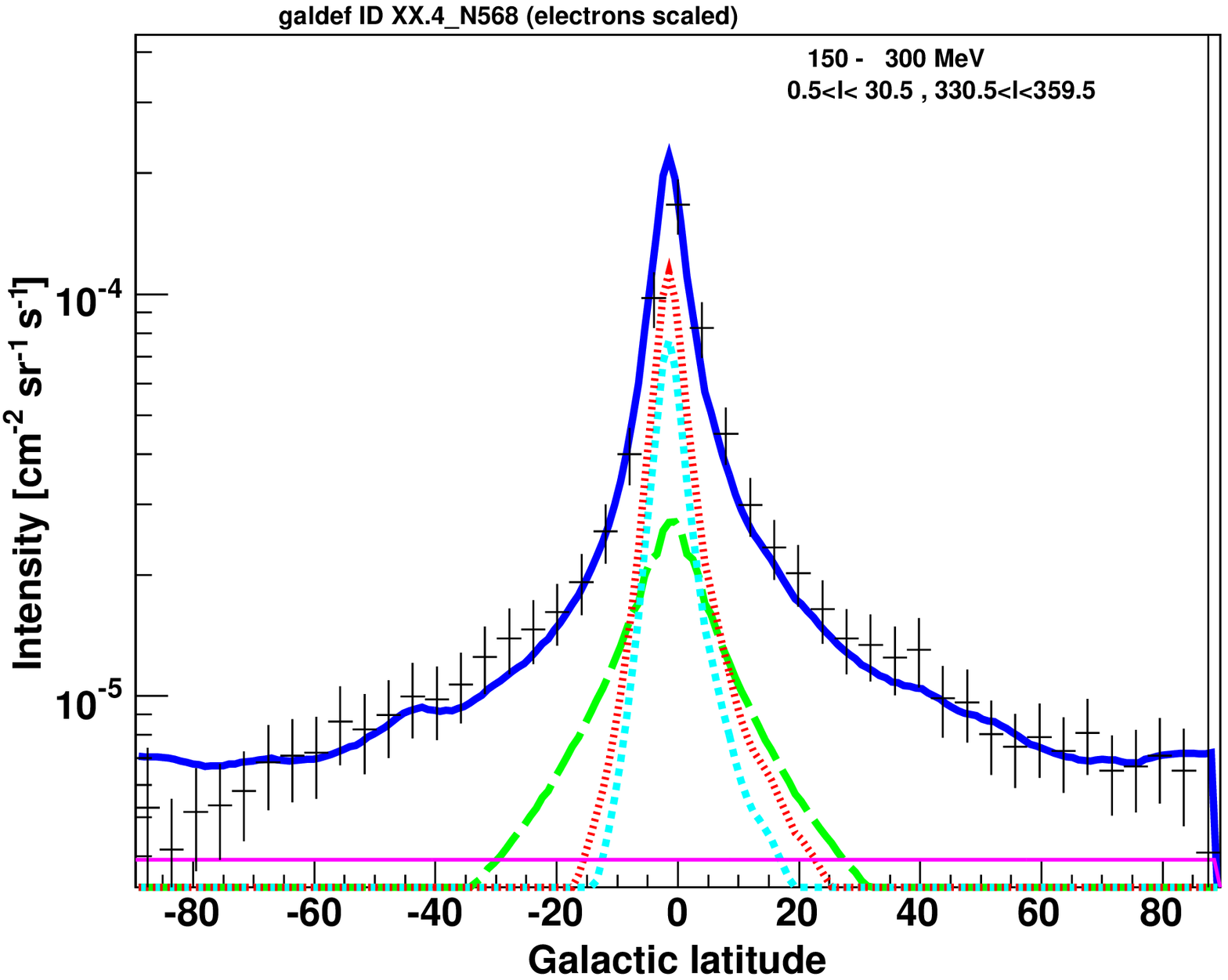}
\includegraphics[width=0.33\textwidth,clip]{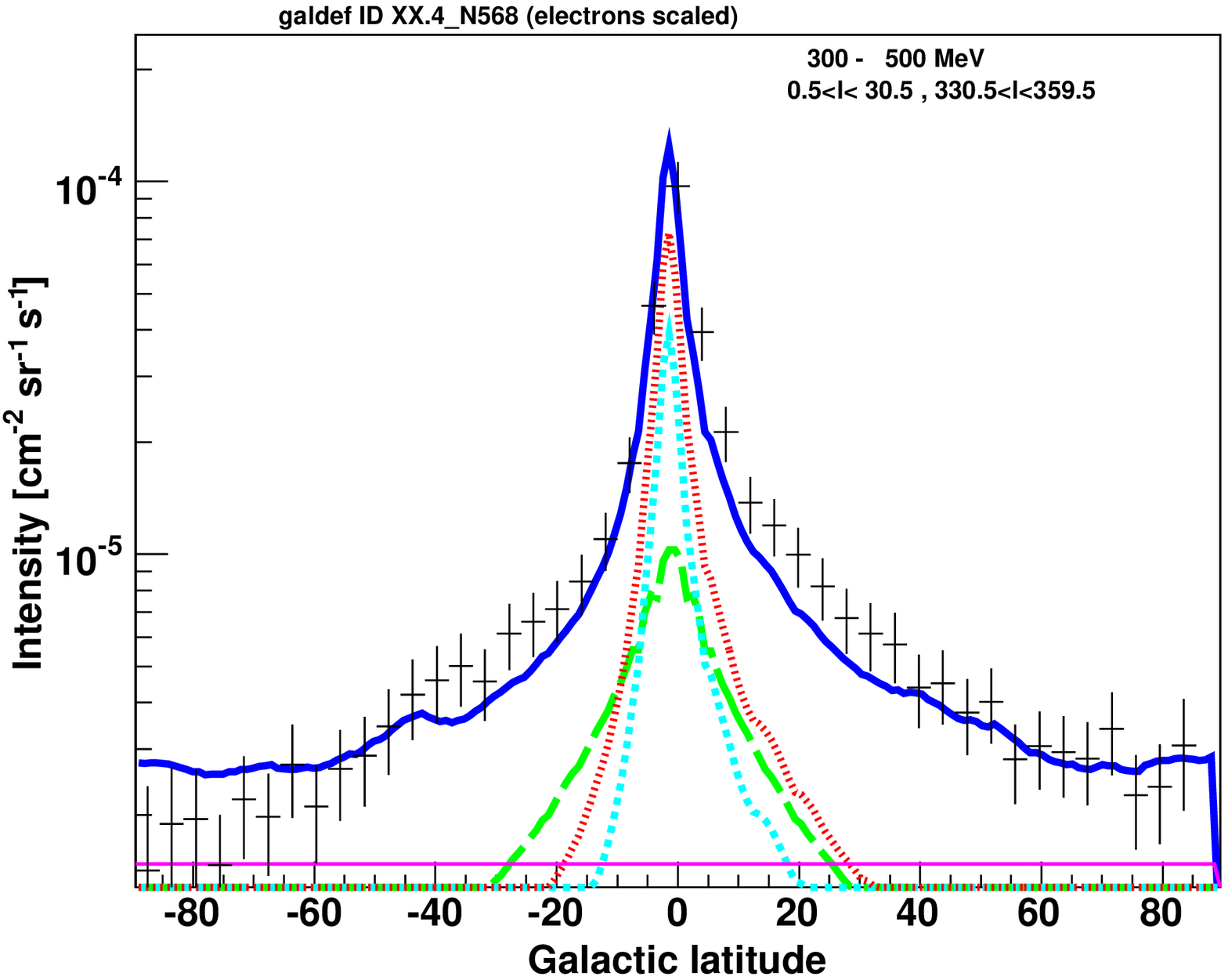}
\caption{Latitude profiles for the inner Galaxy ($|l|<30.5$) for the aPM with the Galactic electron density increased by a factor 1.5 compared to EGRET data between 100 and 500 MeV. Line coding as in Fig.\ref{f_conv_aPM_Regions_1.5}.}\label{f_conv_aPM_lat1.5}
\end{figure*}

\begin{figure*}[]
\includegraphics[width=0.33\textwidth,clip]{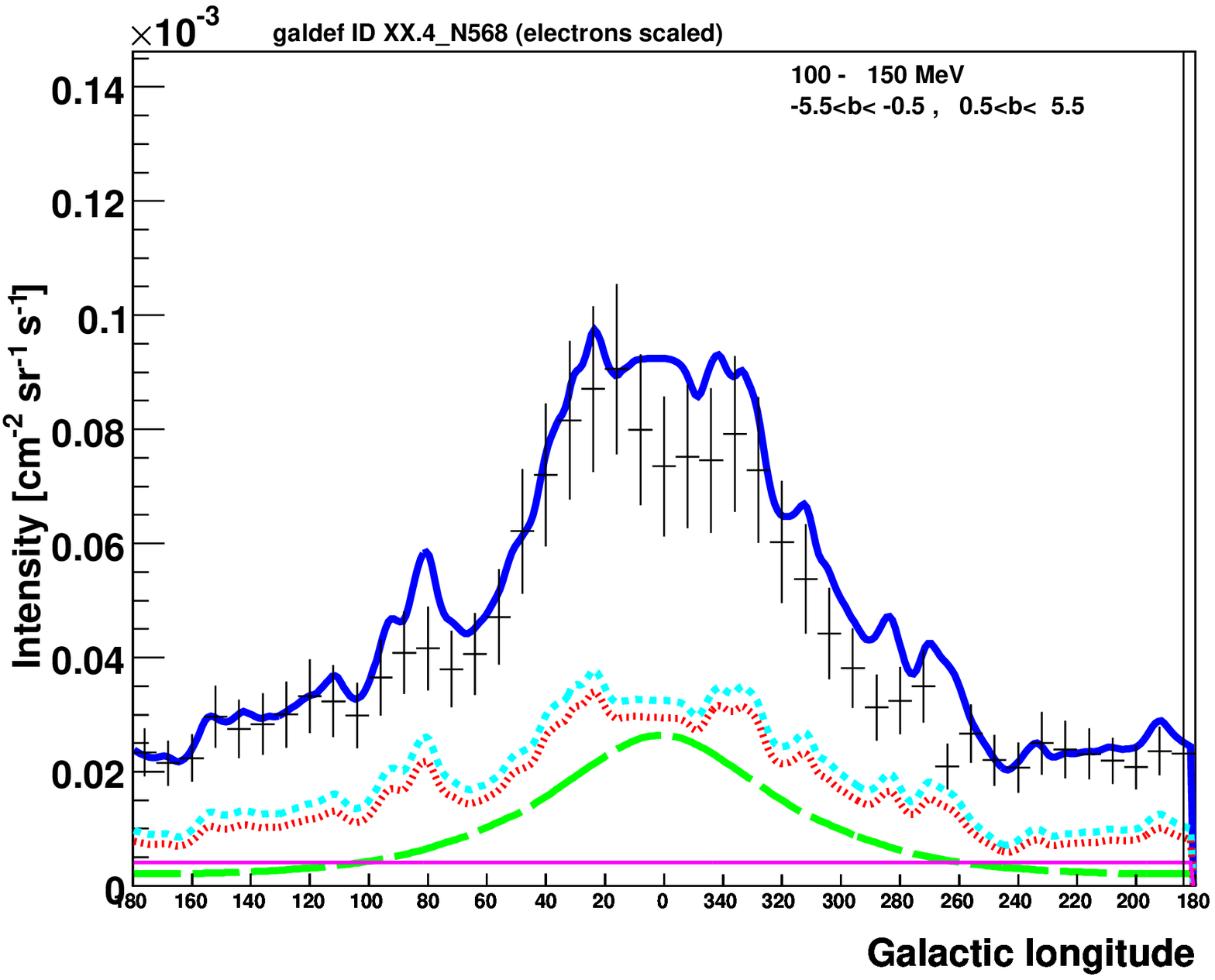}
\includegraphics[width=0.33\textwidth,clip]{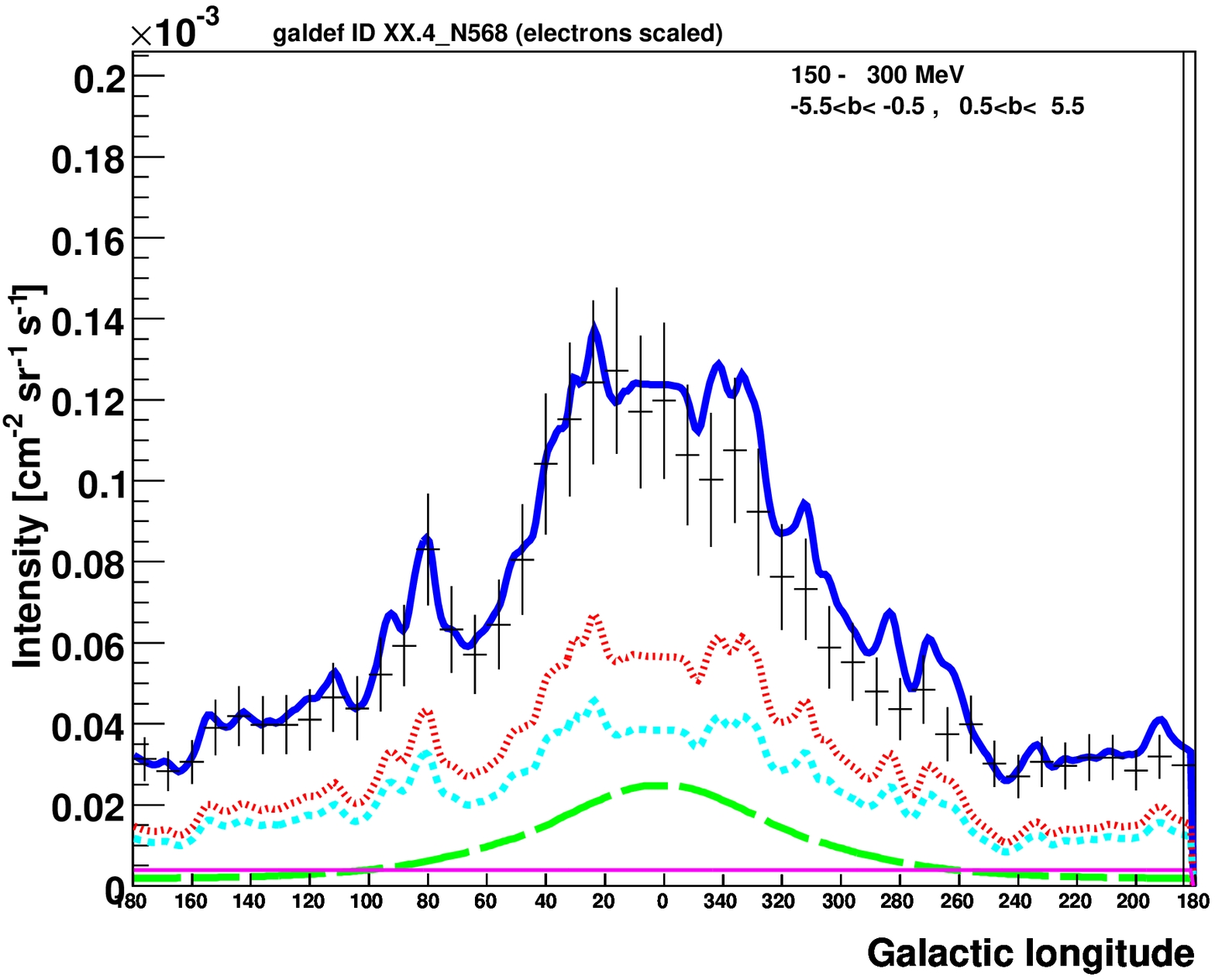}
\includegraphics[width=0.33\textwidth,clip]{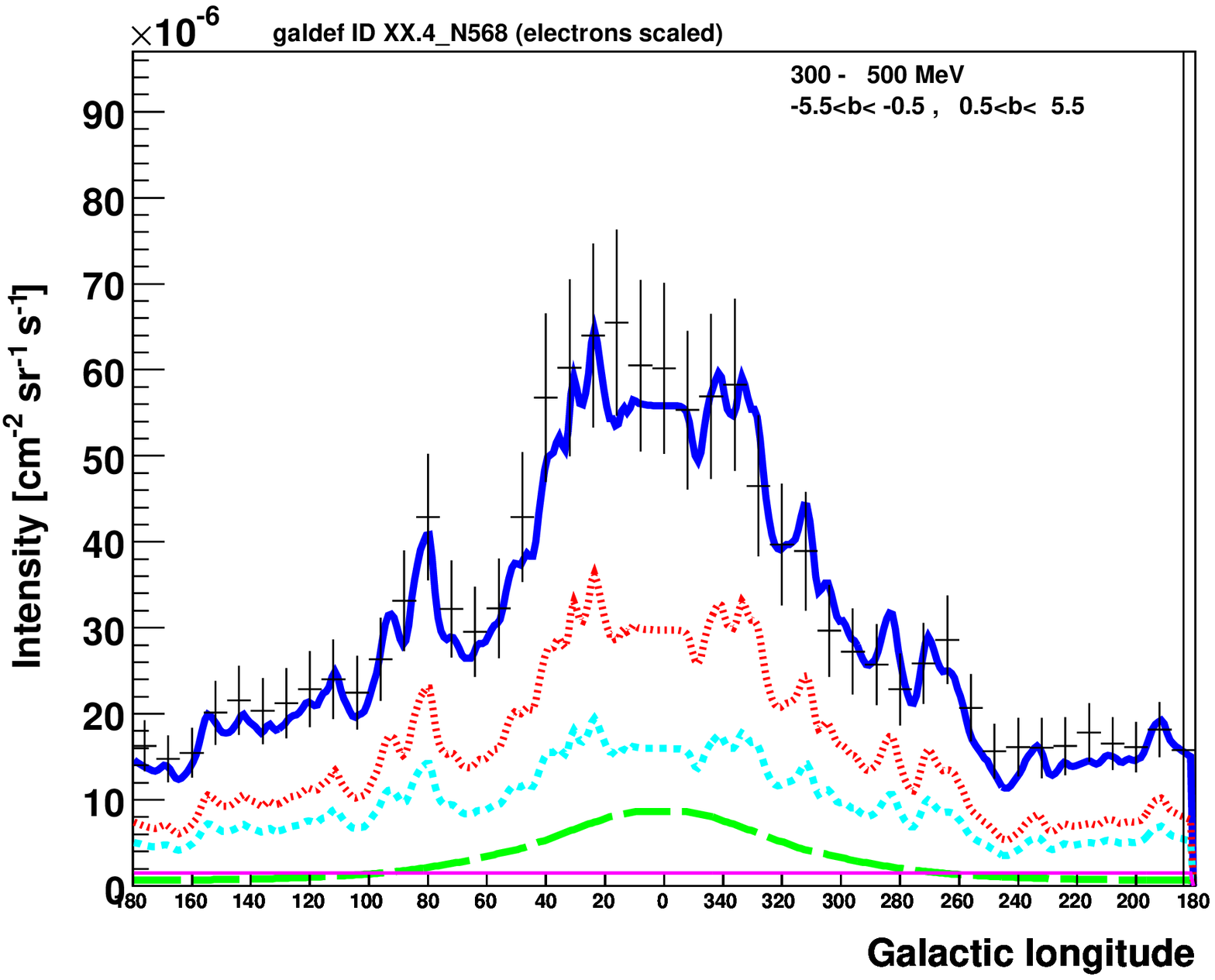}
\caption{Longitude profiles for the Galactic disk ($|b|<5.5$) for the aPM with the Galactic electron density increased by a factor 1.5 compared to EGRET data between 100 and 500 MeV. Line coding as in Fig.\ref{f_conv_aPM_Regions_1.5}.}\label{f_conv_aPM_long1.5}
\end{figure*}

The diffuse Galactic $\gamma$-rays are shown in Fig. \ref{f_conv_aPM_Regions}, the regions used in
this figure are the regions as introduced in \citep{Strong:2005zn}. The latitude and longitude ranges are given in Table \ref{table3}.
The $\gamma$-ray prediction of the aPM in general is similar to the prediction of an isotropic model with the same source distribution and constant $X_{CO}$ (shown in Fig. \ref{f_conv_aPM_Regions} as the dotted blue line). For region A (GC) the flux in the isotropic model is somewhat larger than for the aPM. This is the result of a too high proton density close to the sources in the absence of convection. For the other regions the $\gamma$-ray flux in the aPM is insignificantly larger than in the isotropic model, which is mainly the result of slight deviations in the propagated proton and electron spectra.
The latitude and
longitude profiles for the inner Galaxy are presented in Figs. \ref{f_conv_aPM_lat} and \ref{f_conv_aPM_long} for $100 ~\mathrm{MeV}<E<500 ~\mathrm{MeV}$. The model shows the same deficiency of IC emission at intermediate latitudes as as the conventional isotropic model, this can also be seen from region D in Fig. \ref{f_conv_aPM_Regions}. Note, that we use a simple power-law extragalactic background model. A more detailed model may yield a better description of the $\gamma$-ray flux from the halo and at intermediate latitudes. However, this extragalactic background model should be determined from the Fermi-LAT data. The longitude distribution at intermediate latitudes, shown in Fig. \ref{long_d}, reveals that the deficiency almost exclusively results from a lack of diffuse $\gamma$-rays from the GC region for both, the aPM and the isotropic model. Since bremsstrahlung and emission from $\pi^0$-decay are rather flat for intermediate latitudes, an increased contribution from IC would improve the model prediction.
Independent of direction the model prediction below 100 MeV is somewhat too low, which also indicates that the contribution from IC is underestimated. The deficiency in IC can be remedied by assuming that the Galactic electron density is somewhat larger than the local electron density, thus leading to a larger contribution from IC at low energies, as has been done previously by \citet{Strong:2004de}.
Figure \ref{f_conv_aPM_Regions_1.5} shows the $\gamma$-ray spectra for a Galactic electron density increased by a factor 1.5, while the Galactic proton density remains unchanged.
Figures \ref{f_conv_aPM_lat1.5} and \ref{f_conv_aPM_long1.5} show the longitude and latitude profiles for the inner Galaxy for an aPM with the Galactic electron density increased by a factor 1.5 and the proton density kept constant. The right side of Fig. \ref{long_d} shows the longitude profile at between 150 and 300 MeV for intermediate latitudes.\\
Whether or not the Galactic electron density can be different from the local electron density strongly depends on the transport parameters and the electron energy losses.  Another possibility would be an untraced component of non-equilibrium gas \citep{Breitschwerdt:2006bs}. In this case an additional bremsstrahlung contribution from electrons in the hot gas would be expected. This contribution would be most significant at large latitudes in the source region, where the relative component of the hot gas blown out by SNs is expected to dominate. This is exactly what is required by the latitude and longitude profiles in Figs. \ref{f_conv_aPM_lat} and \ref{long_d}.\\
Alternatively an increase in the proton density above the plane in the inner Galaxy would lead to a more pronounced peak in the $\pi^0$ emission at intermediate latitudes. This could be achieved by a smaller diffusion coefficient in this region which would be well motivated by the decrease of the regular magnetic field above the disk. However, this will not improve the model prediction below 100 MeV, so that additionally a softer electron spectrum would be required.\\
As discussed earlier the aPM is almost completely independent of the position of the boudary condition, i.e. the halo size, provided, that the boundary is positioned well outside the diffusion zone limited by $z_c$. In this model it is therefore possible to examine the impact of additional IC emission from an extended halo with $z_h$=100 kpc without loosening the constraints on local CRs. The gain in additional IC at intermediate latitudes is at the \% level and thus negligible in the total flux.  An increased halo therefore will not improve the $\gamma$-ray emission from the halo significantly. An additional untraced gas component in the source region, a larger Galactic electron density or possibly an additional electron population appears to be required both in the isotropic model and the aPM.


\subsection{Soft $\gamma$-ray gradient}\label{ss_gradient}
\begin{figure*}
\includegraphics[width=0.33\textwidth,clip]{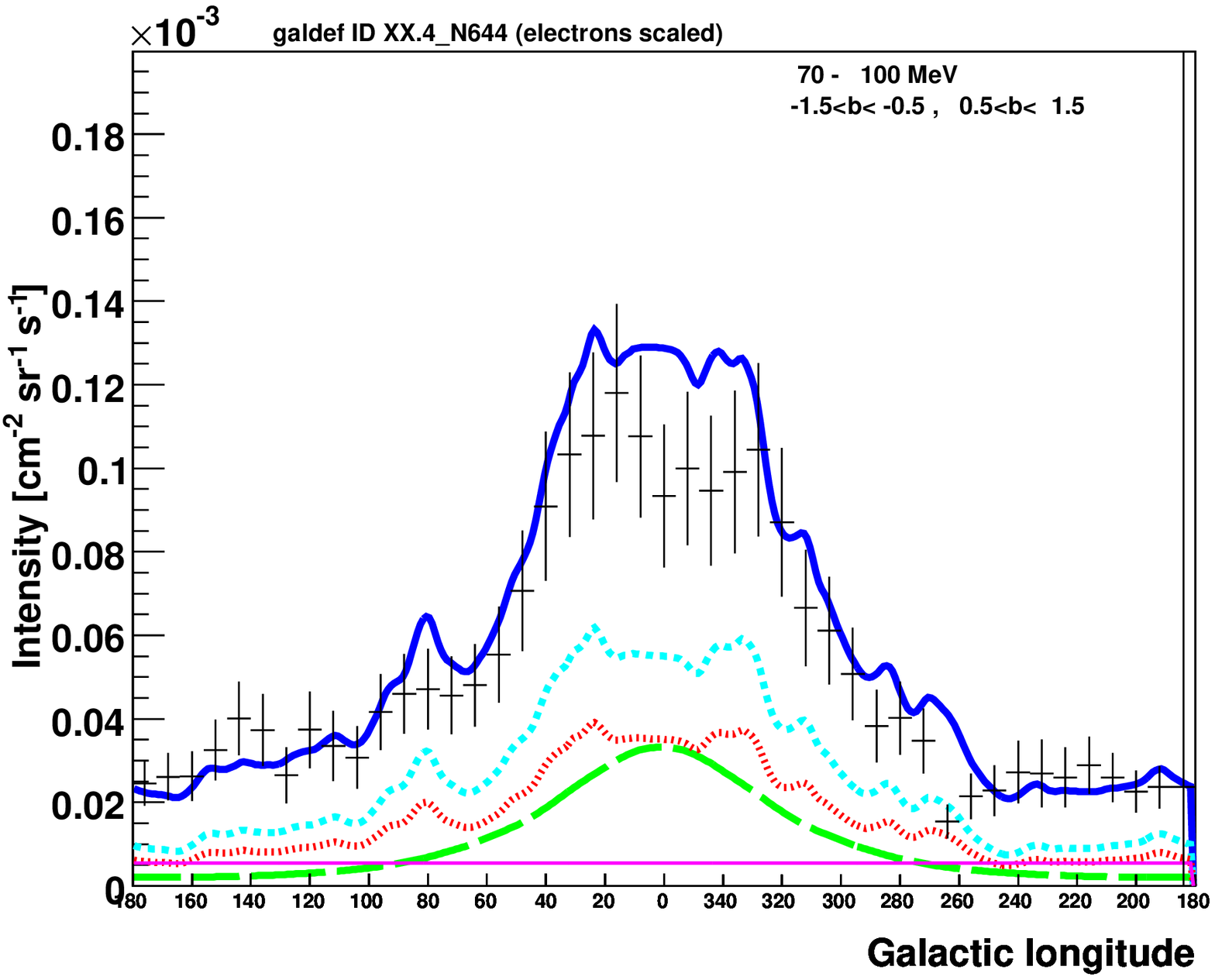}
\includegraphics[width=0.33\textwidth,clip]{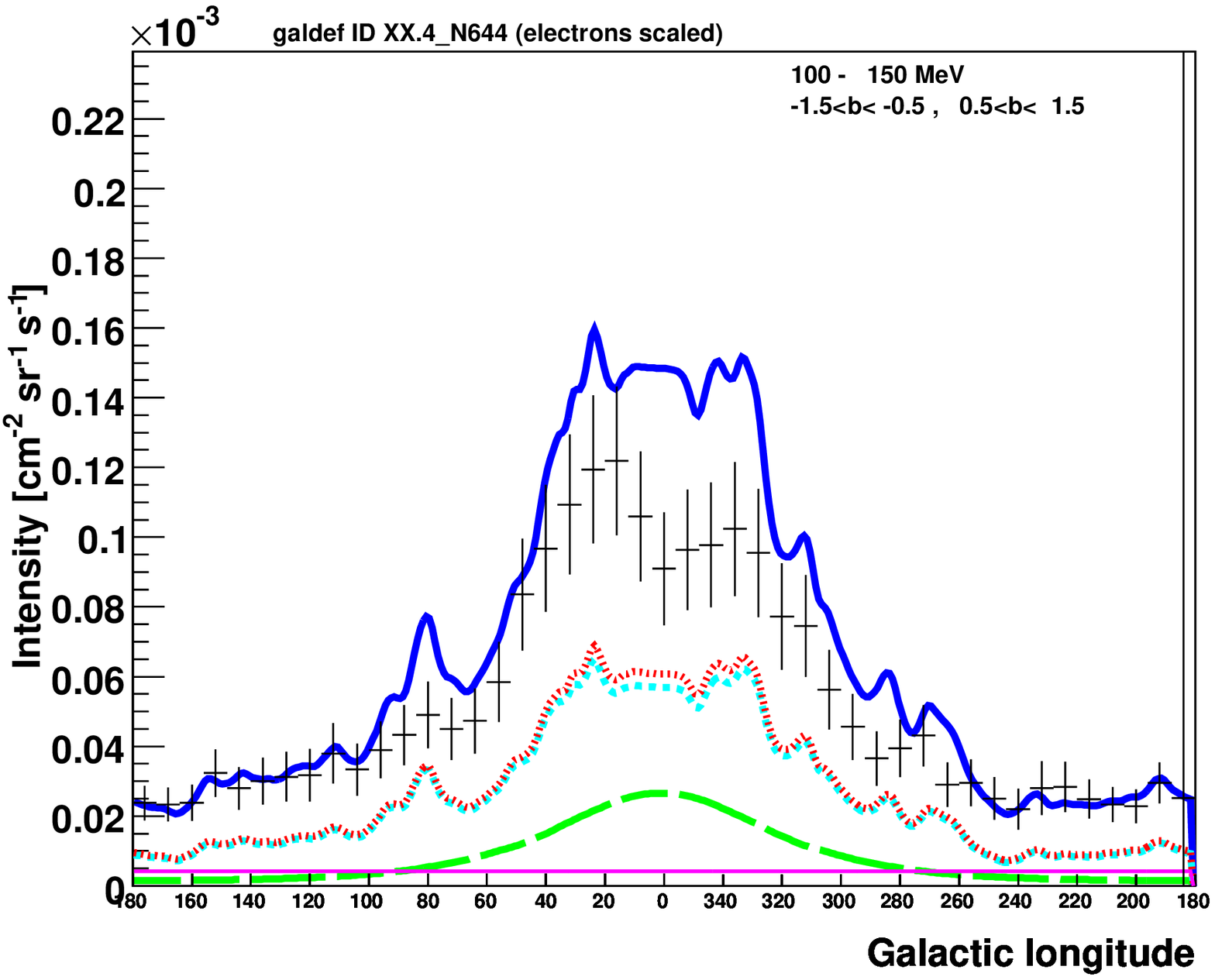}
\includegraphics[width=0.33\textwidth,clip]{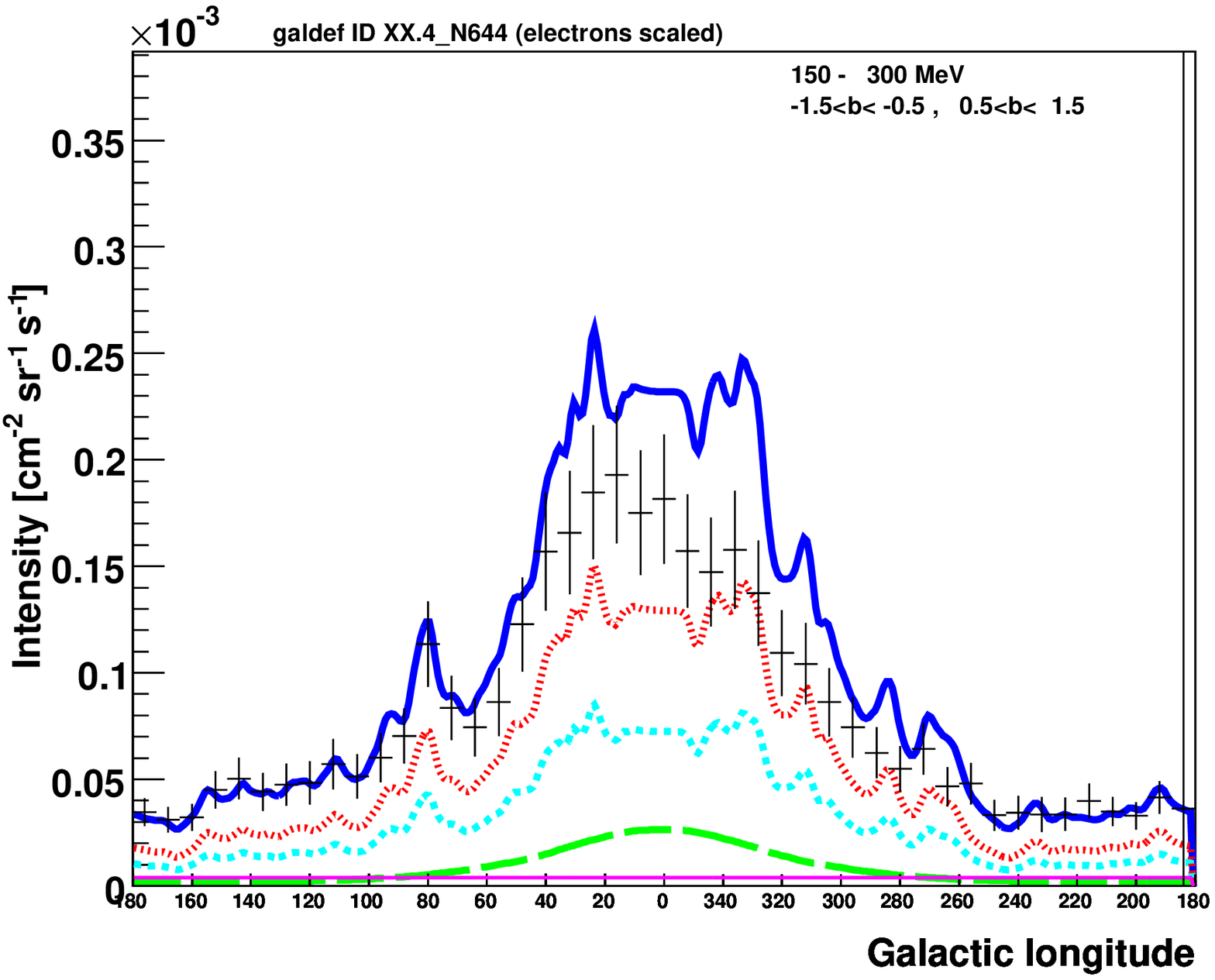}\\
\includegraphics[width=0.33\textwidth,clip]{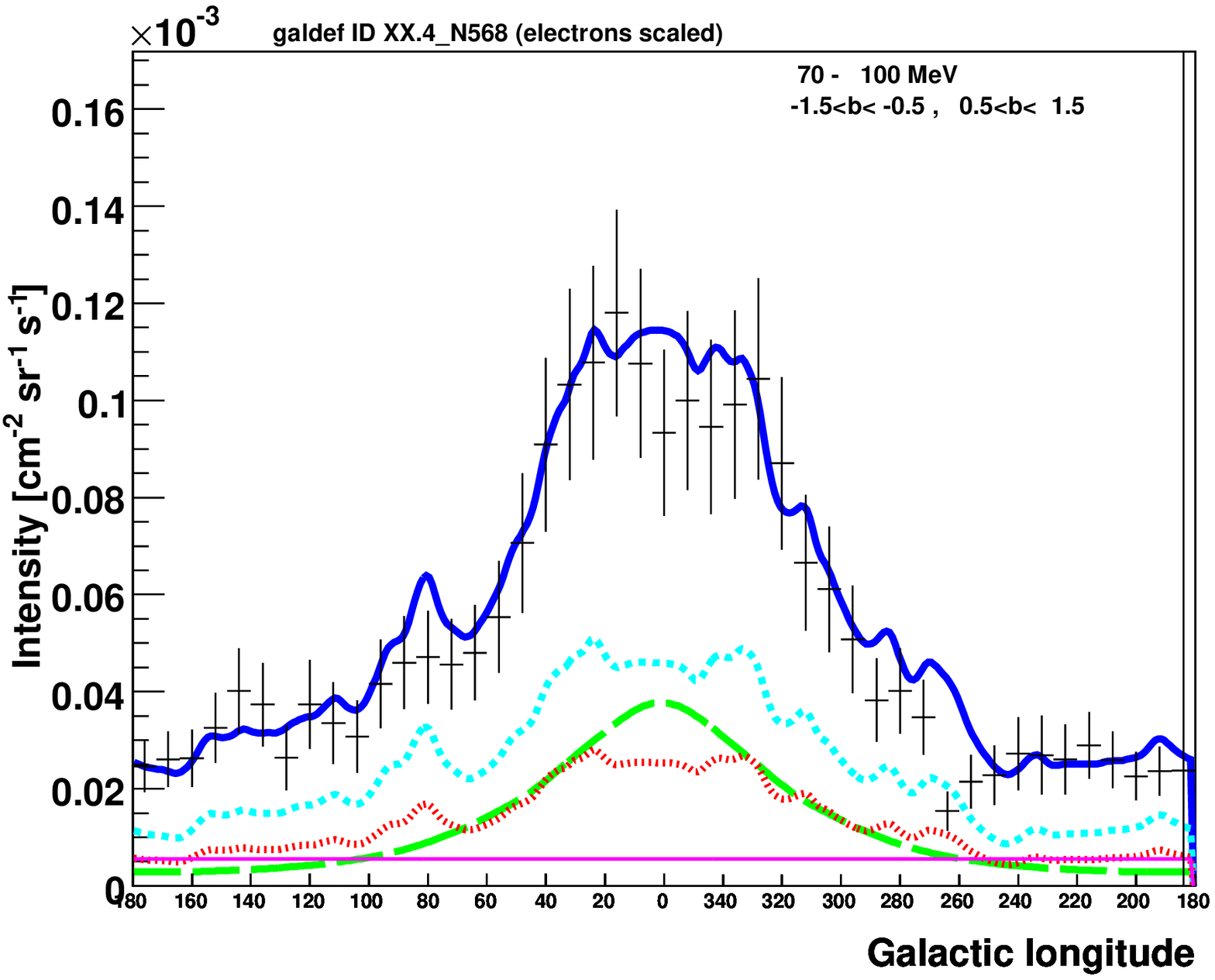}
\includegraphics[width=0.33\textwidth,clip]{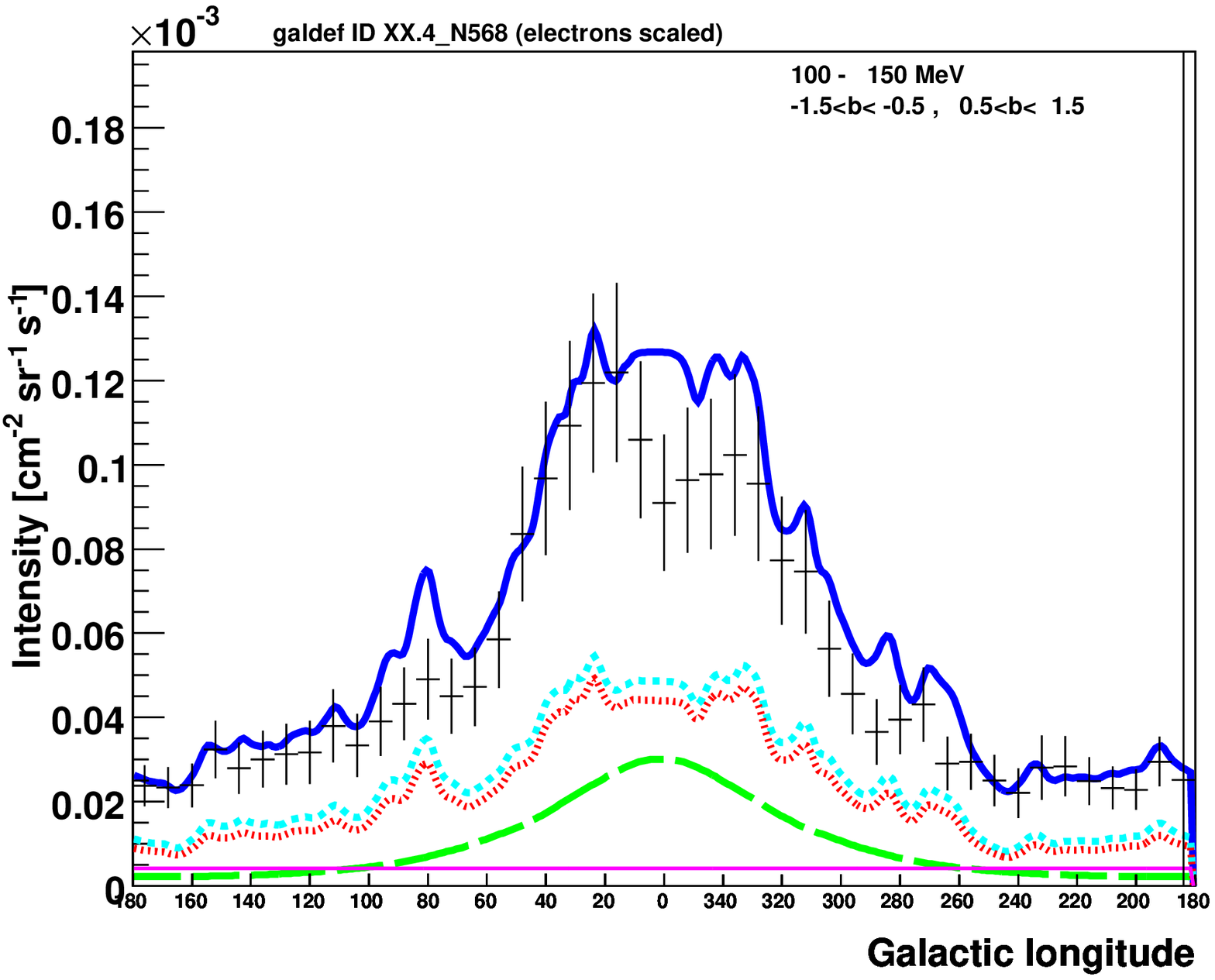}
\includegraphics[width=0.33\textwidth,clip]{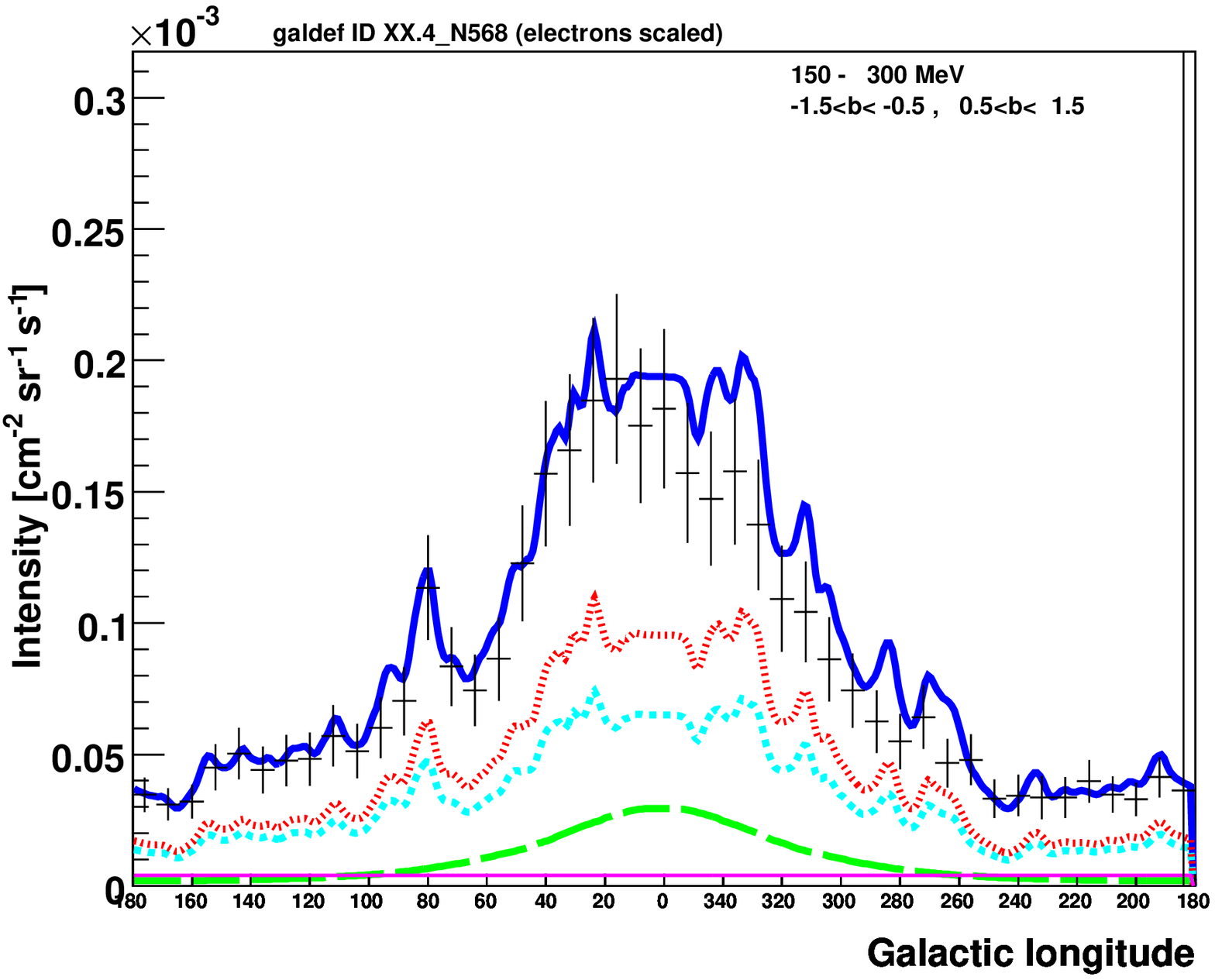}\\

\caption{ Longitude profile for the disk ($1.5 \geq |b|$) for an isotropic model ({\bf top row}) and an aPM ({\bf bottom row}). For both models the Galactic electron density is increased by a factor 1.5 to meet the requirements from $\gamma$-rays at intermediate latitudes.}\label{f_gamma_gradient}
\end{figure*}

The aPM, as well as the isotropic models, require a somewhat increased Galactic electron density. In the isotropic case this leads to a pronounced peak in the Galactic IC emission from the GC, as can be seen from the top row in Fig. \ref{f_gamma_gradient}. As a possible solution a significantly flatter source distribution has been proposed by \citet{Strong:1998pw}, shown in Fig. \ref{f_source_zc} as the black full line, which reduces the contribution from $\pi^0$-decay and bremsstrahlung in the GC. However, this source distribution has to be chosen {\it ad hoc} according to what is expected from $\gamma$-rays. An increase in the $X_{CO}$ scaling factor towards the outer Galaxy is a more likely explanation, but the gradients required by the EGRET data are on the limit of what is expected from the increase in metallicity \citep{Strong:2004td}.\\
As mentioned previously the self-consistent Galactic wind calculations
by \citet{Breitschwerdt:2002vs} predict a softer $\gamma$-ray gradient, because the CR escape time varies with Galactocentric radius depending on the local source strength. Enhanced particle injection by the sources therefore results in enhanced CR escape and thus smoothens the propagated CR distribution. $\gamma$-rays from $\pi_0$-decay predominantly originate from GeV protons. Figure \ref{f_p_distro2} shows the radial distribution for 1-5 GeV protons in an aPM and an isotropic model with $z_h=4~\mathrm{kpc}$, both models use the SNR distribution as the source distribution. 
In the isotropic model the propagated proton distribution still resembles the strong peak of the source distribution, which will lead to problems in the $\gamma$-ray production rate unless a strong increase in $X_{CO}$ is assumed.
Protons in an aPM are significantly flatter than protons in the isotropic model and a constant $X_{CO}$ scaling factor is almost consistent with the observed flat $\gamma$-ray gradient, even if the Galactic electron density is increased to match the IC emission from intermediate latitudes. The bottom row of Fig. \ref{f_gamma_gradient} shows the longitude profile for the Galactic disk in this case. The improvement due to the flatter profile of emission from $\pi^0$-decay and bremsstrahlung is clearly visible, but above 100 MeV a slight excess in emission from $\pi^0$-decay and bremsstrahlung is still visible. A fine-tuned radial dependence of the convection velocity or a rather soft gradient in $X_{CO}$ appears to be compatible with the data.

\subsection{Radio emission in an aPM}\label{ss_radio}
We checked that the electron energy losses via synchrotron radiation are reasonably well described by calculating the radio emission in longitude an latitude and comparison to the 408 MHz data from the \citet{Haslam:1982ha} sky map.

\begin{figure}
\includegraphics[width=0.5\textwidth,clip]{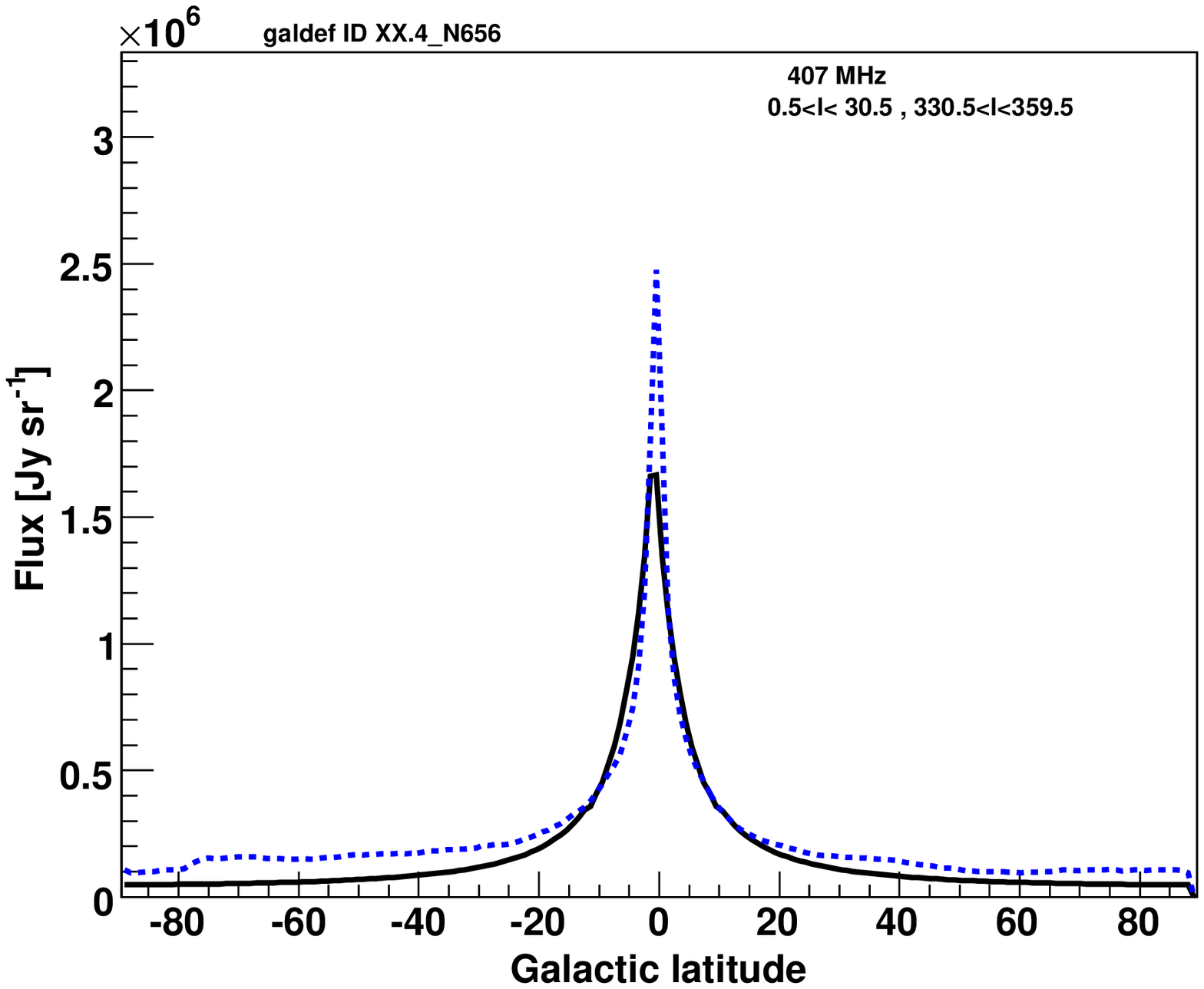}\\
\includegraphics[width=0.5\textwidth,clip]{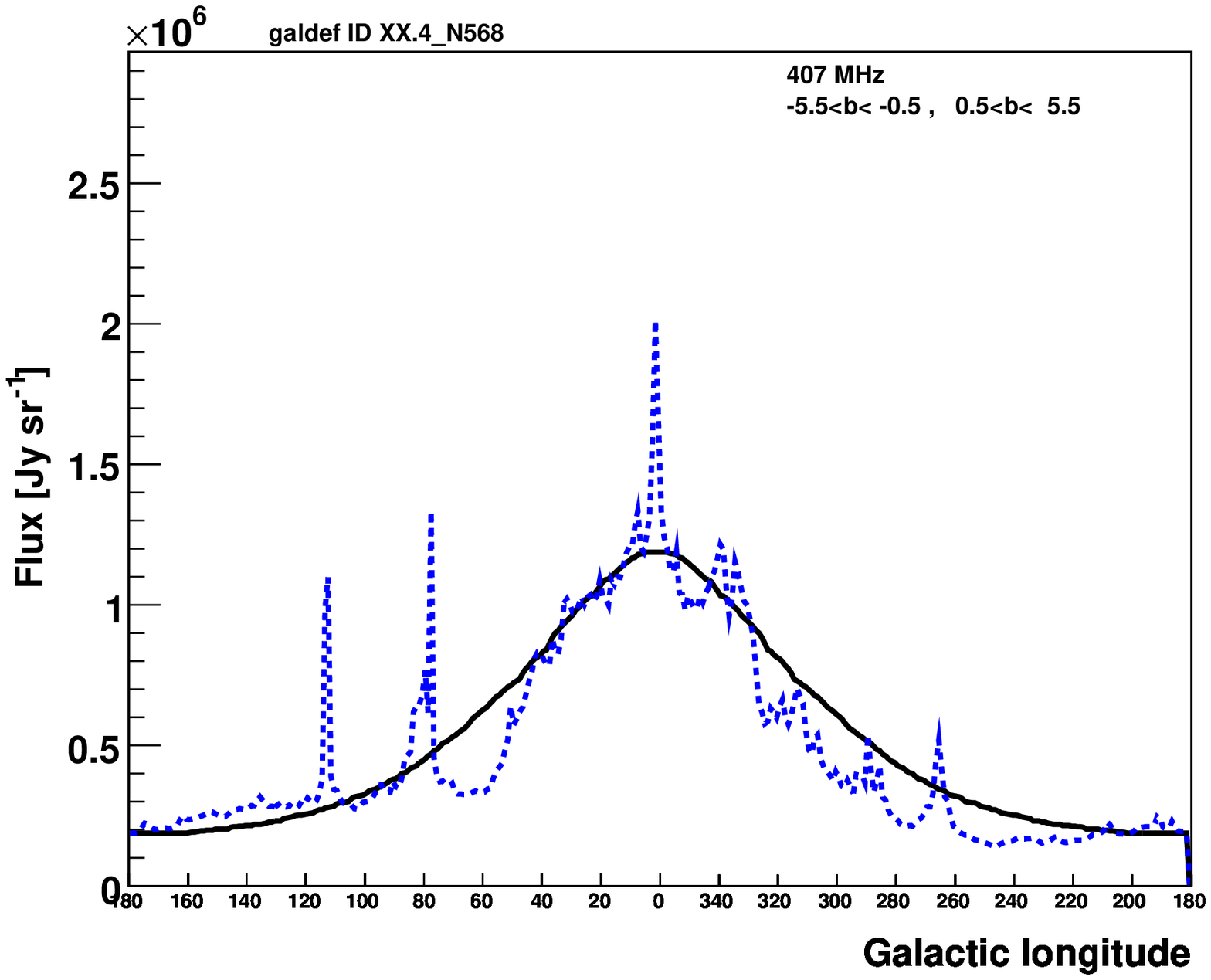}\\
\caption{Synchrotron latitude profile ({\bf top}) and synchrotron longitude profile ({\bf bottom}) at 408 MHz an an aPM. Data (blue dotted): \citet{Haslam:1982ha}.}\label{f_conv_aPM_synchl}
\end{figure}

Following \citet{Moskalenko:1998id} we use the following parameterization of the total regular magnetic field for the calculation of the electron energy losses:
\begin{eqnarray}
B_{reg}(R)= B_{0}exp[\frac{-R-R_{0}}{R_{B}}]exp[-\frac{|z|}{z_B}],\label{e_B}
\end{eqnarray}
with $R_B=10~\mathrm{kpc}$ and $z_B=0.2~\mathrm{kpc}$.
We choose $B_{0}=6.5~\mu\mathrm{G}$ in order to best reproduce the \citet{Haslam:1982ha} all sky map. Figure \ref{f_conv_aPM_synchl} shows the latitude and longitude profile of
synchrotron radiation in an aPM at 408 MHz and the
synchrotron spectrum. 
\\

\begin{figure*}
\includegraphics[width=0.42\textwidth,clip]{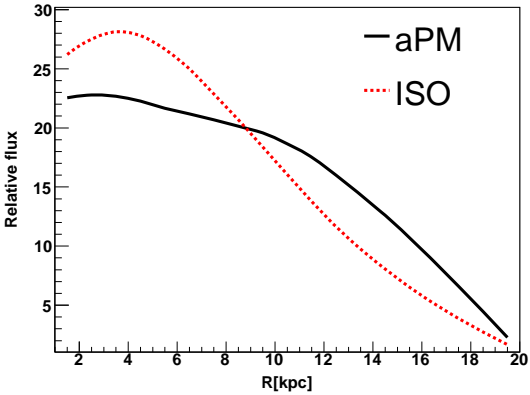}
\caption{Radial distribution of GeV protons for the aPM and an isotropic model. The source distribution is in both cases that of \citet{Case:1998qg}.}
\label{long_d_100}\label{f_p_distro2}
\end{figure*}

\section{Conclusion}\label{s_conc}
A new model for Galactic cosmic ray transport is presented, which allows for significant convective transport
as expected from the Galactic winds deduced from the x-ray data from the ROSAT satellite. The model has been realized by modifying the publicly available GALPROP code, which up to now allowed only isotropic transport.
Isotropic transport models, which feature globally constant transport parameters, can only accommodate negligible convection speeds, since with convection the CRs are driven away from the disc, thus  not returning often enough to the disk to produce secondary particles.

A possible way to allow for large convection is to allow the diffusion in the halo to be different from the diffusion in the disk, i.e. to give up isotropic propagation with globally constant transport parameters.
The GALPROP code was modified in the following way:
\begin{itemize}
\item
the Galactic winds were assumed to be proportional to the CR source distribution, which was taken to be the SNR distribution
\item the mean free path of CRs - and therefore the diffusion coefficient - in the halo was assumed to increase linear with the distance from the disk. Such a choice is motivated  if the scattering centers are produced by the CRs themselves (via the irregular magnetic field components in the plasma) and the density of CRs drops approximately linearly in the halo.
\end{itemize}
Fixing the magnitude of the convection speed to the wind speeds suggested by the ROSAT data, the increase in diffusion coefficient in the halo can be fitted from the amount of secondary production (from B/C ratio) and the residence time of CRs (from the cosmic clocks, in this case the $^{10}Be/^9Be$ ratio).
It is shown that such a  model is consistent with all available CR data, including not only the ROSAT data on
convective winds, but also the large bulge/disk ratio of the positron annihilation line as observed  by the INTEGRAL satellite. In isotropic propagation models one has to introduce a new source of MeV positrons in the bulge, like e.g. dark matter annihilation or tunneling the positrons from the disk to the bulge via regular magnetic fields in the halo. In the anisotropic model presented here the large B/D ratio is naturally explained by the energy independent convective transport of low energy positrons from the disk to the halo, where there are no electrons to annihilate with. Convective transport is absent in the bulge, because the gravitational potential is too strong there to launch Galactic winds.

An additional interesting feature of the present model is the smooth transition to free escape of CRs, because of the increase in mean free path with increasing distances from the disk. Therefore the boundary condition can be moved to infinity in contrast to isotropic propagation models, where the boundary condition is fine-tuned to get the correct residence time of CRs inside the Galaxy.

\begin{acknowledgements}
We thank V. Zhukov for fruitful discussions.
We would also like to thank I.V. Moskalenko and A.W. Strong for sharing their GALPROP routines with the community. Use was made of the CR database provided by A.W. Strong (http://www.mpe.mpg.de/$\sim$aws/propagate.html).
We acknow\-ledge the use of the Legacy Archive for Microwave Background Data Analysis (LAMBDA). Support for LAMBDA is provided by the NASA Office of Space Science. 
\end{acknowledgements}


\appendix

\section{Crank-Nicholson coefficients for R-dependent convection
}\label{a_conv}
The propagation method used in GALPROP can be found in \citet{Strong:1998pw}. For the numerical solution of the transport equation the Crank-Nicholson implicit method is used \citep{Press:1992}.
Following the notation of the GALPROP explanatory supplement \citep{GalpropMan} we find the Crank-Nicholson coefficients for R-dependent convection in z-direction to be
\begin{eqnarray}
\frac{\alpha_1}{\Delta t}=\frac{V(R_j,z_{i-1})}{z_i-z_{i-1}},
\frac{\alpha_2}{\Delta t}=\frac{V(R_j,z_{i})}{z_i-z_{i-1}},
\frac{\alpha_3}{\Delta t}=0
\end{eqnarray}
for $z>0$ and
\begin{eqnarray}
\frac{\alpha_1}{\Delta t}=0,
\frac{\alpha_2}{\Delta t}=\frac{V(R_j,z_{i})}{z_{i+1}-z_{i}};
\frac{\alpha_3}{\Delta t}=\frac{V(R_j,z_{i+1})}{z_{i+1}-z_{i}}
\end{eqnarray}
for $z<0$.\\
For transport in momentum space the coefficients read
\begin{eqnarray}
\frac{\alpha_1}{\Delta t}=0\nonumber
\end{eqnarray}
\begin{eqnarray}
\frac{\alpha_2}{\Delta t}=\frac{V(R_j,z_{i+1})-V(R_j,z_{i})}{z_{i+1}-z_{i}}\frac{p_i}{3(p_{i+1}-p_i)}
\end{eqnarray}
\begin{eqnarray}
\frac{\alpha_3}{\Delta t}=\frac{V(R_j,z_{i+1})-V(R_j,z_{i})}{z_{i+1}-z_{i}}\frac{p_{i+1}}{3(p_{i+1}-p_i)}\nonumber
\end{eqnarray}
for $z>0$ and
\begin{eqnarray}
\frac{\alpha_1}{\Delta t}=0\nonumber
\end{eqnarray}
\begin{eqnarray}
\frac{\alpha_2}{\Delta t}=\frac{V(R_j,z_{i-1})-V(R_j,z_{i})}{z_{i}-z_{i-1}}\frac{p_i}{3(p_{i+1}-p_i)}
\end{eqnarray}
\begin{eqnarray}
\frac{\alpha_3}{\Delta t}=\frac{V(R_j,z_{i-1})-V(R_j,z_{i})}{z_{i}-z_{i-1}}\frac{p_{i+1}}{3(p_{i+1}-p_i)}\nonumber
\end{eqnarray}
for $z<0$.
\section{Crank-Nicholson coefficients for anisotropic diffusion}\label{a_diff}
Following the notation of the GALPROP explanatory supplement \citep{GalpropMan} we find the Crank-Nicholson
coefficients for the R and z dependent diffusion coefficients $D_{RR}(R,z,p)$ and $D_{zz}(R,z,p)$ to be
\begin{eqnarray}
\frac{\alpha_1}{\Delta t}=\frac{D_{RR}(R_i, z_j, p_j)}{(R_{i+1}-R_i)(R_i-R_{i-1})}-\frac{D_{RR}(R_i, z_j, p_j)}{R_i(R_i-R_{i-1})}-\nonumber \\
-\frac{D_{RR}(R_{i+1}, z_j, p_j)-D_{RR}(R_{i-1},z_j,p_j)}{(R_{i+1}-R_{i-1})^2},\nonumber
\end{eqnarray}
\begin{eqnarray}
\frac{\alpha_2}{\Delta t}=
\frac{D_{RR}(R_i, z_j, p_j)}{(R_{i+1}-R_i)(R_{i+1}-R_{i})}+\frac{D_{RR}(R_i, z_j, p_j)}{(R_{i+1}-R_{i})(R_i-R_{i-1})},
\end{eqnarray}
\begin{eqnarray}
\frac{\alpha_3}{\Delta t}=\frac{D_{RR}(R_{i+1}, z_j, p_j)-D_{RR}(R_{i-1}, z_j, p_j)}{(R_{i+1}-R_{i-1})^2}-\nonumber\\
-\frac{D_{RR}(R_i, z_j, p_j)}{R_i(R_{i+1}-R_{i-1})}+\frac{D_{RR}(R_{i+1}, z_j, p_j)}{(R_{i+1}-R_{i})^2} \nonumber
\end{eqnarray}
 for transport in $R$ direction and
\begin{eqnarray}
\frac{\alpha_1}{\Delta t}=\frac{D_{zz}(R_j, z_i, p_j)}{(z_{i+1}-z_i)(z_i-z_{i-1})}-\frac{D_{zz}(R_j, z_{i+1}, p_j)-D_{zz}(R_j,z_{i-1})}{(z_{i+1}-z_{i-1})^2},\nonumber
\end{eqnarray}
\begin{eqnarray}
\frac{\alpha_2}{\Delta t}=\frac{D_{zz}(R_j, z_i, p_j)}{(z_{i+1}-z_i)^2}-\frac{D_{zz}(R_j, z_{i}, p_j)}{(z_{i+1}-z_{i})(z_i-z_{i-1})},
\end{eqnarray}
\begin{eqnarray}
\frac{\alpha_3}{\Delta t}=\frac{D_{zz}(R_j, z_{i+1}, p_j)-D_{zz}(R_j,z_{i-1},p_j)}{(z_{i+1}-z_{i-1})^2}+\frac{D_{zz}(R_j, z_{i}, p_j)}{(z_{i+1}-z_{i})^2}\nonumber
\end{eqnarray}
for transport in z direction.
In the limit of an equidistant grid and constant diffusion our Crank-Nicholson coefficients agree with those used in \citep{GalpropMan} except for a factor 2 in $\alpha_1$ and $\alpha_3$ for transport along $R$. However, deriving these coefficients for the case of an equidistant grid and constant diffusion from their Eq. 25 we find  
 \begin{eqnarray}
\frac{\alpha_1}{\Delta t}=D_{RR}\frac{R_i-\Delta R}{R_i (\Delta R)^2},\nonumber
\end{eqnarray}
and
\begin{eqnarray}
\frac{\alpha_3}{\Delta t}=D_{RR}\frac{R_i+\Delta R}{R_i (\Delta R)^2}\nonumber
\end{eqnarray}
in agreement with our coefficients.
The coefficients for diffusive reacceleration remain unchanged, because here only derivatives in momentum space occur. However, if $D_{zz}>D_{RR}$ one has to keep in mind that $v_{\alpha}$ has to be considered an effective parameter, scaled with the anisotropy in diffusion, e.g. $\bar{v}_{\alpha}=v_{\alpha}\cdot \sqrt{D_{zz}/D_{RR}}$

\bibliographystyle{aa.bst}
\bibliography{arxiv_aPM}
\end{document}